\documentclass{jfm}

\usepackage{amsmath} 
\usepackage{mathtools}
\usepackage{xfrac} 
\usepackage{bm} 
\usepackage{verbatim}
\usepackage{IEEEtrantools} 
	\IEEEeqnarraydefcolsep{0}{\leftmargini}
\usepackage{graphicx} 
\usepackage{caption} 
\usepackage{subcaption} 
\usepackage{placeins} 
\usepackage{epstopdf}
\usepackage{array}
\usepackage[usenames,dvipsnames]{xcolor}

\IEEEeqnarraydefcolsep{0}{\leftmargini}

\newcommand{\mbs}[1]{\boldsymbol{#1}}
\newcommand{\mrm}[1]{\mathrm{#1}}

\makeatletter
\newcommand*{\getlength}[1]{\strip@pt#1}
\makeatother
\newlength{\myFigBoxWidth}
\newlength{\myFigWidth}
\newlength{\myFigHeight}
\newlength{\myFigBoxHeight}
\newlength{\xCoord}
\newlength{\yCoord}
\newlength{\myPictureHeight}
\newlength{\myHalfTextWidth}
\newlength{\myHalfBoxWidth}

\newcommand{\placeTwoSubfigures}[6]{%
\setlength{\unitlength}{1pt}%
\setlength{\myFigBoxWidth}{0.5\textwidth}%
\setlength{\myFigWidth}{\myFigBoxWidth - #4 - #5}%
\setlength{\myFigHeight}{#3\myFigWidth}%
\setlength{\myFigBoxHeight}{\myFigHeight + #6}%
\begin{picture}(\getlength{\textwidth},\getlength{\myFigBoxHeight})%
    \setlength{\xCoord}{0pt + #4}%
    \setlength{\yCoord}{0pt}%
    \put(\getlength{\xCoord}, \getlength{\yCoord}){\includegraphics[width=\myFigWidth]{#1}}%
    \setlength{\xCoord}{\myFigBoxWidth + #4}%
    \put(\getlength{\xCoord}, \getlength{\yCoord}){\includegraphics[width=\myFigWidth]{#2}}%
    \setlength{\xCoord}{0pt}%
    \setlength{\yCoord}{\myFigBoxHeight - \heightof{\Huge{b)}}}%
    \put(\getlength{\xCoord}, \getlength{\yCoord}) {a) }%
    \setlength{\xCoord}{\myFigBoxWidth}%
    \put(\getlength{\xCoord}, \getlength{\yCoord}) {b) }%
\end{picture}%
}

\newcommand{\placeThreeSubfiguresDown}[7]{%
\setlength{\unitlength}{1pt}%
\setlength{\myFigBoxWidth}{0.5\textwidth}%
\setlength{\myFigWidth}{\myFigBoxWidth - #5 - #6}%
\setlength{\myFigHeight}{#4\myFigWidth}%
\setlength{\myFigBoxHeight}{\myFigHeight + #7}%
\setlength{\myPictureHeight}{2\myFigBoxHeight}%
\setlength{\myHalfTextWidth}{0.5\textwidth}%
\setlength{\myHalfBoxWidth}{0.5\myFigBoxWidth}%
\begin{picture}(\getlength{\textwidth},\getlength{\myPictureHeight})%
    \setlength{\xCoord}{\myHalfTextWidth - \myHalfBoxWidth + #5}%
    \setlength{\yCoord}{0pt + \myFigBoxHeight}%
    \put(\getlength{\xCoord}, \getlength{\yCoord}){\includegraphics[width=\myFigWidth]{#1}}%
    \setlength{\xCoord}{0pt + #5}%
    \setlength{\yCoord}{0pt}%
    \put(\getlength{\xCoord}, \getlength{\yCoord}){\includegraphics[width=\myFigWidth]{#2}}%
    \setlength{\xCoord}{\myFigBoxWidth + #5}%

    \put(\getlength{\xCoord}, \getlength{\yCoord}){\includegraphics[width=\myFigWidth]{#3}}%
    \setlength{\xCoord}{\myHalfTextWidth - \myHalfBoxWidth}%
    \setlength{\yCoord}{\myFigBoxHeight + \myFigBoxHeight - \heightof{\Huge{a)}}}%
    \put(\getlength{\xCoord}, \getlength{\yCoord}) {a) }%
    \setlength{\xCoord}{0pt}%
    \setlength{\yCoord}{\myFigBoxHeight - \heightof{\Huge{b)}}}%
    \put(\getlength{\xCoord}, \getlength{\yCoord}) {b) }%
    \setlength{\xCoord}{\myFigBoxWidth}%
    \put(\getlength{\xCoord}, \getlength{\yCoord}) {c) }%
\end{picture}%
}

\newcommand{\placeThreeSubfiguresLine}[8]{%
\setlength{\unitlength}{1pt}%
\setlength{\myFigBoxWidth}{#4}%
\setlength{\myFigWidth}{\myFigBoxWidth - #6 - #7}%
\setlength{\myFigHeight}{#5\myFigWidth}%
\setlength{\myFigBoxHeight}{\myFigHeight + #8}%
\setlength{\myPictureHeight}{3\myFigBoxHeight}%
\setlength{\myHalfTextWidth}{0.5\textwidth}%
\setlength{\myHalfBoxWidth}{0.5\myFigBoxWidth}%
\begin{picture}(\getlength{\textwidth},\getlength{\myPictureHeight})%
    \setlength{\xCoord}{0pt + \myHalfTextWidth - \myHalfBoxWidth + #6}%
    \setlength{\yCoord}{0pt + \myFigBoxHeight + \myFigBoxHeight}%
    \put(\getlength{\xCoord}, \getlength{\yCoord}){\includegraphics[width=\myFigWidth]{#1}}%
    \setlength{\yCoord}{0pt + \myFigBoxHeight}%
    \put(\getlength{\xCoord}, \getlength{\yCoord}){\includegraphics[width=\myFigWidth]{#2}}%
    \setlength{\yCoord}{0pt}%
    \put(\getlength{\xCoord}, \getlength{\yCoord}){\includegraphics[width=\myFigWidth]{#3}}%
    \setlength{\xCoord}{0pt + \myHalfTextWidth - \myHalfBoxWidth}%
    \setlength{\yCoord}{\myFigBoxHeight + \myFigBoxHeight + \myFigBoxHeight - \heightof{\Huge{a)}}}%
    \put(\getlength{\xCoord}, \getlength{\yCoord}) {a) }%
    \setlength{\yCoord}{\myFigBoxHeight + \myFigBoxHeight - \heightof{\Huge{b)}}}%
    \put(\getlength{\xCoord}, \getlength{\yCoord}) {b) }%
    \setlength{\yCoord}{\myFigBoxHeight - \heightof{\Huge{c)}}}%
    \put(\getlength{\xCoord}, \getlength{\yCoord}) {c) }%
\end{picture}%
}

\newcommand{\placeFourSubfigures}[8]{%
\setlength{\unitlength}{1pt}%
\setlength{\myFigBoxWidth}{0.5\textwidth}%
\setlength{\myFigWidth}{\myFigBoxWidth - #6 - #7}%
\setlength{\myFigHeight}{#5\myFigWidth}%
\setlength{\myFigBoxHeight}{\myFigHeight + #8}%
\setlength{\myPictureHeight}{2\myFigBoxHeight}%
\begin{picture}(\getlength{\textwidth},\getlength{\myPictureHeight})%
    \setlength{\xCoord}{0pt + #6}%
    \setlength{\yCoord}{0pt + \myFigBoxHeight}%
    \put(\getlength{\xCoord}, \getlength{\yCoord}){\includegraphics[width=\myFigWidth]{#1}}%
    \setlength{\xCoord}{\myFigBoxWidth + #6}%
    \put(\getlength{\xCoord}, \getlength{\yCoord}){\includegraphics[width=\myFigWidth]{#2}}%
    \setlength{\xCoord}{0pt + #6}%
    \setlength{\yCoord}{0pt}%
    \put(\getlength{\xCoord}, \getlength{\yCoord}){\includegraphics[width=\myFigWidth]{#3}}%
    \setlength{\xCoord}{\myFigBoxWidth + #6}%
    \put(\getlength{\xCoord}, \getlength{\yCoord}){\includegraphics[width=\myFigWidth]{#4}}%
    \setlength{\xCoord}{0pt}%
    \setlength{\yCoord}{\myFigBoxHeight + \myFigBoxHeight - \heightof{\Huge{a)}}}%
    \put(\getlength{\xCoord}, \getlength{\yCoord}) {a) }%
    \setlength{\xCoord}{\myFigBoxWidth}%
    \put(\getlength{\xCoord}, \getlength{\yCoord}) {b) }%
    \setlength{\xCoord}{0pt}%
    \setlength{\yCoord}{\myFigBoxHeight - \heightof{\Huge{c)}}}%
    \put(\getlength{\xCoord}, \getlength{\yCoord}) {c) }%
    \setlength{\xCoord}{\myFigBoxWidth}%
    \put(\getlength{\xCoord}, \getlength{\yCoord}) {d) }%
\end{picture}%
}

\newcommand{\placeSixSubfigures}[6]{%
    \def\myFigA{#1}%
    \def\myFigB{#2}%
    \def\myFigC{#3}%
    \def\myFigD{#4}%
    \def\myFigE{#5}%
    \def\myFigF{#6}%
    \placeSixSubfiguresContinued
}
\newcommand{\placeSixSubfiguresContinued}[4]{%
\setlength{\unitlength}{1pt}%
\setlength{\myFigBoxWidth}{0.5\textwidth}%
\setlength{\myFigWidth}{\myFigBoxWidth - #2 - #3}%
\setlength{\myFigHeight}{#1\myFigWidth}%
\setlength{\myFigBoxHeight}{\myFigHeight + #4}%
\setlength{\myPictureHeight}{3\myFigBoxHeight}%
\begin{picture}(\getlength{\textwidth},\getlength{\myPictureHeight})%
    \setlength{\xCoord}{0pt + #2}%
    \setlength{\yCoord}{0pt + \myFigBoxHeight + \myFigBoxHeight}%
    \put(\getlength{\xCoord}, \getlength{\yCoord}){\includegraphics[width=\myFigWidth]{\myFigA}}%
    \setlength{\xCoord}{\myFigBoxWidth + #2}%
    \put(\getlength{\xCoord}, \getlength{\yCoord}){\includegraphics[width=\myFigWidth]{\myFigB}}%
    \setlength{\xCoord}{0pt + #2}%
    \setlength{\yCoord}{0pt + \myFigBoxHeight}%
    \put(\getlength{\xCoord}, \getlength{\yCoord}){\includegraphics[width=\myFigWidth]{\myFigC}}%
    \setlength{\xCoord}{\myFigBoxWidth + #2}%
    \put(\getlength{\xCoord}, \getlength{\yCoord}){\includegraphics[width=\myFigWidth]{\myFigD}}%
    \setlength{\xCoord}{0pt + #2}%
    \setlength{\yCoord}{0pt}%
    \put(\getlength{\xCoord}, \getlength{\yCoord}){\includegraphics[width=\myFigWidth]{\myFigE}}%
    \setlength{\xCoord}{\myFigBoxWidth + #2}%
    \put(\getlength{\xCoord}, \getlength{\yCoord}){\includegraphics[width=\myFigWidth]{\myFigF}}%
    \setlength{\xCoord}{0pt}%
    \setlength{\yCoord}{\myFigBoxHeight + \myFigBoxHeight + \myFigBoxHeight - \heightof{\Huge{a)}}}%
    \put(\getlength{\xCoord}, \getlength{\yCoord}) {a) }%
    \setlength{\xCoord}{\myFigBoxWidth}%
    \put(\getlength{\xCoord}, \getlength{\yCoord}) {b) }%
    \setlength{\xCoord}{0pt}%
    \setlength{\yCoord}{\myFigBoxHeight + \myFigBoxHeight - \heightof{\Huge{c)}}}%
    \put(\getlength{\xCoord}, \getlength{\yCoord}) {c) }%
    \setlength{\xCoord}{\myFigBoxWidth}%
    \put(\getlength{\xCoord}, \getlength{\yCoord}) {d) }%
    \setlength{\xCoord}{0pt}%
    \setlength{\yCoord}{\myFigBoxHeight - \heightof{\Huge{c)}}}%
    \put(\getlength{\xCoord}, \getlength{\yCoord}) {e) }%
    \setlength{\xCoord}{\myFigBoxWidth}%
    \put(\getlength{\xCoord}, \getlength{\yCoord}) {f) }%
\end{picture}%
}

\shorttitle{Momentum balance of particle-laden flows}
\shortauthor{E. Biegert, B. Vowinckel and E. Meiburg}

\title{Momentum balance of a laminar flow over a bed of particles}

\author{E. Biegert\aff{1}
  \corresp{\email{ebiegert@engineering.ucsb.edu}},
  B. Vowinckel\aff{1}
 \and E. Meiburg\aff{1}}

\affiliation{\aff{1}Department of Mechanical Engineering, University of California Santa Barbara,
Santa Barbara, CA 93106, USA
}

\begin{document}

\maketitle





\begin{abstract}
We develop a framework for analyzing the momentum balance of laminar particle-laden flows based on immersed boundary methods, which solve the Navier-Stokes equations and resolve the particle surfaces. This framework differs from previous studies by explicitly accounting for the fluid inside the particles, which is a by-product of the immersed boundary method, allowing us to close the momentum balance for the flow around a single rolling sphere.  We then compute a momentum balance of a laminar Poiseuille flow over a dense bed of particles, finding that the stresses remain in equilibrium even during unsteady flow conditions.  While previous studies have focused on stresses for the streamwise momentum balance, the present approach also allows us to evaluate stress balances in the vertical direction, which are necessary to understand the role that collisions and hydrodynamic drag play during dilation and contraction of particle beds.  While our analysis accounts for the fluid and particle phases separately, we attempt to establish a momentum balance for the fluid/particle mixture, but find that it does not completely close locally due to collision stresses not being resolved across the particle diameter.  However, we find a correlation between the local shear rate and the gap in the mixture balance, which can potentially be used to close the balance for the mixture.
\end{abstract}


\section{Introduction}

Understanding and predicting the behavior of a granular sediment bed exposed to a shear flow is essential for a number of applications in chemical and environmental engineering. Apart from the more obvious but nevertheless difficult task of predicting sediment transport rates \citep[e.g.][]{Bathurst2007, Frey2011, Lajeunesse2010}, it is believed that the nonlinear response of the sediment to the forces exerted by the fluid can result in the sudden mobilization of the entire sediment bed and trigger disastrous mudslides events \citep{Prancevic2018,Takahashi1978}, or enhance the propagation speed of turbidity currents as the shear stress leads to the erosion of particles, which further enhances the density difference between the current and the clear-water ambient \citep{Meiburg2010}.

The importance of these issues has prompted a number experimental investigations of sediment exposed to different flow types. \cite{Houssais2016} conducted studies of a laminar linear Couette flow, \cite{Aussillous2013} presented results for a laminar pressure-driven flow, \cite{Capart2011} provided benchmark data for intense bed-load in turbulent open-channel flow, and \cite{Revil-Baudard2015} studied sheet flows in which entire sediment layers are mobilized. These studies provide valuable insight into the the bulk behavior of the fluid-sediment mixture, although experimental limitations make it difficult to obtain information on continuous particle trajectories, and to measure the time-resolved individual forces acting on the particles. For the purpose of developing continuum-type constitutive models, however, it is highly desirable to obtain such information. Perhaps the two most popular approaches in this regard are the $\mu(I)$-rheology for viscous flows \citep{Cassar2005}, which has been calibrated for neutrally-buoyant spheres in a pressure-controlled rheometer \citep{Boyer2011}, and the kinetic theory for turbulent flows \citep{Hsu2004}, which has been tested against steady-state sheet flow experiments. The situation has been less clear for bed-load transport and bed-morphology evolution that is decoupled from the fluid time-scales, an issue that has recently been addressed by studies employing the Double-Averaging Methodology \citep[DAM;][]{Nikora2013}.

To test, validate, and enhance these frameworks, highly resolved data are needed with a degree of detail that is difficult to obtain experimentally. A starting point of a rigorous analysis should be the full description of the momentum balance and the resulting stress budget of the fluid-particle mixture. This, however, has proved to be challenging task due to the nontrivial coupling of the continuous fluid phase on the one hand, and the disperse particle-phase on the other \citep{Ouriemi2009a}. Nevertheless, this analysis will be crucial for measuring the effective granular stress required to characterize the rheology of the sediment bed. Recently, several numerical studies of particle-resolving Direct Numerical Simulations (DNS) based on the Immersed Boundary Method (IBM) have been carried out that couple the two phases and obtain the stresses within particle-laden flows in various ways. However, these studies were not specifically designed to decompose the stress budget into its different components, nor have they been employed to formulate new, or compare against existing constitutive models. For example, \cite{Kidanemariam2013} included a stress balance for turbulent particle-resolved flows to justify a statistical steady-state by evaluating average velocity profiles. \cite{Picano2015} considered a momentum balance for the shear stress of a turbulent flow laden with neutrally-buoyant particles. They employed the stress balance developed by \cite{Zhang2010}, which is based on averaging volumes containing many particles. Due to the neutrally-buoyant particles, however, a sediment layer did not form. \cite{Vowinckel2017a, Vowinckel2017b} developed momentum balances for double-averaged turbulent flows over granular beds, which also require averaging volumes containing many particles. These latter two studies did not analyze the interfacial stresses coupling the fluid stress to the granular stress.

The present study addresses this issue in detail. We develop a momentum balance for laminar flows, whose terms can be computed in a straightforward manner. We begin by considering the simple scenario of a single particle in a shear flow, before moving on to the more complex situation of a thick sediment bed consisting of thousands of particles that is fully or partially in motion. We develop a framework that will allow us to carefully analyze the components contributing to the stress balance of the fluid and the particle. We apply our analysis to the data generated by grain-resolving DNS using the IBM \citep{Biegert2017a,Biegert2017b}. After validating the concept for the single-grain case, we compute the stress budget for a computational scenario that is very similar to the experimental setup of \cite{Aussillous2013}. Preliminary comparisons of our simulation results with experimental data for identical flow rates were presented in \cite{Biegert2017a}, so that we can here cover a wider parameter range by systematically varying the flow rate.

The paper is structured as follows. We briefly review our numerical technique in section~\ref{sec:methods} and describe the computational setups in section~\ref{sec:setup}. Subsequently, we present the derivation of the stress budgets for the fluid and the particle phase in section~\ref{sec:balance}. Finally, results are presented for the single-particle case as well as for the entire sediment bed with complex rheology in section~\ref{sec:results}.

\section{Equations of motion and methods} \label{sec:methods}

The particle-laden flows of interest require us to solve the Navier-Stokes equation
\begin{equation} \label{eq:NS}
\rho_f \left(\frac{\partial{\mbs{u}}}{\partial{t}}+\nabla\cdot(\mbs{u}\mbs{u})\right) = \nabla \cdot \mbs{\tau} + \mbs{f}_b + \mbs{f}_\mathit{IBM} ,
\end{equation}
where $\mbs{u}$ denotes the fluid velocity, $t$ is time, and $\rho_f$ indicates the fluid density. The fluid stress tensor is given by $\mbs{\tau} = -p \mbs{I} + \mu_f (\nabla \mbs{u} + (\nabla \mbs{u})^T)$, where $p$ represents the pressure with the hydrostatic component subtracted out, $\mbs{I}$ is the identity matrix, and $\mu_f$ denotes the dynamic viscosity of the fluid. The right-hand side includes the volume forces $\mbs{f}_b$ and $\mbs{f}_\mathit{IBM}$, the former a source term used to create the pressure gradient driving the flow and the latter an immersed boundary force used to enforce the no-slip condition on the particle surface. We remark that IBMs solve \eqref{eq:NS} everywhere in the domain, including inside the particles, so that they effectively assume that the particles are filled with fluid.  This fluid within the particles represents a technicality of the IBM and does not have a physical significance; the effect on the fluid surrounding the particles is the same as if the particles were solid.  However, further below we will discuss the importance of this technicality for determining the forces that the fluid and particle phases exert on each other.

We solve for the particle translational velocity, $\mbs{u}_p$,
\begin{equation} \label{eq:p_translational}
m_p \frac{\text{d}\mbs{u}_p}{\text{d} t} = \int\limits_{\Gamma^p} \mbs{\tau}^+ \cdot \mbs{n}^-\, {\mrm{d}A}
+ \int\limits_{\Omega^-} \mbs{f}_b \,{\mrm{d}V}
+ V_p( \rho_p-\rho_f ) \mbs{g} + \mbs{F}_{c,p} ,
\end{equation}
and angular velocity, $\mbs{\omega}_p$,
\begin{equation} \label{eq:p_rotational}
I_p \frac{ \mrm{d}\mbs{\omega}_p}{\mrm{d} t} = \int\limits_{\Gamma^p} \mbs{r}\times(\mbs{\tau}^+\cdot\mbs{n}^-)\,{\mrm{d}A} + \mbs{T}_{c,p} ,
\end{equation}
where $m_p$ is the particle mass, $I_p$ the particle moment of inertia, $V_p$ the particle volume, $\rho_p$ the particle density, and $\mbs{g}$ the gravitational acceleration. The fluid acts on the particles through the hydrodynamic stress tensor $\mbs{\tau}^+$, where $\mbs{r}$ represents the vector from the particle center to a point on the surface $\Gamma^p$, and $\mbs{n}^-$ is the unit normal vector pointing outwards from that point.  The body force, $\mbs{f}_b$, also acts on the particle volume, denoted by $\Omega^-$.  The net force and torque acting on the particle center of mass due to collisions are given by $\mbs{F}_{c,p}$ and $\mbs{T}_{c,p}$, respectively.
The collision model we implement includes a normal contact force to prevent particles from overlapping, a tangential contact force to account for friction, and a lubrication force to account for subgrid hydrodynamic forces.

We solve the equations of motion for the fluid and particles on a cubic finite difference mesh ($h = \Delta x = \Delta y = \Delta z$) using our code described in \citet{Biegert2017a}, which was validated against experiments involving settling spheres, dry and immersed particle-wall collisions, and Poiseuille flows over particle beds. The numerical treatment is based on the IBM of \citet{Uhlmann2005} and the particle-fluid coupling of \citet{Kempe2012a}, which is stable for a larger range of particle/fluid density ratios. In the present work we explicitly introduce into the equations of motion $\mbs{f}_b$, the body force acting on the fluid that acts as a source term for generating a pressure gradient.  Note that this term is included for both the fluid momentum, \eqref{eq:NS}, and the particle momentum, \eqref{eq:p_translational}. We evaluate the collision forces and torques according to \citet{Biegert2017a}, where we combined and modified existing collision models. The resulting collision model involves normal contact forces, $\mbs{F}_{n,p}$, frictional contact forces, $\mbs{F}_{t,p}$, and lubrication forces, $\mbs{F}_{l,p}$, to provide the total collision force
\begin{equation} \label{eq:p_collision}
\mbs{F}_{c,p} = \mbs{F}_{n,p} + \mbs{F}_{t,p} + \mbs{F}_{l,p} + \mbs{F}_{f,p} ,
\end{equation}
where the respective forces account for the collective collisions with all other particles, e.g.
\begin{equation}
\mbs{F}_{n,p} = \sum_{q,q\neq p}^{N_p} \mbs{F}_{n,pq} ,
\end{equation}
where $\mbs{F}_{n,pq}$ is the normal contact force acting on particle $p$ from particle $q$. As will be described below, in some of our simulations we employ fixed particles acting as a rough lower wall, cf. section~\ref{sec:bed_mom}. This is accomplished by the ``fixed particle force" $\mbs{F}_{f,p}$ required to hold a fixed particle $p$ in place, which is equal and opposite to the hydrodynamic and other collision forces acting on the fixed particle.

\begin{figure}
\centering
\includegraphics[width=0.5\textwidth]{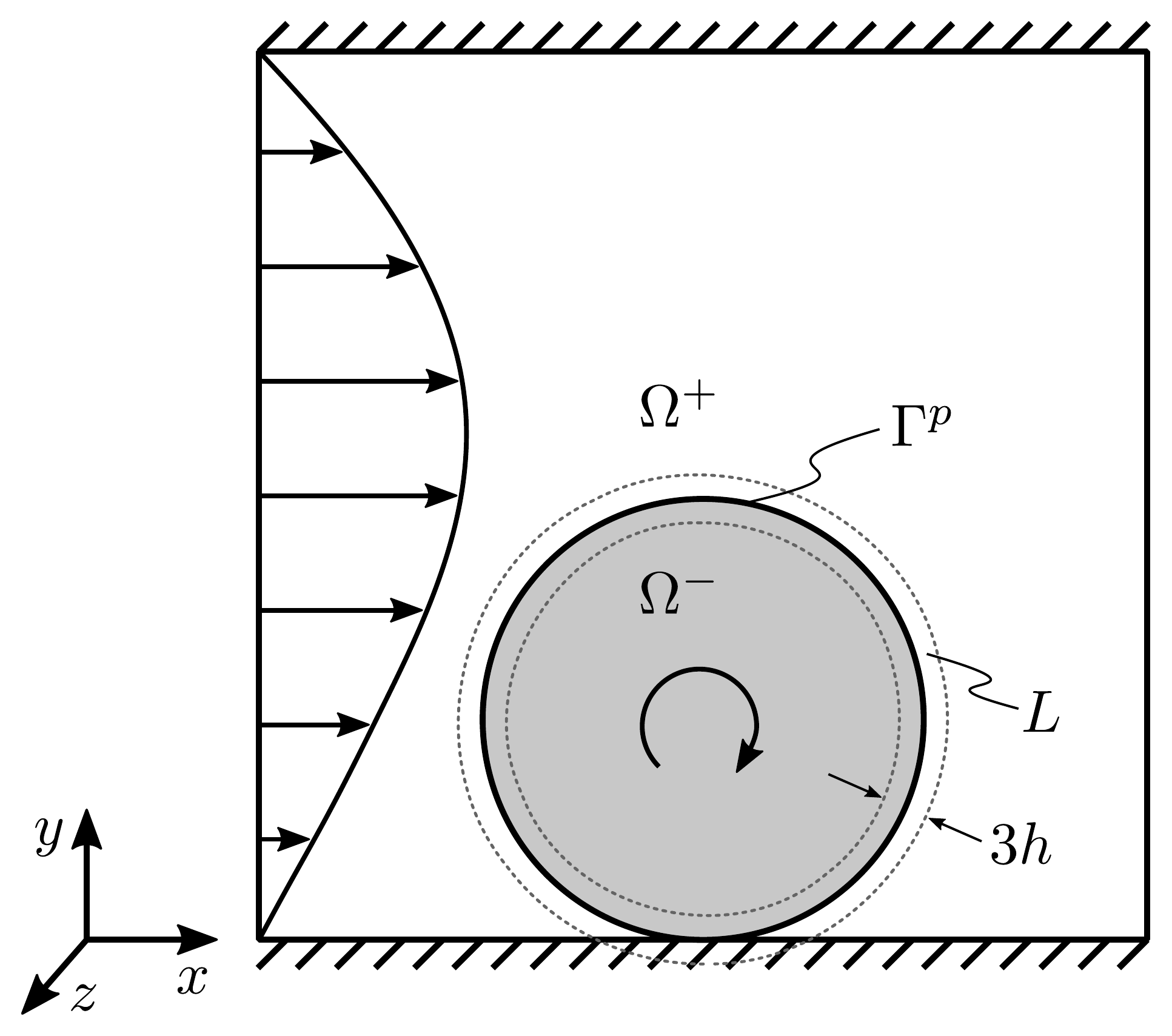}
\caption{Setup for analyzing the stress balance for a single particle rolling in a pressure-driven Poiseuille flow.  Regions of interest are $\Omega^+$, the fluid region outside the particle, $\Omega^-$, the fluid region inside the particle, $\Gamma^p$, the surface interface between the two fluid regions, and $L$, the fluid region surrounding $\Gamma^p$ having thickness $3h$.  Region $L$ represents the volume over which the IBM delta function acts, where $h$ is the fluid grid spacing.}  \label{fig:setup}
\end{figure}

The hydrodynamic force in \eqref{eq:p_translational} and torque \eqref{eq:p_rotational} can be difficult to evaluate accurately, so we implement the following procedure based on the work of \citet{Tschisgale2017}. The fluid domains outside ($\Omega^+$) and inside ($\Omega^-$) of the particles are separated by the particle interface, $\Gamma^p$, as shown in figure~\ref{fig:setup}.  The immersed boundary force leads to a jump condition between the fluid stresses inside, $\mbs{\tau}^-$, and outside, $\mbs{\tau}^+$, the particle
\begin{equation} \label{eq:stress_jump}
-\int\limits_{L} \mbs{f}_\mathit{IBM}\, {\mrm{d}V} = \int\limits_{\Gamma^p} \mbs{\tau}^+ \cdot \mbs{n}^-\, {\mrm{d}A} - \int\limits_{\Gamma^p} \mbs{\tau}^- \cdot \mbs{n}^-\, {\mrm{d}A} ,
\end{equation}
where $L$ denotes the shell volume surrounding the particle surface, whose thickness is determined by the width of the Dirac delta function used for the IBM. The Navier-Stokes equation \eqref{eq:NS} governs the motion of the fluid inside the particles, whose integral form can be written as
\begin{equation} \label{eq:mat_deriv}
\int\limits_{\Gamma^p} \mbs{\tau}^- \cdot \mbs{n}^-\, {\mrm{d}A} = \frac{\mrm{d}}{\mrm{d}t} \int\limits_{\Omega^-} \rho_f \mbs{u} \,{\mrm{d}V} - \int\limits_{\Omega^-} \mbs{f}_b \,{\mrm{d}V} ,
\end{equation}
where we do not include $\mbs{f}_\mathit{IBM}$, which only acts at the fluid/particle interface.  Thus, using \eqref{eq:stress_jump} and \eqref{eq:mat_deriv}, \eqref{eq:p_translational} becomes
\begin{equation} \label{eq:p_translational_final}
m_p \frac{\text{d}\mbs{u}_p}{\text{d} t} = \underbrace{\frac{\mrm{d}}{\mrm{d}t} \int\limits_{\Omega^-} \rho_f \mbs{u} \,{\mrm{d}V}}_{\mbs{F}_{I,p}}
- \underbrace{\int\limits_{L} \mbs{f}_\mathit{IBM}\, {\mrm{d}V}}_{\mbs{F}_{\mathit{IBM},p}}
+ \underbrace{V_p( \rho_p-\rho_f ) \mbs{g}}_{\mbs{F}_{g,p}} + \mbs{F}_{c,p} ,
\end{equation}
and \eqref{eq:p_rotational} becomes
\begin{equation} \label{eq:p_rotational_final}
I_p \frac{ \mrm{d}\mbs{\omega}_p}{\mrm{d} t} = \underbrace{\frac{\mrm{d}}{\mrm{d}t} \int\limits_{\Omega^-} \rho_f \mbs{r} \times \mbs{u} \,{\mrm{d}V}}_{\mbs{T}_{I,p}}
- \underbrace{\int\limits_{L} \mbs{r} \times \mbs{f}_\mathit{IBM}\, {\mrm{d}V}}_{\mbs{T}_{\mathit{IBM},p}} + \mbs{T}_{c,p} .
\end{equation}
Here we define $\mbs{F}_{I,p}$ and $\mbs{T}_{I,p}$ to be the inertial force and torque, $\mbs{F}_{\mathit{IBM},p}$ and $\mbs{T}_{\mathit{IBM},p}$ the IBM force and torque, and $\mbs{F}_{g,p}$ the buoyancy force acting on particle $p$. \citet{Kempe2012a} demonstrate the importance of $\mbs{F}_{I,p}$ and $\mbs{T}_{I,p}$ for capturing transient particle motions, i.e. that $\mbs{F}_{\mathit{IBM},p}$ and $\mbs{T}_{\mathit{IBM},p}$ alone do not account for the full effects of the IBM acting on the particles. Note that the body force $\mbs{f}_b$ drops out of the particle momentum equations \eqref{eq:p_translational_final} and \eqref{eq:p_rotational_final}. Thus, forcing the fluid inside the particles with $\mbs{f}_b$ implicitly accounts for the effects of this body force on the particles through $\mbs{f}_\mathit{IBM}$.


\section{Simulation setup} \label{sec:setup}

We will apply our stress balance framework to two different configurations.  The first one will involve the flow around a single rolling sphere, where we can obtain a true steady-state flow.  This simple case will thus test the framework under ideal conditions and illustrate its ability to resolve stresses on the particle scale.  The second configuration will be a more realistic case involving the flow over a bed of thousands of particles.  This complex case will illustrate the type of information this framework can provide for general particle-laden flows and how it can be useful for future studies.

\subsection{Single rolling particle} \label{sec:single_setup}

\begin{figure}
\placeTwoSubfigures{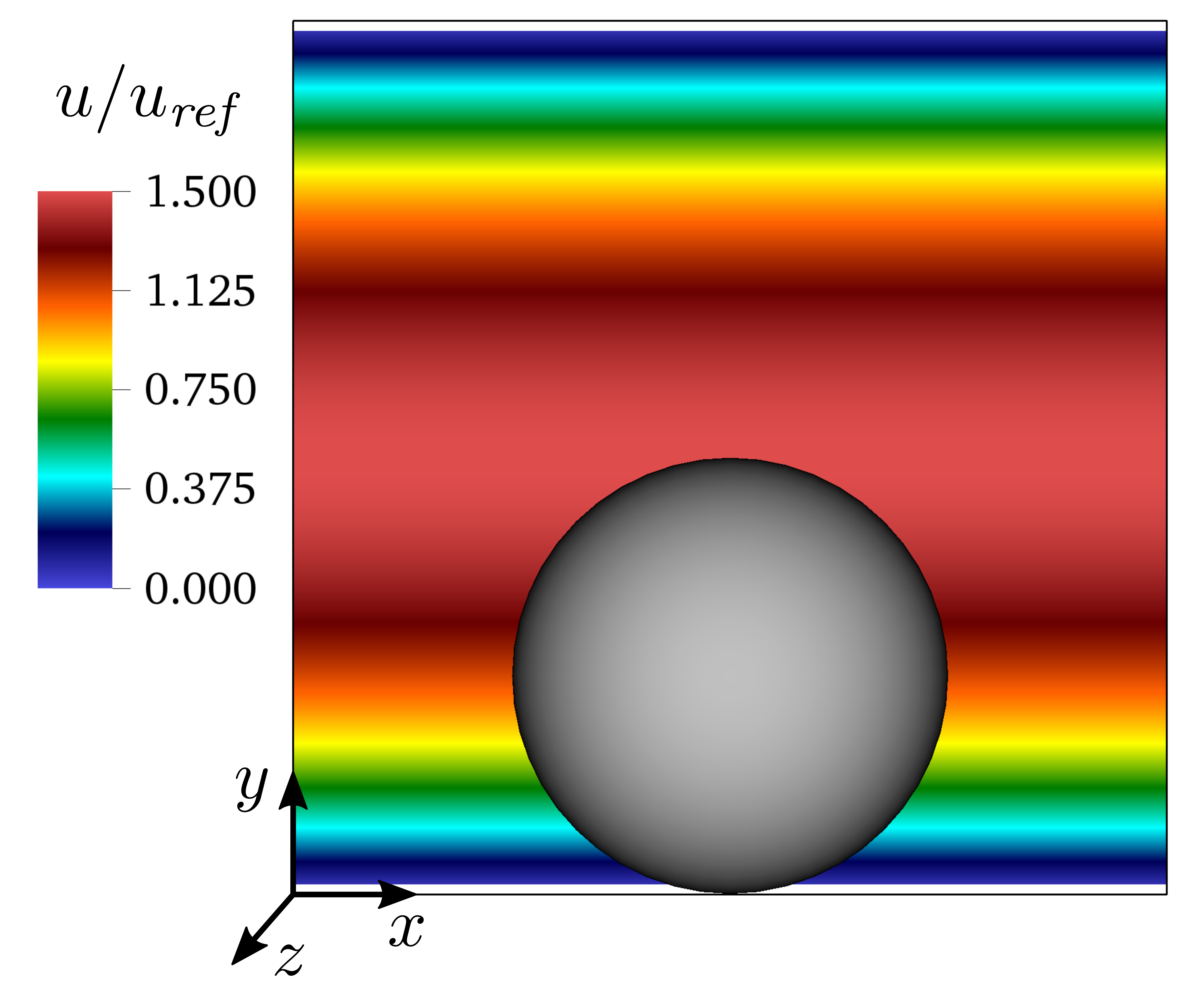}{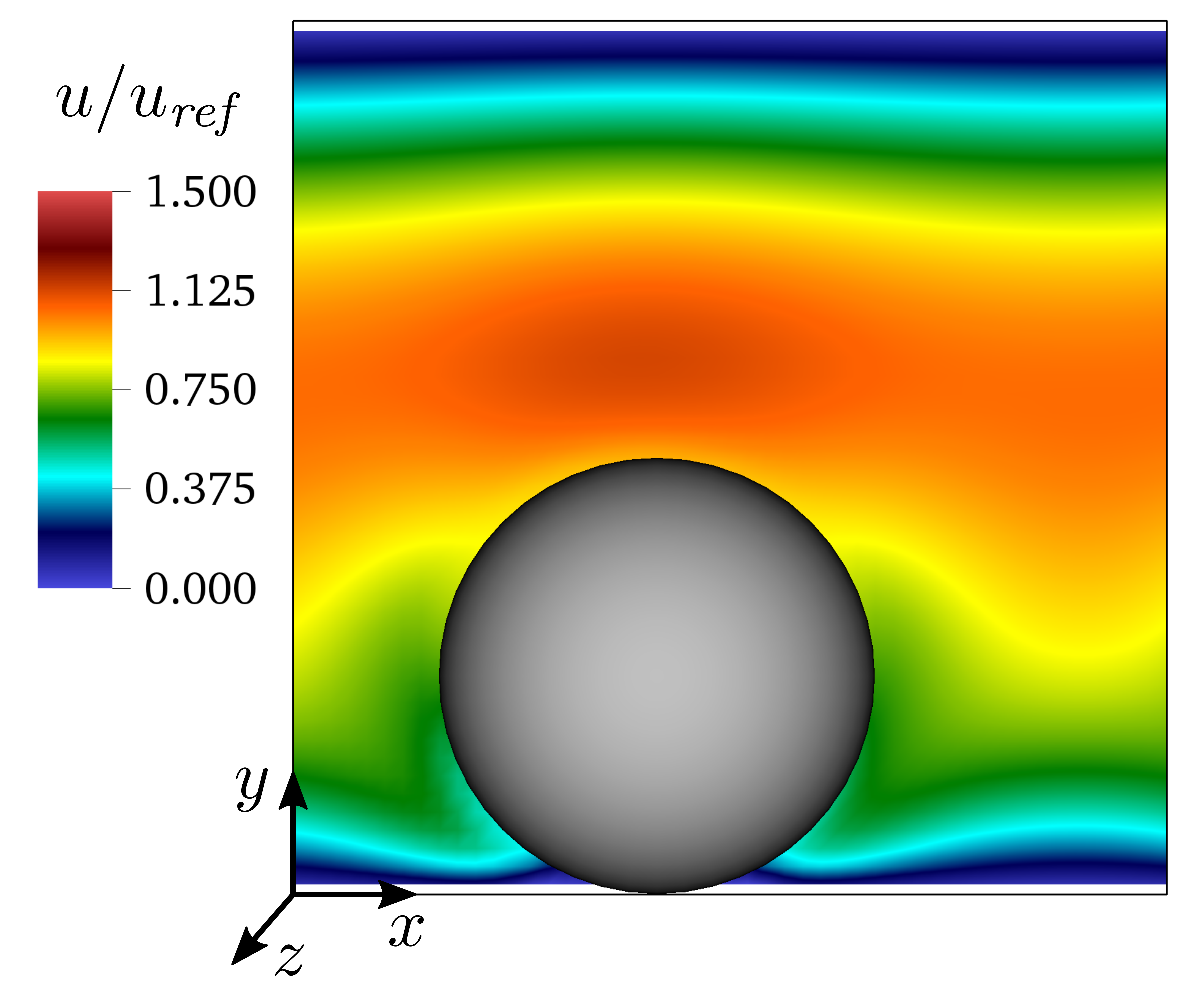}{0.8247}{5pt}{5pt}{10pt}
\caption{Fluid flow field along the particle center plane ($z/L_z=0.5$) at (a) $t/t_\mathit{ref}=0$ and (b) $t/t_\mathit{ref}=4$ for the simulation of a single rolling particle.  Pseudocolor indicates fluid velocity in the $x$-direction.}
\label{fig:single_start_end}
\end{figure}

\begin{table}
\centering
\begin{tabular}{ll}
$Ga$				& 8.29	\\
$Re_\mathit{ref}$	& 10	\\
$\rho_p / \rho_f$	& 2.1	\\
Timestep	& $\mrm{CFL}=0.5$ \\
Domain size ($L_x/D_p \times L_y/D_p \times L_z/D_p$)
	& $2 \times 2 \times 2$	\\
Domain grid size ($L_x/h \times L_y/h \times L_z/h$)
	& $48 \times 48 \times 48$	\\
Domain boundary conditions
	& p $\times$ ns $\times$ p	\\
Particle resolution, $D_p/h$	& 24	\\
Coarse-graining grid size, $h^{cg}/h$	& 1	\\
Coarse-graining width, $w/h$	& 16
\end{tabular}
\caption{Simulation parameters for a single rolling particle.  Boundary conditions are periodic (p) and no-slip (ns).  Coarse-graining parameters are defined in Appendix~\ref{sec:cg}.}
\label{tab:single_parameters}
\end{table}

One key feature of our analysis is that it should work just as well for a single particle as for a large number of particles.  As a simple case, we consider a single sphere rolling along the bottom of a channel with a pressure-driven flow.
In the absence of the particle, there would be a laminar Poiseuille flow with a bulk (average) velocity of
$u_\mathit{ref} = -L_y^2 / (12 \mu_f) f_{b,x}$
and a Reynolds number of
$Re_\mathit{ref} = \rho_f u_\mathit{ref} L_y/\mu_f = 10$,
where $L_y$ is the channel height.
The presence of the particle, however, changes the bulk velocity and Reynolds number.  The domain size is two particle diameters in each of the $x$-, $y$-, and $z$-directions, discretized with 24 grid cells per particle diameter, so that the particle has a significant influence on the flow field.  We provide the other parameters associated with this simulation in table~\ref{tab:single_parameters}, where we define the Galileo number to be $Ga = \rho_f \sqrt{(\rho_p / \rho_f - 1) g D_p^3} / \mu_f$.  The collision parameters $\zeta_\mathit{n,min}$, $e_\mathit{dry}$, $\mu_k$, and $\mu_s$, which are not listed in table~\ref{tab:single_parameters}, are the same as those in \citet{Biegert2017a}.

We initialize the velocity field with the reference Poiseuille parabolic profile, shown in figure~\ref{fig:single_start_end}a. The particle starts with a translational and rotational velocity obtained from averaging the initial flow field within its volume. At this low Reynolds number, the flow remains laminar, but takes time to develop because the presence of the particle constricts and slows the flow.  We run the simulation until time $t = 4 t_\mathit{ref}$, where $t_\mathit{ref} = L_y / u_\mathit{ref}$, by which the particle, rolling along the lower wall, had slowed to a constant streamwise velocity, as shown in figure~\ref{fig:single_start_end}b. We therefore consider the flow to be in a steady state from the reference frame of the particle. We remark that the simulation is carried out in the laboratory reference frame. Furthermore, due to the assumption of periodic boundary conditions in the streamwise and spanwise directions, the simulation effectively considers a periodic array of rolling spheres.

\subsection{Sheared bed of particles} \label{sec:bed_setup}

\begin{figure}
\placeTwoSubfigures{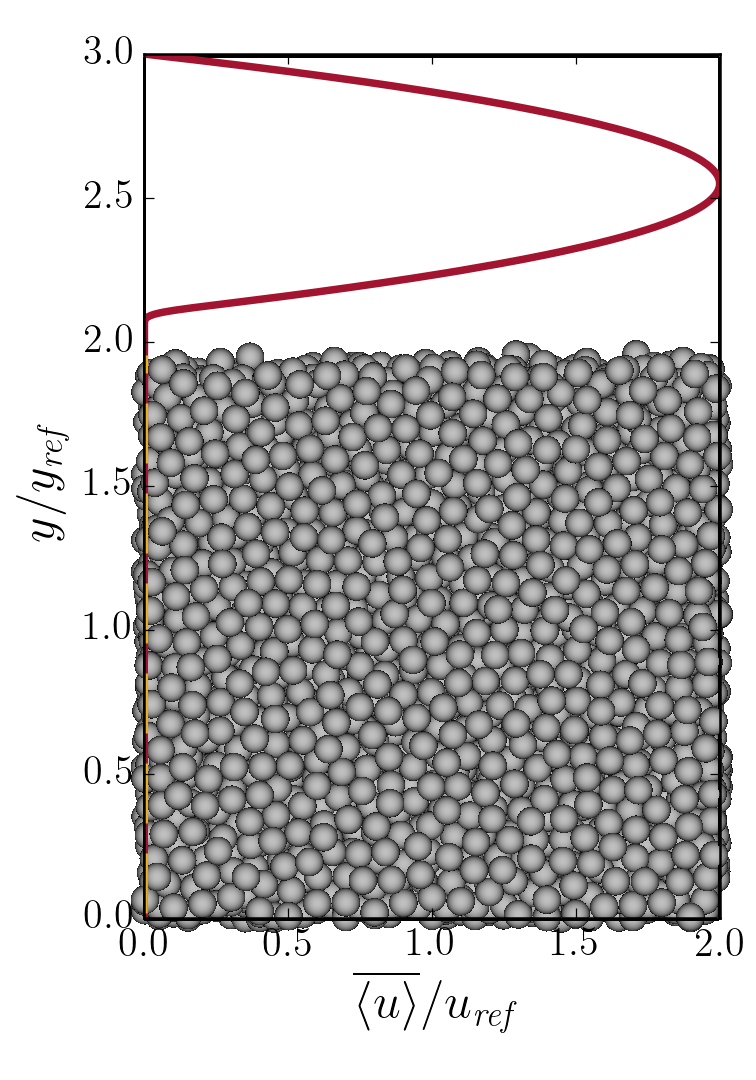}{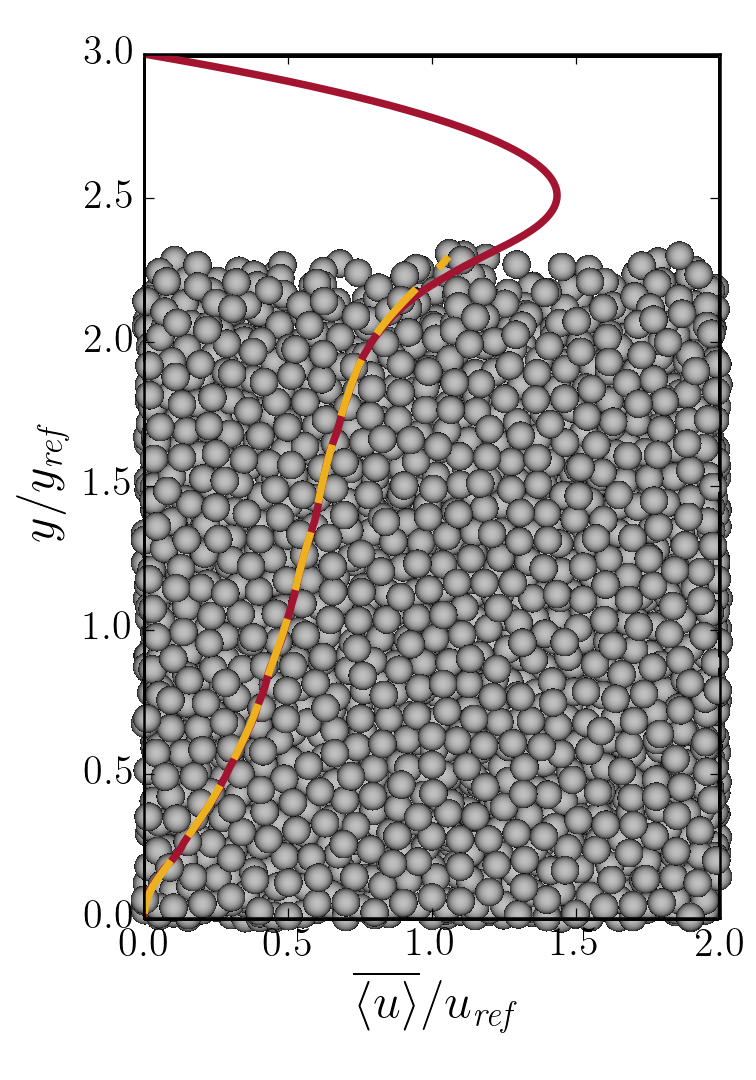}{1.4286}{0pt}{10pt}{0pt}
\caption{Bed configuration at (a) $t/t_\mathit{base}=0$ and (b) $t/t_\mathit{base}=10$ for simulation Re67 listed in table~\ref{tab:bed_runs}.  Lines indicate the average streamwise ($x$-direction) fluid velocity (dark gray, red online) and particle velocity (light gray, yellow online), where the horizontal average is defined by \eqref{eq:horizontal_average} and the time average is defined by \eqref{eq:time_average}.}
\label{fig:bed_start_end}
\end{figure}

\begin{table}
\centering
\begin{tabular}{ll}
$Ga$				& 0.850	\\
$\rho_p / \rho_f$	& 2.1	\\
Timestep		& $\mrm{CFL}=0.5$ \\
Domain size ($L_x/D_p \times L_y/D_p \times L_z/D_p$)
	& $20 \times 30 \times 10$	\\
Domain grid size ($L_x/h \times L_y/h \times L_z/h$)
	& $512 \times 768 \times 256$	\\
Domain boundary conditions
	& p $\times$ ns $\times$ p	\\
Initial $h_f / D_p$	& 10.0	\\
Particle resolution, $D_p/h$	& 25.6	\\
Coarse-graining grid size, $h^{cg}/h$	& 8	\\
Coarse-graining width, $w/h$	& 24
\end{tabular}
\caption{Simulation parameters for the pressure-driven flow over a bed of particles.}
\label{tab:bed_parameters}
\end{table}

\begin{table}
\centering
\begin{tabular}{lccc}
Simulation run	& $Re_\mathit{ref}$	& $t_\mathit{sim}/t_\mathit{base}$	& $t_\mathit{avg}/t_\mathit{base}$ \\
Re67	& 66.7	& $[0.00,10.00]$	& $-$ \\
Re17	& 16.7	& $[10.00,47.20]$	& $[16.00,47.20]$ \\
Re33	& 33.3	& $[10.00,58.80]$	& $[44.00,52.15]$ \\
Re8	& 8.33	& $[47.20,92.05]$	& $[77.00,92.05]$
\end{tabular}
\caption{Simulation parameters for different runs of the pressure-driven flow over a bed of particles. The Reynolds number is based on the reference case, $Re_\mathit{ref} = \rho_f u_\mathit{ref} y_\mathit{ref} / \mu_f$.  The individual simulation is run for the duration $t_\mathit{sim}$, and the momentum balance is analyzed by using time-averaged data over the interval $t_\mathit{avg}$.}
\label{tab:bed_runs}
\end{table}

We are ultimately interested in understanding flows involving many (thousands or more) particles.  In the present work, we consider a setup very similar to the one in \citet{Biegert2017a}, which involves a pressure-driven flow over a bed of particles. The domain has dimensions $20D_p \times 30D_p \times 10D_p$ and is discretized with 25.6 grid cells per particle diameter. We generate the bed by allowing 4,339 monodisperse particles to settle under gravity, without the influence of the surrounding fluid, onto a layer of 200 fixed particles whose centers randomly vary in height above the bottom wall within a range of $D_p$, providing an irregular roughness \citep{Jain2017}. The resulting bed fills the domain to about a height of $h_p \approx 20D_p$ from the bottom wall, where $h_p$ is the particle bed height, leaving a gap of about $10D_p$ between the top wall and the top of the particle bed, as shown in figure~\ref{fig:bed_start_end}a.

We again employ a predefined Poiseuille flow in the clear fluid region above the sediment bed as a reference case for the simulation. We define the reference length, $y_\mathit{ref} = 10D_p = L_y/3$, to be one-third of the domain height, or the intended clear-fluid height above the particle bed. That is, if the particle bed were to remain motionless, the reference case would represent the fluid flow fairly accurately. The reference velocity, $u_\mathit{ref} = -y_\mathit{ref}^2 f_{b,x} / (12 \mu_f)$, represents the average fluid velocity of the reference case. Finally, we define the reference stress, $\sigma_\mathit{ref} = -y_\mathit{ref} f_{b,x} /2$, to be the wall stress for the reference case.

We are interested in studying the bed in different states, ranging from a few moving particle layers to having the entire bed mobilized. This is accomplished by enforcing different volumetric flow rates, governed by the volume force $\mbs{f}_b$. As it can take a long time for a simulation to reach a steady state when initialized from rest, we found it to be more efficient to obtain a steady state by starting from the final time of a previous simulation with a larger flow rate, and modifying the volume force along the following lines. We initialize the flow by applying a large pressure gradient that mobilizes the entire bed, as described by run Re67 in table~\ref{tab:bed_runs}. By the end of this simulation, the bed has dilated to a height of $h_p/y_\mathit{ref} \approx 2.3$, and the particles just above the fixed layer at the bottom of the domain are moving, as shown in figure~\ref{fig:bed_start_end}b. After this initialization phase, the imposed pressure gradient is reduced to produce simulations Re17 and Re33. Re8 is carried out by continuing Re17 with an even lower imposed pressure gradient. As described in more detail in section~\ref{sec:bed_mom}, this procedure allows us to quickly reach a steady state for run Re17, although not for runs Re33 and Re8.

In contrast to the single rolling sphere case, the steady-state configuration for the moving bed is steady only in a time-averaged sense because particle collisions and positions continuously fluctuate.  We therefore define the time average of a quantity $\theta$ to be
\begin{equation} \label{eq:time_average}
\overline{\theta} = \frac{1}{t_\mathit{avg,2}-t_\mathit{avg,1}} \int_{t_\mathit{avg,1}}^{t_\mathit{avg,2}} \theta \, \mrm{d}t ,
\end{equation}
where we present the values for $t_\mathit{avg,1}$ and $t_\mathit{avg,2}$ in table~\ref{tab:bed_runs}. These time-averaging windows were chosen to capture the steady-state results when possible. For those simulations that had not reached a quasisteady state, the averaging windows were chosen to capture as large a time span as possible for as similar a particle flux as possible. To compare the temporal evolution of the simulations directly to each other in section~\ref{sec:bed_time}, we employ a single set of characteristic quantities to nondimensionalize the velocity, time and stress, in the form of $u_\mathit{base}=u_\mathit{ref}(\mrm{Re67})$, $t_\mathit{base}=y_\mathit{ref}/(1.5u_\mathit{base})$ and $\sigma_\mathit{base}=\sigma_\mathit{ref}(\mrm{Re67})$.

\section{Theoretical stress balance} \label{sec:balance}

\subsection{Fluid phase balance}

\begin{figure}
\placeTwoSubfigures{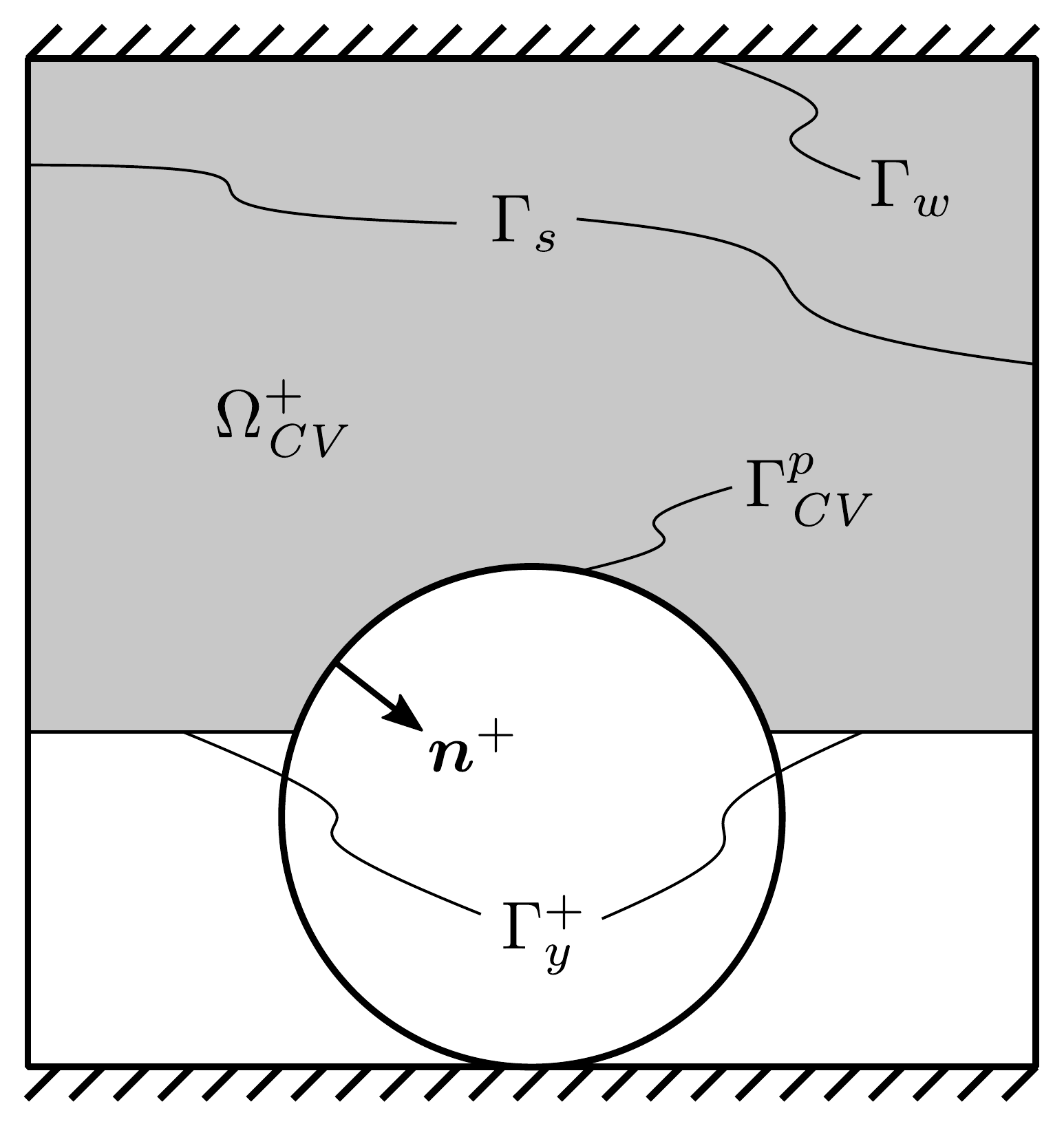}{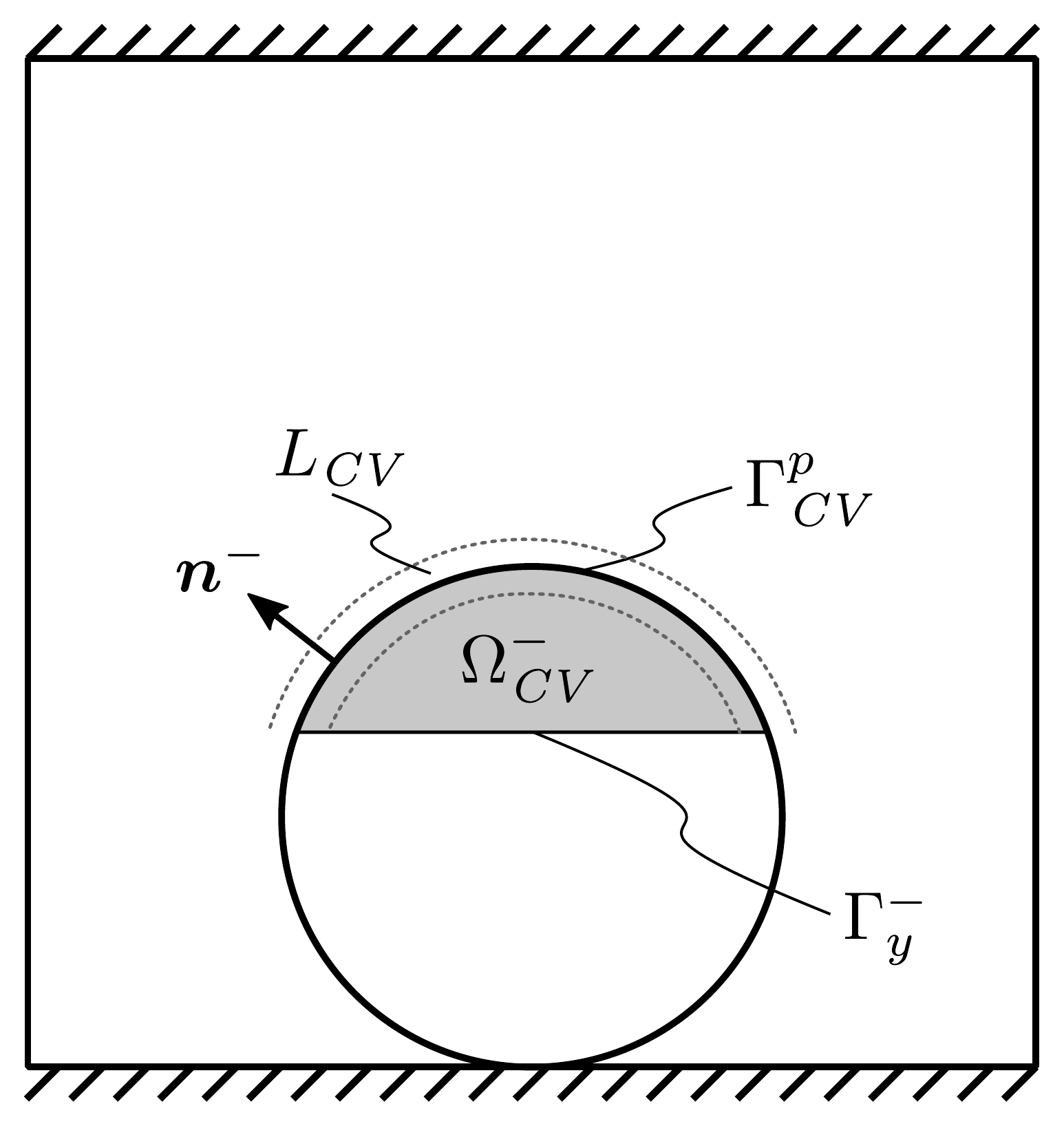}{1.058}{10pt}{10pt}{0pt}
\caption{Shaded control volumes for (a) the fluid surrounding the particle, and (b) the fluid within the particle.}
\label{fig:control_volume}
\end{figure}

\begin{figure}
\centering
\includegraphics[width=0.75\textwidth]{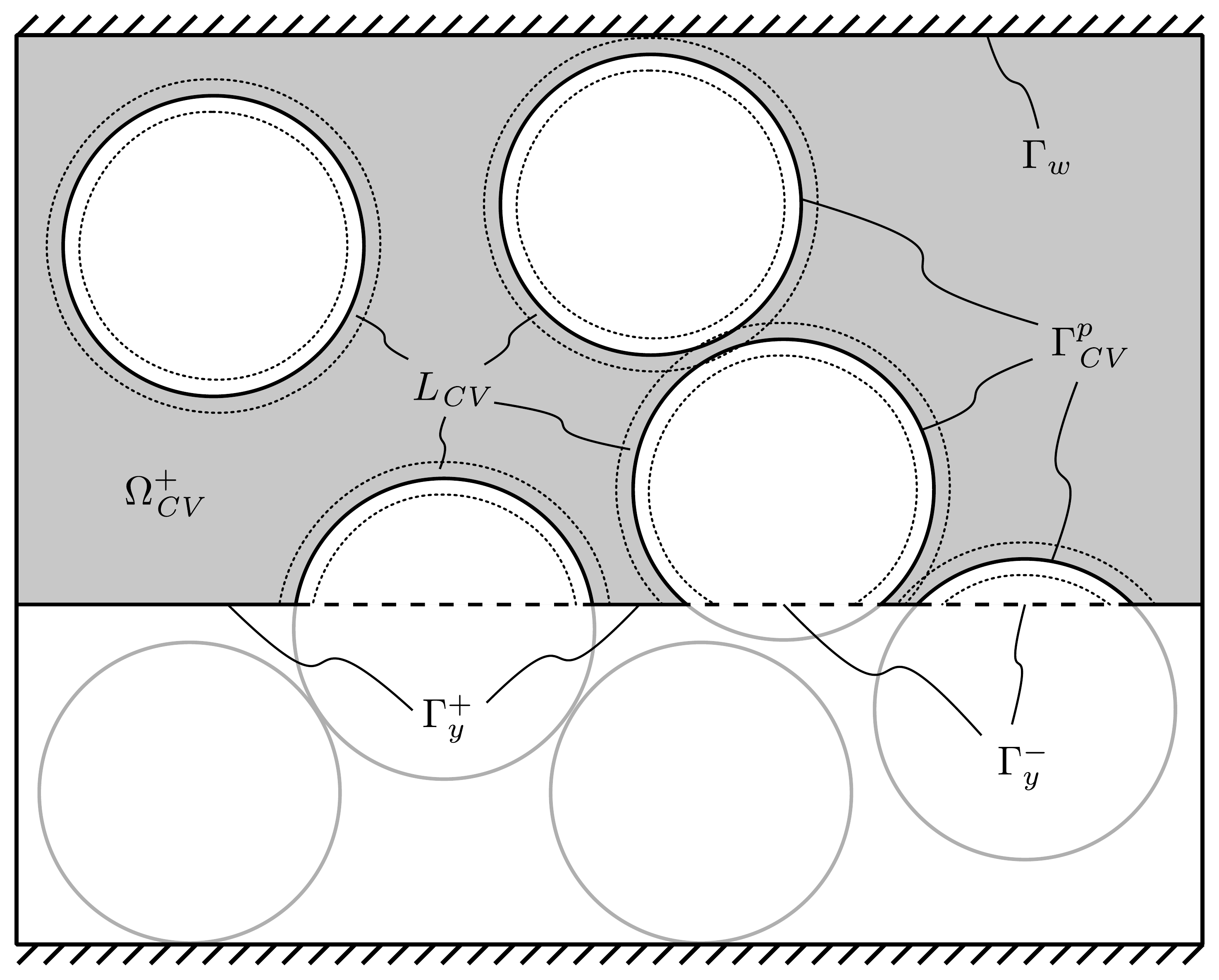}
\caption{Shaded control volume for the general case involving multiple particles. All of the volumes and surfaces required for \eqref{eq:mom_fluid} are indicated.}  \label{fig:control_volume_general}
\end{figure}

In analyzing the momentum balance of the fluid/particle system, we will initially look separately at the fluid and particle phases. Later, we will combine these two components in order to obtain the momentum balance for the fluid/particle mixture, which then  implicitly accounts for the two-way interactions between the particles and the fluid. We begin by investigating the fluid phase alone, excluding the volume occupied by the particles, as well as the inter-particle forces. We do account for the effect of the particles on the fluid, however, through the stress the particles impart on the fluid at their boundaries. We conduct our stress analysis in an integral sense, using a control volume $\Omega_\mathit{CV}^+$ that extends from the top wall to an arbitrary height $y$ in the vertical dimension, encompasses the entire domain in the streamwise $x$- and spanwise $z$-directions, and excludes the volume within particles. Figure~\ref{fig:control_volume}a illustrates the control volume for the case of a single particle, whereas figure~\ref{fig:control_volume_general} shows the case involving many particles.  We write the integral form of \eqref{eq:NS} over this fluid control volume as
\begin{equation} \label{eq:NS_int1}
\int\limits_{\Omega_\mathit{CV}^+} \rho_f \frac{\partial{\mbs{u}}}{\partial{t}} \,\mrm{d}V +
\int\limits_{\Omega_\mathit{CV}^+} \rho_f \nabla\cdot(\mbs{u}\mbs{u}) \,\mrm{d}V =
\int\limits_{\Omega_\mathit{CV}^+} \nabla \cdot \mbs{\tau} \,\mrm{d}V +
\int\limits_{\Omega_\mathit{CV}^+} \mbs{f}_b \,\mrm{d}V ,
\end{equation}
where we did not include the IBM force term from \eqref{eq:NS} because the fluid stress at the fluid/particle interface accounts for the effects of the particles. We do, however, include the forcing term $\mbs{f}_b$, which represents the background pressure gradient employed to drive the flow. Application of the divergence theorem then gives
\begin{equation} \label{eq:NS_int2}
\int\limits_{\Omega_\mathit{CV}^+} \rho_f \frac{\partial{\mbs{u}}}{\partial{t}} \,\mrm{d}V +
\int\limits_{\Gamma_\mathit{CV}^+} \rho_f (\mbs{u}\mbs{u}) \cdot \mbs{n}^+ \,\mrm{d}A =
\int\limits_{\Gamma_\mathit{CV}^+} \mbs{\tau}^+ \cdot \mbs{n}^+ \,\mrm{d}A +
\int\limits_{\Omega_\mathit{CV}^+} \mbs{f}_b \,\mrm{d}V ,
\end{equation}
where $\mbs{n}^+$ denotes the normal vector pointing outwards from $\Omega_\mathit{CV}^+$, and $\mbs{\tau}^+$ represents the stress tensor of the fluid outside the particle.  The boundary of $\Omega_\mathit{CV}^+$ is denoted by $\Gamma_\mathit{CV}^+$, which is composed of the surfaces $\Gamma_\mathit{CV}^+ = \Gamma_w \cup \Gamma_s \cup \Gamma_y^+ \cup \Gamma_\mathit{CV}^p$, as shown in figure~\ref{fig:control_volume}a.  Note that surface $\Gamma_s$ encompasses the periodic boundaries in both the $x$- and $z$-directions.
These control volumes are time-dependent, i.e. $\Omega_\mathit{CV}^+ = \Omega_\mathit{CV}^+(t)$ and $\Gamma_\mathit{CV}^+ = \Gamma_\mathit{CV}^+(t)$. We can consider a steady state to be one in which the particle rolls along the lower wall at a constant speed. In such a case, the time-dependent term in \eqref{eq:NS_int2}, which is in the laboratory reference frame, cancels out with the advective term along $\Gamma_\mathit{CV}^p$, as explained in Appendix~\ref{sec:time_derivative}.  For the situation involving many particles in motion, however, the fluid volume will continue to evolve, never reaching a true instantaneous steady-state. In this case, we apply time-averaging to eliminate the time-dependent term.
Due to the periodic boundary conditions on the $x$- and $z$-boundaries, all of the terms along $\Gamma_s$ cancel out. Furthermore, the upper wall imposes a no-flux condition, i.e. $(\mbs{u}\mbs{u}) \cdot \mbs{n}^+ = 0$ at $\Gamma_w$. Thus, we can simplify \eqref{eq:NS_int2} to
\begin{equation} \label{eq:NS_int3}
\int\limits_{\Gamma_y^+} \rho_f (\mbs{u}\mbs{u}) \cdot \mbs{n}^+ \,\mrm{d}A =
\int\limits_{\Gamma_w \cup \Gamma_y^+ \cup \Gamma_\mathit{CV}^p} \mbs{\tau}^+ \cdot \mbs{n}^+ \,\mrm{d}A +
\int\limits_{\Omega_\mathit{CV}^+} \mbs{f}_b \,\mrm{d}V .
\end{equation}

All of these terms are straightforward to calculate, except for the fluid stress at the particle surface.  However, we can evaluate this term indirectly using the IBM force, as was done to obtain the particle equations of motion \eqref{eq:p_translational_final} and \eqref{eq:p_rotational_final}.  That is, the IBM force acts as a jump in stress between the fluid outside and the fluid inside the particle
\begin{equation} \label{eq:IBM_CV_jump}
\int\limits_{\Gamma_\mathit{CV}^p} \mbs{\tau}^+ \cdot \mbs{n}^+\, {\mrm{d}A}
= \int\limits_{L_\mathit{CV}} \mbs{f}_\mathit{IBM}\, {\mrm{d}V}
- \int\limits_{\Gamma_\mathit{CV}^p} \mbs{\tau}^- \cdot \mbs{n}^-\, {\mrm{d}A} ,
\end{equation}
where we are careful to distinguish between $\mbs{n}^+$, the outward surface normal for the volume $\Omega_\mathit{CV}^+$, and $\mbs{n}^-$, the outward surface normal for the volume $\Omega_\mathit{CV}^-$, which point in opposite directions. Hence, in order to determine the force the particle imparts on the fluid, our analysis requires us to account for the fluid inside the particle. To evaluate $\mbs{\tau}^- \cdot \mbs{n}^-$, we can evaluate the momentum balance on the fluid inside the particle, shown in figure~\ref{fig:control_volume}b.  The integral form of the Navier-Stokes equations together with the divergence theorem give us
\begin{equation} \label{eq:NS_in_int1}
\int\limits_{\Omega_\mathit{CV}^-} \rho_f \frac{\partial{\mbs{u}}}{\partial{t}} \,\mrm{d}V
+ \int\limits_{\Gamma_\mathit{CV}^-} \rho_f (\mbs{u}\mbs{u}) \cdot \mbs{n}^- \,\mrm{d}A
= \int\limits_{\Gamma_\mathit{CV}^-} \mbs{\tau}^- \cdot \mbs{n}^- \,\mrm{d}A
+ \int\limits_{\Omega_\mathit{CV}^-} \mbs{f}_b \,\mrm{d}V ,
\end{equation}
where $\Gamma_\mathit{CV}^- = \Gamma_\mathit{CV}^p \cup \Gamma_y^-$.  The first term cancels out the convective term along $\Gamma_\mathit{CV}^p$ just as it did for the fluid outside the particle, which is explained in Appendix~\ref{sec:time_derivative}.
Then, \eqref{eq:NS_in_int1} reduces to
\begin{equation} \label{eq:NS_in_int3}
\int\limits_{\Gamma_y^-} \rho_f (\mbs{u}\mbs{u}) \cdot \mbs{n}^- \,\mrm{d}A
= \int\limits_{\Gamma_\mathit{CV}^p} \mbs{\tau}^- \cdot \mbs{n}^- \,\mrm{d}A
+ \int\limits_{\Gamma_y^-} \mbs{\tau}^- \cdot \mbs{n}^- \,\mrm{d}A
+ \int\limits_{\Omega_\mathit{CV}^-} \mbs{f}_b \,\mrm{d}V.
\end{equation}
Using \eqref{eq:NS_in_int3} together with \eqref{eq:IBM_CV_jump}, we obtain
\begin{equation} \label{eq:NS_in_int4}
\int\limits_{\Gamma_\mathit{CV}^p} \mbs{\tau}^+ \cdot \mbs{n}^+\, {\mrm{d}A}
= \int\limits_{L_\mathit{CV}} \mbs{f}_\mathit{IBM}\, {\mrm{d}V}
- \int\limits_{\Gamma_y^-} \rho_f (\mbs{u}\mbs{u}) \cdot \mbs{n}^- \,\mrm{d}A
+ \int\limits_{\Gamma_y^-} \mbs{\tau}^- \cdot \mbs{n}^- \,\mrm{d}A
+ \int\limits_{\Omega_\mathit{CV}^-} \mbs{f}_b \,\mrm{d}V .
\end{equation}
Finally, combining \eqref{eq:NS_in_int4} and \eqref{eq:NS_int3} gives
\begin{IEEEeqnarray}{r} \label{eq:mom_fluid}
\underbrace{
\int\limits_{\Gamma_w} \mbs{\tau}^+ \cdot \mbs{n}^+ \,\mrm{d}A
+ \int\limits_{\Omega_\mathit{CV}} \mbs{f}_b \,\mrm{d}V
}_\text{External force}
= \underbrace{
-\int\limits_{\Gamma_y^+} \mbs{\tau}^+ \cdot \mbs{n}^+ \,\mrm{d}A
+ \int\limits_{\Gamma_y^+} \rho_f (\mbs{u}\mbs{u}) \cdot \mbs{n}^+ \,\mrm{d}A
}_\text{Fluid force} \nonumber\\
\underbrace{
- \int\limits_{L_\mathit{CV}} \mbs{f}_\mathit{IBM}\, {\mrm{d}V}
- \int\limits_{\Gamma_y^-} \mbs{\tau}^- \cdot \mbs{n}^- \,\mrm{d}A
+ \int\limits_{\Gamma_y^-} \rho_f (\mbs{u}\mbs{u}) \cdot \mbs{n}^- \,\mrm{d}A
}_\text{Particle force} ,
\end{IEEEeqnarray}
where $\Omega_\mathit{CV} = \Omega_\mathit{CV}^+ \cup \Omega_\mathit{CV}^-$.
The left-hand side of \eqref{eq:mom_fluid} contains the external forces acting on the control volume via the top wall, $\Gamma_w$, and the body force applied to the whole volume, $\Omega_\mathit{CV}$. These external forces are balanced by fluid and particle forces within and at the lower boundary of the control volume.  The particle force shown above represents the force the particles exert on the fluid phase together with the body force acting on the particle phase
\begin{equation} \label{eq:particle_force}
\text{Particle force} = \int\limits_{\Gamma_\mathit{CV}^p} \mbs{\tau}^+ \cdot \mbs{n}^- \,\mrm{d}A +
\int\limits_{\Omega_\mathit{CV}^-} \mbs{f}_b \,\mrm{d}V .
\end{equation}
Comparing this relationship to the particle equation of motion \eqref{eq:p_translational}, we see that, if the particle acceleration is negligible, the particle force is balanced by the particle weight and collision forces.  Thus, the particle force also represents the portion of the momentum balance that is supported by the particle weight and collision forces.

The relationship given by \eqref{eq:mom_fluid} is valid for the case of a single sphere moving at a constant velocity in a flow.  For a more general situation involving multiple particles moving relative to one another, the method used to eliminate the time derivate and one of the convective terms cannot be used.  Instead, we will average \eqref{eq:mom_fluid} in time and demonstrate that this time-averaged relationship closes for the case involving multiple particles, which is illustrated in figure~\ref{fig:control_volume_general}.
The fluid force consists of pressure and viscous stresses as well as convective momentum transport, all of which act at the lower boundary outside the particles, $\Gamma_y^+$.  The particle force consists of the IBM force, which acts throughout the control volume over $L_\mathit{CV}$, and convective and fluid stresses, which act at the lower boundary inside the particles, $\Gamma_y^-$.  Note that the fluid inside the particles is only considered for those particles cut by the control volume; for particles wholly inside the control volume, the IBM force alone accounts for the effect of the particles acting on the fluid.

\subsubsection{Fluid phase momentum in the $x$-direction}

We now consider the momentum balance over the control volume in the $x$-direction.
At the top wall, $\Gamma_w$, the pressure does not contribute to the $x$-momentum, and the vertical velocity, $v$, is zero, so that only $\mu_f \partial u/\partial y$ contributes to the fluid stress.  At the lower boundary, $\Gamma_y$, the pressure again does not play a role, but we keep the complete viscous terms and convective terms for generality.  Due to the periodic boundaries, $\int_{\Gamma_y} \partial v /\partial x \, \mrm{d}A = 0$, but the integrals of this quantity in the separate domains $\Gamma_y^+$ and $\Gamma_y^-$ can be nonzero, so we leave the expression in the more general form
%
\begin{IEEEeqnarray}{r} \label{eq:momx_fluid}
\underbrace{
\int\limits_{\Gamma_w} \mu_f \frac{\partial u}{\partial y} \,\mrm{d}A
+ \int\limits_{\Omega_\mathit{CV}} f_{b,x} \,\mrm{d}V
}_\text{External force}
= \underbrace{
\int\limits_{\Gamma_y^+} \mu_f \left(\frac{\partial u}{\partial y} + \frac{\partial v}{\partial x}\right) \,\mrm{d}A
- \int\limits_{\Gamma_y^+} \rho_f uv \,\mrm{d}A
}_\text{Fluid force} \nonumber\\
\underbrace{
- \int\limits_{L_\mathit{CV}} f_{\mathit{IBM,x}}\, {\mrm{d}V}
+ \int\limits_{\Gamma_y^-} \mu_f \left(\frac{\partial u}{\partial y} + \frac{\partial v}{\partial x}\right) \,\mrm{d}A
- \int\limits_{\Gamma_y^-} \rho_f uv \,\mrm{d}A
}_\text{Particle force} .
\end{IEEEeqnarray}

It is important to note that here we are explicitly separating the stresses arising from the fluid and particle phases.  We could consider all the viscous and convective terms acting along both $\Gamma_y^+$ and $\Gamma_y^-$ to be the fluid stress terms and likewise consider only the $f_{IBM}$ term to be the particle stress, as was done in \citet{Kidanemariam2017}.  However, while this method may be accurate in recovering the overall stress, it may not be accurate in apportioning the stress between the fluid and particle phases (unless the viscous and convective stresses within the particles are negligible).

Dividing by the horizontal area of the domain and using the definition of the horizontal average,
\begin{equation} \label{eq:horizontal_average}
\left< \theta \right> = \frac{1}{L_x L_z} \int_0^{L_z} \int_0^{L_x} \theta \,\mrm{d}x \, \mrm{d}z ,
\end{equation}
we can rewrite \eqref{eq:momx_fluid} as
\begin{IEEEeqnarray}{r} \label{eq:fx_stress}
\underbrace{
\underbrace{\mu_f \left<\left.\frac{\partial u}{\partial y}\right|_{L_y}\right>}_{\displaystyle \sigma_\mathit{Evisc,x}}
\:\underbrace{+\: f_{b,x} (L_y - y)}_{\displaystyle \sigma_\mathit{Ebody,x}}
}_\text{External stress}
= \underbrace{
\underbrace{\mu_f \left<\gamma\left.\left(\frac{\partial u}{\partial y} + \frac{\partial v}{\partial x} \right)\right|_y \right>}_{\displaystyle \sigma_\mathit{Fvisc,x}}
\:\underbrace{-\: \rho_f \left<\left.\gamma uv\right|_y \right>}_{\displaystyle \sigma_\mathit{Fconv,x}}
}_\text{Fluid stress} \nonumber\\
\underbrace{
\underbrace{-\: \int_y^{L_y} \left<f_{\mathit{IBM,x}}\right> {\mrm{d}y}}_{\displaystyle \sigma_\mathit{PIBM,x}}
\:\underbrace{+\: \mu_f \left<\phi\left.\left(\frac{\partial u}{\partial y} + \frac{\partial v}{\partial x}\right)\right|_y\right>}_{\displaystyle \sigma_\mathit{Pvisc,x}}
\:\underbrace{-\: \rho_f \left<\left.\phi uv\right|_y \right>}_{\displaystyle \sigma_\mathit{Pconv,x}}
}_\text{Particle stress} ,
\end{IEEEeqnarray}
where $\gamma$ is an indicator function for the fluid volume fraction ($\gamma = 1$ outside the particle and $\gamma = 0$ inside the particle) and $\phi$ is an indicator function for the particle volume fraction ($\phi = 1-\gamma$), in line with the volume-averaging approach of \citet{Nikora2013}. We have also used the fact that $\mu_f$, $\rho_f$, and $f_{b,x}$ are constant throughout the domain. The external stress consists of $\sigma_\mathit{Evisc,x}$, the viscous stress at the top wall, and $\sigma_\mathit{Ebody,x}$, the stress from the body force acting throughout the control volume. This component is of particular importance since this quantity is needed for various continuum model closures such as those for $\mu(I)$-rheology \citep[e.g,][]{Boyer2011}, kinetic theory \citep[e.g.,][]{Hsu2004}, or effective viscosity \citep[e.g.,][]{Stickel2005}. The fluid stress consists of $\sigma_\mathit{Fvisc,x}$, the viscous stress, and $\sigma_\mathit{Fconv,x}$, the convective stress, both of which are evaluated outside the particles at height $y$.  The particle stress consists of $\sigma_\mathit{PIBM,x}$, the IBM stress, $\sigma_\mathit{Pvisc,x}$, the viscous stress, and $\sigma_\mathit{Pconv,x}$, the convective stress, the latter two of which are evaluated inside the particles at height $y$.

\subsubsection{Fluid phase momentum in the $y$-direction}

For the $y$-momentum component, the pressure, in addition to the viscous stress, contributes to the fluid stress tensor at the boundaries $\Gamma_w$ and $\Gamma_y^+$, but only the $vv$ component contributes to the convective term, reducing \eqref{eq:mom_fluid} to the following
\begin{IEEEeqnarray}{r} \label{eq:momy_fluid}
\underbrace{
-\int\limits_{\Gamma_w} p \,\mrm{d}A
+ \int\limits_{\Gamma_w} 2\mu_f \frac{\partial v}{\partial y} \,\mrm{d}A
}_\text{External force}
= \underbrace{
-\int\limits_{\Gamma_y^+} p \,\mrm{d}A
+ \int\limits_{\Gamma_y^+} 2\mu_f \frac{\partial v}{\partial y} \,\mrm{d}A
- \int\limits_{\Gamma_y^+} \rho_f vv \,\mrm{d}A
}_\text{Fluid force} \nonumber\\
\underbrace{
- \int\limits_{L_\mathit{CV}} f_{\mathit{IBM,y}}\, {\mrm{d}V}
- \int\limits_{\Gamma_y^-} p \,\mrm{d}A
+ \int\limits_{\Gamma_y^-} 2\mu_f \frac{\partial v}{\partial y} \,\mrm{d}A
- \int\limits_{\Gamma_y^-} \rho_f vv \,\mrm{d}A
}_\text{Particle force} .
\end{IEEEeqnarray}
$f_{b,y} = 0$ because the externally imposed pressure gradient acts only in the streamwise direction. On the left-hand side, the external force consists of the pressure and viscous stress acting at the top wall. This force is balanced on the right-hand side by the fluid force, consisting of the fluid pressure, viscous stress, and convective transport outside the particles at the lower boundary of the control volume, and the particle force, consisting of the IBM force throughout the control volume as well as the pressure, viscous stress, and convective fluid transport within the particles cut by the lower wall of the control volume. Again, dividing by the horizontal domain area and applying the spatial averaging operator \eqref{eq:horizontal_average}, we can reduce \eqref{eq:momy_fluid} to
\begin{IEEEeqnarray}{r} \label{eq:fy_stress}
\underbrace{
\underbrace{-\left<\left.p\right|_{L_y}\right>}_{\displaystyle \sigma_\mathit{Epres,y}}
\:\underbrace{+\: 2\mu_f \left<\left.\frac{\partial v}{\partial y}\right|_{L_y} \right>}_{\displaystyle \sigma_\mathit{Evisc,y}}
}_\text{External stress}
= \underbrace{
\underbrace{-\left<\left.\gamma p\right|_y \right>}_{\displaystyle \sigma_\mathit{Fpres,y}}
\:\underbrace{+\: 2\mu_f \left<\left.\gamma\frac{\partial v}{\partial y}\right|_y \right>}_{\displaystyle \sigma_\mathit{Fvisc,y}}
\:\underbrace{-\: \rho_f \left<\left.\gamma vv\right|_y \right>}_{\displaystyle \sigma_\mathit{Fconv,y}}
}_\text{Fluid stress} \nonumber\\
\underbrace{
\underbrace{- \int_y^{L_y} \left<f_{\mathit{IBM,y}}\right> {\mrm{d}y}}_{\displaystyle \sigma_\mathit{PIBM,y}}
\:\underbrace{-\: \left<\left.\phi p\right|_y \right>}_{\displaystyle \sigma_\mathit{Ppres,y}}
\:\underbrace{+\: 2\mu_f \left<\left.\phi\frac{\partial v}{\partial y}\right|_y \right>}_{\displaystyle \sigma_\mathit{Pvisc,y}}
\:\underbrace{-\: \rho_f \left<\left.\phi vv\right|_y \right>}_{\displaystyle \sigma_\mathit{Pconv,y}}
}_\text{Particle stress} .
\end{IEEEeqnarray}
The external stress consists of $\sigma_\mathit{Epres,y}$, the average pressure at the top wall, and $\sigma_\mathit{Evisc,y}$, the viscous stress at the top wall. The fluid stress is composed of $\sigma_\mathit{Fpres,y}$, the pressure, $\sigma_\mathit{Fvisc,y}$, the viscous stress, and $\sigma_\mathit{Fconv,y}$, the convective stress, all of which are evaluated outside the particles at height $y$. The particle stress consists of $\sigma_\mathit{PIBM,y}$, the IBM stress, $\sigma_\mathit{Ppres,y}$, the pressure, $\sigma_\mathit{Pvisc,y}$, the viscous stress, and $\sigma_\mathit{Pconv,y}$, the convective stress, the latter three of which are evaluated inside the particles at height $y$.

\subsection{Particle phase balance}

Although the fluid stress analysis accounts for the effects of the particles through the IBM force, we can also perform an analysis on the particle phase by itself in order to ensure that the particle momentum also closes and to try to bridge the two balances into a single one for the mixture as a whole.  Additionally, rheological descriptions require information about the particle pressure, which we can only obtain by analyzing the particle phase.  We can apply the coarse-graining method to \eqref{eq:p_translational_final} as described in Appendix~\ref{sec:cg} to obtain
\begin{equation} \label{eq:p_int1}
\mbs{a}^{cg} = \mbs{F}_I^{cg} + \mbs{F}_\mathit{IBM}^{cg} + \mbs{F}_g^{cg} + \mbs{F}_c^{cg} ,
\end{equation}
where
\begin{equation}
\mbs{a}^{cg}(\mbs{x},t) = \sum_{p=1}^{N_p} m_p \frac{\mrm{d}\mbs{u}_p}{\mrm{d}t} \, \mathcal{W}(\mbs{x}-\mbs{x}_p)
\end{equation}
is the coarse-grained local particle acceleration, and
\begin{equation}
\mbs{F}_\mathit{IBM}^{cg} = \sum_{p=1}^{N_p} \mbs{F}_{\mathit{IBM},p} \, \mathcal{W}(\mbs{x}-\mbs{x}_p)
\end{equation}
is the coarse-grained IBM force (likewise for the other forces acting on the particle center of mass).  The coarse-graining function, $\mathcal{W}(\mbs{r})$, spreads the particle-centered quantities onto an Eulerian mesh, allowing us to treat them as a continuum field.  Note that the coarse-graining function is scaled by $w^{-3}$, as shown in \eqref{eq:cg_function}, where $w$ is the coarse-graining length scale.  Thus, $\mbs{F}_\mathit{IBM}^{cg}$ represents a force per unit volume.

Similar to the fluid momentum balance, we can analyze the coarse-grained particle forces within a control volume spanning the entire domain in the streamwise and spanwise directions and extending from the top wall to an arbitrary height $y$.  Integrating \eqref{eq:p_int1} over this volume, we obtain
\begin{equation} \label{eq:p_int2}
\int\limits_{\Omega_\mathit{CV}} \mbs{a}^{cg} \,\mrm{d}V = \int\limits_{\Omega_\mathit{CV}} \left(\mbs{F}_I^{cg} + \mbs{F}_\mathit{IBM}^{cg} + \mbs{F}_g^{cg} + \mbs{F}_c^{cg} \right) \,\mrm{d}V .
\end{equation}
We can again apply the averaging operator to recast \eqref{eq:p_int2} as a line integral in the wall-normal direction
\begin{equation} \label{eq:p_int3}
\int_y^{L_y} \left<\mbs{a}^{cg}\right> \,\mrm{d}y = \int_y^{L_y} \left(\left<\mbs{F}_I^{cg}\right> + \left<\mbs{F}_\mathit{IBM}^{cg}\right> + \left<\mbs{F}_g^{cg}\right> + \left<\mbs{F}_c^{cg}\right> \right) \,\mrm{d}y .
\end{equation}
If the particles are in a steady state, either naturally or through double-averaging, then the acceleration term vanishes. We decompose the equation into its $x$-component
\begin{equation} \label{eq:px_stress}
\underbrace{\int_y^{L_y} \left<F_{I,x}^{cg}\right> \,\mrm{d}y
+ \int_y^{L_y} \left<F_\mathit{IBM,x}^{cg}\right> \,\mrm{d}y}_\text{Hydrodynamic stress}
+ \underbrace{\int_y^{L_y} \left<F_{c,x}^{cg}\right> \,\mrm{d}y}_\text{Collision stress}
= 0 ,
\end{equation}
where the gravitational force is zero, and the $y$-component
\begin{equation} \label{eq:py_stress}
\underbrace{-\int_y^{L_y} \left<F_{g}^{cg}\right> \,\mrm{d}y}_\text{Bed weight}
= \underbrace{\int_y^{L_y} \left<F_{I,y}^{cg}\right> \,\mrm{d}y
+ \int_y^{L_y} \left<F_\mathit{IBM,y}^{cg}\right> \,\mrm{d}y}_\text{Hydrodynamic stress}
+ \underbrace{\int_y^{L_y} \left<F_{c,y}^{cg}\right> \,\mrm{d}y}_\text{Collision stress} .
\end{equation}

\subsection{Mixture balance}

Instead of considering the fluid and particle phases separately, we could combine them into a single mixture.  For example, in the $x$-direction, equating the particle stress on the fluid in \eqref{eq:fx_stress} to the hydrodynamic stress on the particles in \eqref{eq:px_stress}, we obtain
\begin{IEEEeqnarray}{r} \label{eq:bx_stress}
\underbrace{
\mu_f \left<\left.\frac{\partial u}{\partial y}\right|_{L_y}\right>
+ f_{b,x} (L_y - y)
}_\text{External stress}
= \underbrace{
\mu_f \left<\gamma\left(\frac{\partial u}{\partial y} + \frac{\partial v}{\partial x} \right)_y \right>
- \rho_f \left<\left.\gamma uv\right|_y \right>
}_\text{Fluid stress} \nonumber\\
\underbrace{- \int_y^{L_y} \left<F_{c,x}^{cg}\right> \,\mrm{d}y}_\text{Collision stress} .
\end{IEEEeqnarray}
This formulation has several advantages over the separate phase balances.  First, the collision information for the particles is generally more readily available from simulation results than the Eulerian IBM data is.  Second, we can reformulate the coarse-grained collision stress as a stress acting over the lower surface of the control volume instead of a force integrated over the volume
\begin{equation} \label{eq:collision_stress_equivalence}
\int_y^{L_y} \left<F_{c,x}^{cg}\right> \,\mrm{d}y = \left<\sigma_{xy}^{cg}\right> ,
\end{equation}
where $\sigma_{xy}^{cg}$ is the $xy$-component of a coarse-grained particle collision stress, such as that defined by \citet{Weinhart2012}
\begin{equation} \label{eq:collision_stress}
\sigma_{ij}^{cg}(\mbs{x}) = \sum_{p=1}^{N_p} \sum_{q=p+1}^{N_p} F_{c,pq,i} \, r_{pq,j} \int_0^1 \mathcal{W}(\mbs{x} - \mbs{x}_p + s\mbs{r}_{pq}) \, \mrm{d}s ,
\end{equation}
where $\mbs{F}_{c,pq}$ is the collision force acting on particle $p$ from particle $q$, and $\mbs{r}_{pq} = \mbs{x}_p - \mbs{x}_q$ points from the center of particle $q$ to the center of particle $p$. The integral effectively spreads the contact force along the line connecting $\mbs{x}_p$ to $\mbs{x}_q$. While the collision force in \eqref{eq:bx_stress} can only provide information in the $x$-, $y$-, and $z$-directions and must be integrated over a volume, the stress tensor in \eqref{eq:collision_stress} can provide more information about shear and normal stresses in the particle phase without averaging over volumes. However, for the momentum balance in the present work, we will focus only on the collision stress presented in \eqref{eq:bx_stress}.


\section{Results} \label{sec:results}

\subsection{Stress balance of a single rolling particle} \label{sec:single_mom}

\subsubsection{Stress balance of the fluid phase in the $x$-direction}
\label{sec:single_momx_fluid}

\begin{figure}
\centering
\includegraphics[width=0.65\textwidth]{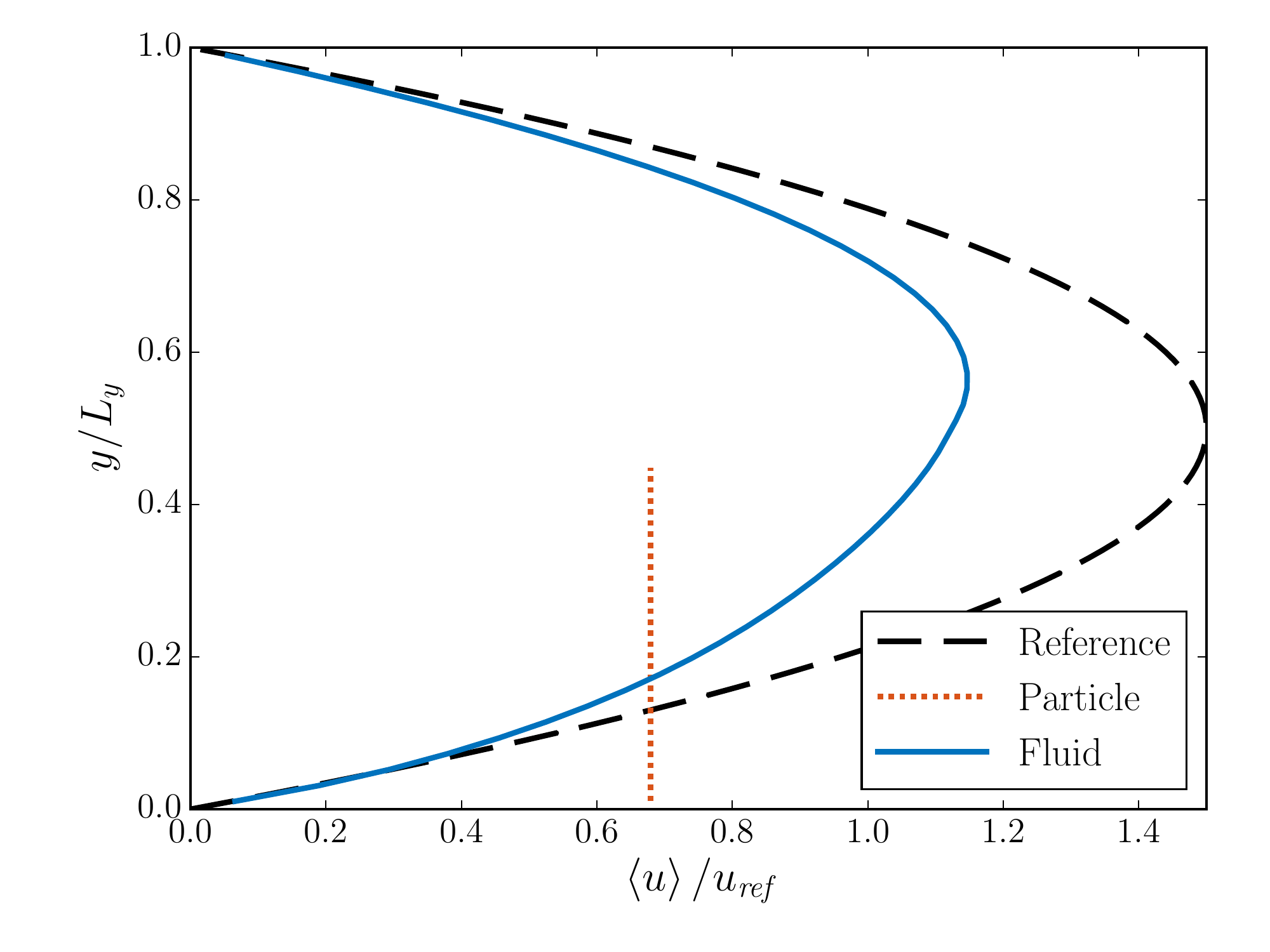}
\caption{Single rolling particle: average fluid velocity profile, given by \eqref{eq:u_fluid}, along with the average coarse-grained particle velocity profile, given by \eqref{eq:cg_u}, and the velocity profile for the reference case.}  \label{fig:single_velocity}
\end{figure}

Having established momentum balance relationships \eqref{eq:fx_stress} \eqref{eq:px_stress}, we will now apply them to the single rolling sphere case described in section~\ref{sec:single_setup}.  In figure~\ref{fig:single_velocity}, we present the horizontally-averaged fluid velocity, defined as
\begin{equation} \label{eq:u_fluid}
u_\mathit{fluid} = \frac{\left<\gamma u\right>}{\left<\gamma\right>} ,
\end{equation}
along with the coarse-grained particle velocity, which in this case represents the translational velocity of the particle. This figure also demonstrates the velocity profile for the reference case, which would result from the pressure-driven flow in the channel if the particle were not present. The particle lags behind the flow, decreasing the fluid velocity profile from the reference case. To understand how this occurs, we turn our attention to the momentum balance in the $x$-direction. Overall, this balance is between the pressure gradient driving the flow in the positive $x$-direction, the viscous stress of the fluid at the walls, and the friction between the particle and the lower wall.  This friction is transmitted to the fluid via hydrodynamic stresses between the particle and the fluid, which we label the ``particle stress" when considering the fluid phase and the ``hydrodynamic stress" when considering the particle phase.

\begin{figure}
\placeThreeSubfiguresDown{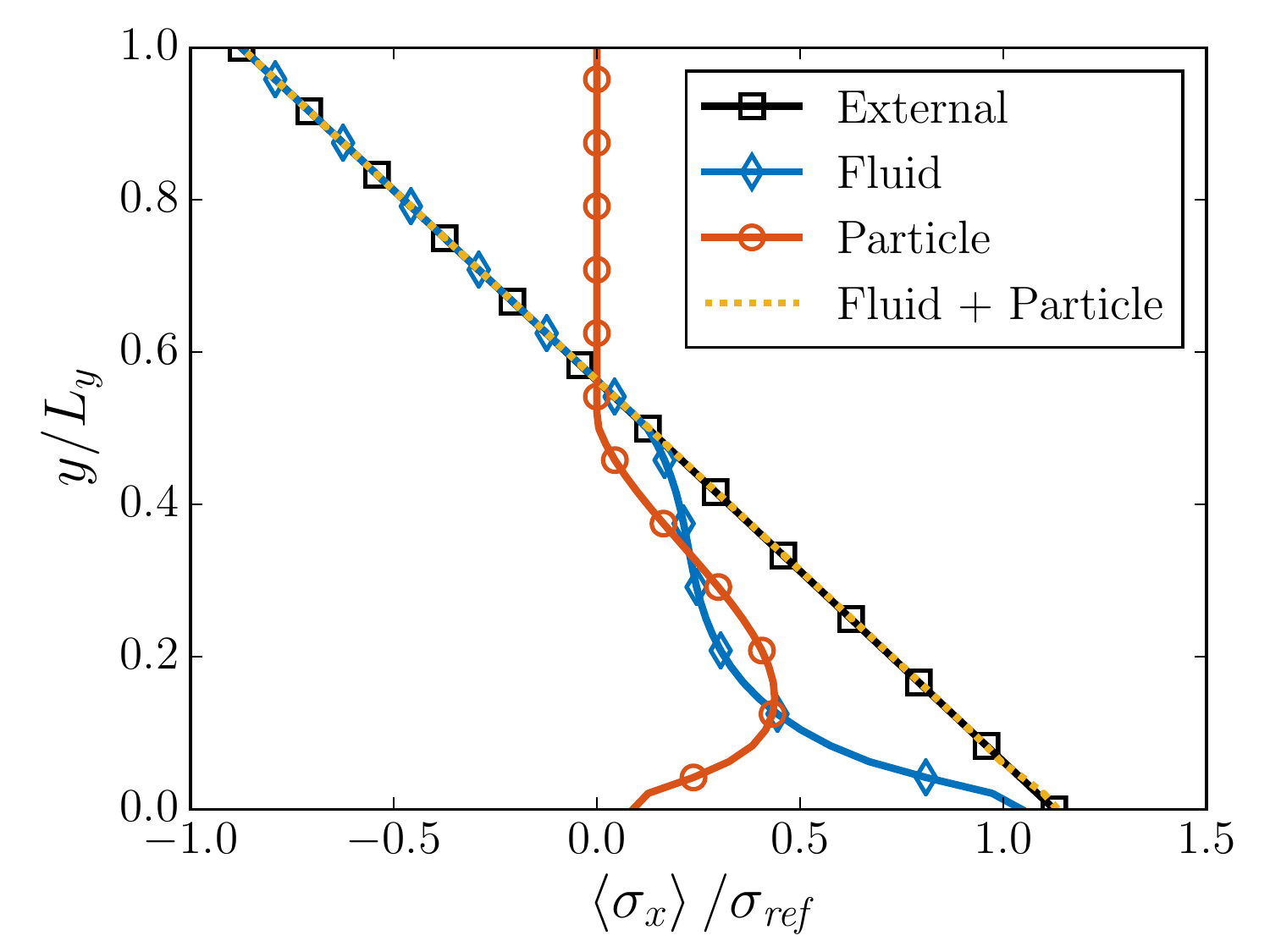}{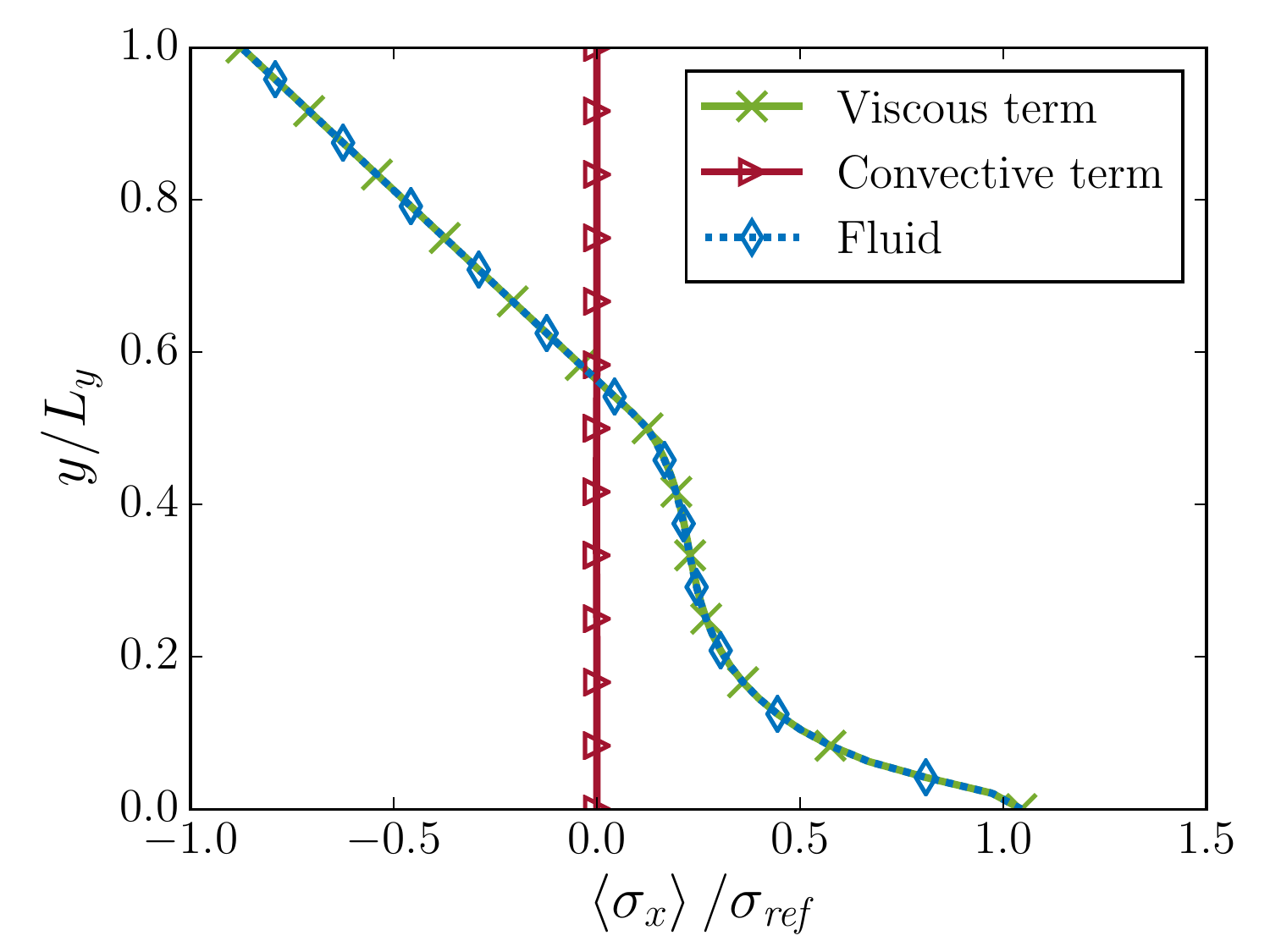}{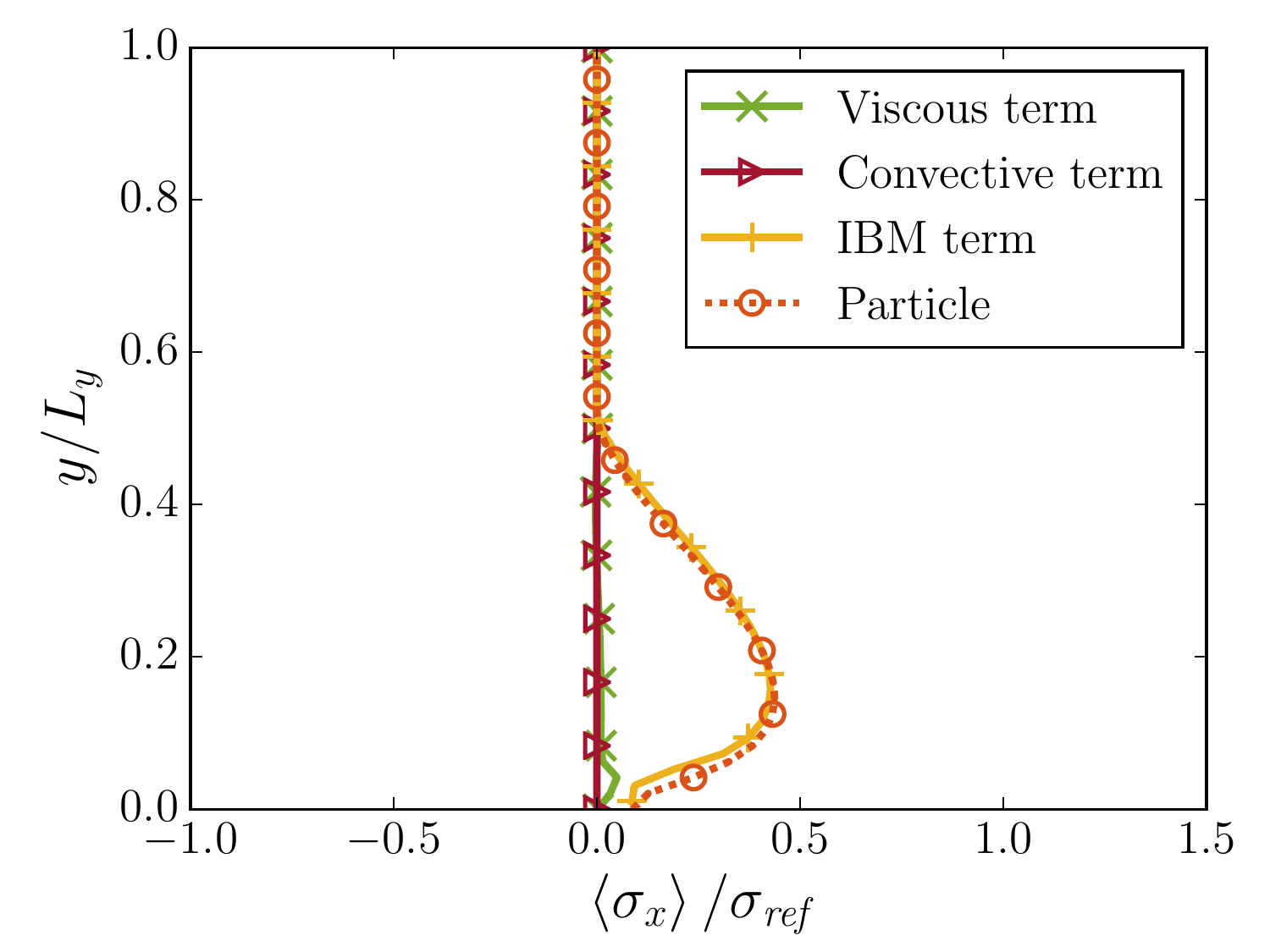}{0.75}{0pt}{0pt}{0pt}
\caption{Single rolling particle: stress balance of the fluid phase in the $x$-direction according to \eqref{eq:fx_stress}.  The components of (a) are further broken down for the (b) fluid stress and (c) particle stress.  As shown in (a), the stress balance is in equilibrium because the sum of the fluid and particles stresses matches the external stress, and both the fluid stress and particle stress contribute significantly in the lower half of the domain. As shown in (b), the viscous term ($\sigma_\mathit{Fvisc,x}$) accounts for the fluid stress. As shown in (c), the IBM term ($\sigma_\mathit{PIBM,x}$) and the viscous term ($\sigma_\mathit{Pvisc,x}$) account for the particle stress.}
\label{fig:single_momx_fluid}
\end{figure}

Figure~\ref{fig:single_momx_fluid} shows the $x$-momentum balance of the fluid phase, given by \eqref{eq:fx_stress}. The stresses are a function of the $y$-coordinate, where each value of $\sigma_x$ corresponds to the control volume extending from the top wall to the $y$-coordinate. The reference stress for these plots is the wall shear stress for the reference case, $\sigma_\mathit{ref} = \mu_f \mrm{d}u/\mrm{d}y|_{y=0}$. Figure~\ref{fig:single_momx_fluid}a shows the instantaneous particle, fluid, and external stresses at $t = 2.5 t_\mathit{ref}$. As expected, the external stress is in equilibrium with the sum of the fluid and particle stresses acting on the horizontal plane located at $y$.  For control volumes above the particle ($y/L_y > 0.5$), the particle stress is zero, and the external stress is balanced entirely by the fluid stress. However, in the lower half of the domain, where the particle is located, the particle stress accounts for most of the stress in the associated control volumes. Note that, in this simulation, the particle diameter fills half the domain in the streamwise and spanwise directions ($L_x = 2D_p$, $L_z = 2D_p$). Thus, the particle has a significant effect on the horizontally-averaged stresses. In contrast, we would expect a single particle in a much larger domain to have a much smaller effect on the flow and likewise to have a much smaller particle stress relative to the fluid stress. Near the lower wall ($y/L_y < 0.1$), a decrease in particle stress and increase in fluid stress indicates a transfer of $x$-momentum back to the fluid. The total drag on the particle, given by the particle stress at the lower wall, is thus only a small fraction of the drag experienced by the upper half of the particle.  This total drag is equivalent to the frictional force the wall exerts on the particle, which will be shown in section~\ref{sec:single_momx_particle}.
%
%

Based on our definition of $\sigma_\mathit{ref}$, in the absence of the particle, the external stress in figure~\ref{fig:single_momx_fluid}a would extend from $\sigma_x/\sigma_\mathit{ref} = -1$ at $y/L_y = 1$ to $\sigma_x/\sigma_\mathit{ref} = 1$ at $y/L_y = 0$.  The presence of the particle causes this curve to shift to the right, decreasing the magnitude of the stress at the top wall and increasing the stress at the bottom wall.  This rightward shift results in a decrease in the fluid velocity in the upper half of the domain, as shown in figure~\ref{fig:single_velocity}.  At the lower wall, the fluid stress is close to $\sigma_x/\sigma_\mathit{ref} = 1$, while the rightward shift in the external stress results from the particle stress at the lower wall.  The collisional friction with the lower wall, therefore, accounts for a large portion of the decrease in the flow rate relative to the reference case, the other portion arising from the constricted flow between the particle and the wall increasing the fluid shear stress.

The fluid stress in figure~\ref{fig:single_momx_fluid}a is further decomposed into its components in figure~\ref{fig:single_momx_fluid}b, which shows the relative contributions from the viscous stress, $\sigma_\mathit{Fvisc,x}$, and the convective stress, $\sigma_\mathit{Fconv,x}$, given in \eqref{eq:fx_stress}.  The convective term is negligible, so that the viscous term alone accounts for the fluid stress.  Though we do not show it here, we also found the $\partial v /\partial x$ term to be negligible.

Likewise, the particle stress in figure~\ref{fig:single_momx_fluid}a is further decomposed into its components in figure~\ref{fig:single_momx_fluid}c, which include the IBM, viscous, and convective stresses, given by $\sigma_\mathit{PIBM,x}$, $\sigma_\mathit{Pvisc,x}$, and $\sigma_\mathit{Pconv,x}$, respectively, in \eqref{eq:fx_stress}. The IBM term dominates, the convective term is negligible, and the viscous term is detectable only near the lower wall. After omitting the negligible terms, \eqref{eq:fx_stress} yields the following balance between the dominant terms in the fluid momentum equation
\begin{IEEEeqnarray}{rCl} \label{eq:fx_stress_simplified}
\underbrace{
\mu_f \left<\left.\frac{\partial u}{\partial y}\right|_{L_y}\right>
+ f_{b,x} (L_y - y)
}_\text{External stress}
&=& \underbrace{
\mu_f \left<\gamma\left.\frac{\partial u}{\partial y}\right|_y \right>
}_\text{Fluid stress} \nonumber\\
&&\underbrace{
- \int_y^{L_y} \left<f_{\mathit{IBM,x}}\right> {\mrm{d}y}
+ \mu_f \left<\phi\left.\frac{\partial u}{\partial y}\right|_y\right>
}_\text{Particle stress} .
\end{IEEEeqnarray}

\subsubsection{Stress balance of the particle phase in the $x$-direction}
\label{sec:single_momx_particle}

\begin{figure}
\centering
\includegraphics[width=0.5\textwidth]{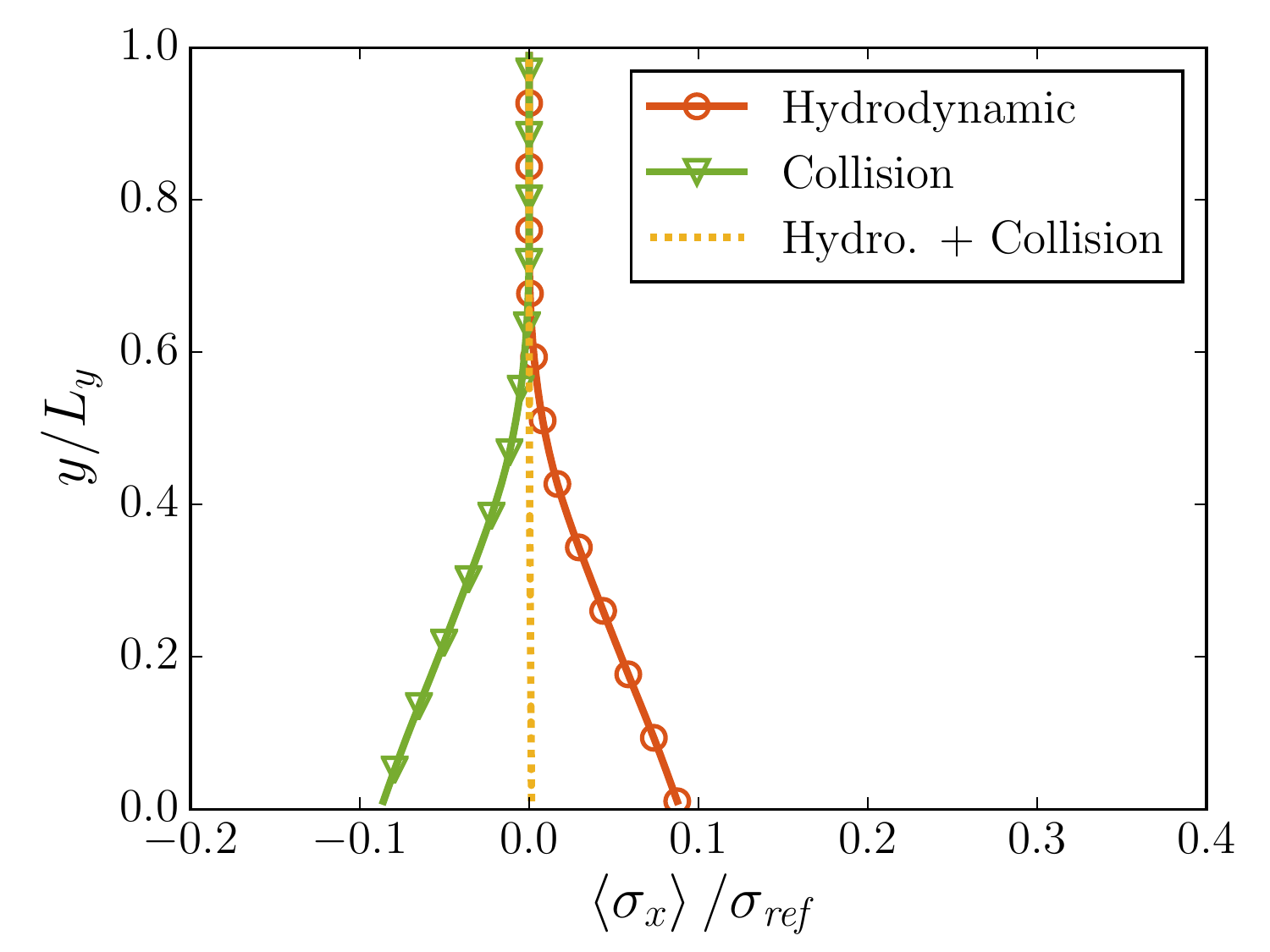}
\caption{Single rolling particle: stress balance of the particle phase in the $x$-direction according to \eqref{eq:px_stress}.  The hydrodynamic stress, which propels the particle in the positive $x$-direction, is in equilibrium with the stress due to the collision between the particle and the lower wall, which slows the particle down.}
\label{fig:single_momx_particle}
\end{figure}

Figure~\ref{fig:single_momx_particle} shows the momentum balance for \eqref{eq:px_stress}, in which the sum of the hydrodynamic and collision stresses is zero. Thus, the hydrodynamic force driving the particle in the positive $x$-direction is balanced by the collision forces between the particle and the lower wall acting in the negative $x$-direction, indicating that the particle is not accelerating. For a single particle, this figure shows only the net hydrodynamic force, $\mbs{F}_{I,p}+\mbs{F}_{\mathit{IBM},p}$, and net collision force, $\mbs{F}_{c,p}$ at the lower wall, smeared by the coarse-graining method. Thus, there exist stresses above the particle diameter ($y/L_y > 0.5$) because the coarse-graining width we chose ($w=0.67D_p$) spreads values beyond the particle radius. For each stress component, the value at the lower wall represents the entire stress (e.g., $\mbs{F}_{c,p} / (L_x \, L_z)$) acting on the particle. At the lower wall, the particle stress in figure~\ref{fig:single_momx_fluid}a matches the hydrodynamic stress in figure~\ref{fig:single_momx_particle}, which is balanced by the collision stress. Therefore, the particle stress at the lower wall in figure~\ref{fig:single_momx_fluid}a represents the stress between the particle and wall due to collision forces.

\subsubsection{Stress balance of the fluid/particle mixture in the $x$-direction}
\label{sec:single_momx_both}

\begin{figure}
\centering
\includegraphics[width=0.5\textwidth]{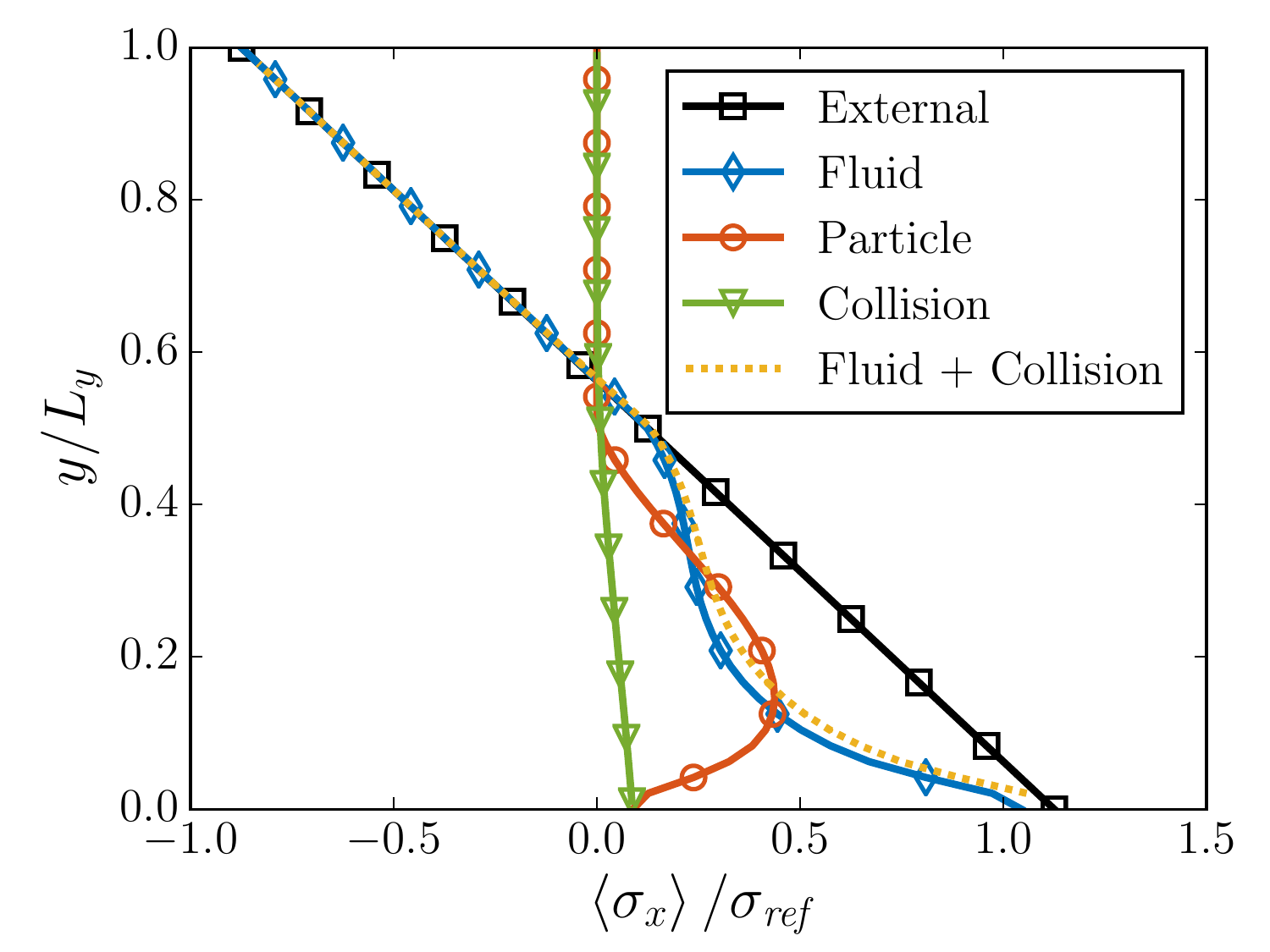}
\caption{Single rolling particle: stress balance in the $x$-direction for the fluid/particle mixture, given by \eqref{eq:bx_stress}.  The sum of the fluid and collision stresses is not in equilibrium with the external stress where the particle is present ($y/L_y < 0.5$), except at the lower wall ($y/L_y=0$).}  \label{fig:single_momx_both}
\end{figure}

We can also consider the momentum balance for the mixture, given by \eqref{eq:bx_stress} and shown in figure~\ref{fig:single_momx_both}. The sum of the fluid and collision stresses matches the external stress in the clear fluid layer above the particle and at the lower wall. However, due to the coarse-graining (smearing) of the collision stress the momentum balance is not closed within the particle region. To understand this imbalance, we have included in this plot the particle stress, which represents the local hydrodynamic interactions that occur along the particle surface, as shown by \eqref{eq:particle_force}. The particle stress and collision stress should be equivalent when the particle acceleration is negligible. However, they match only at the lower wall because the coarse-graining method distributes the collision stress over the volume, and the collision stress matches the net hydrodynamic stress acting on the particle center of mass, which does not account for local variations along its surface. Thus, due to the coarse-graining of the collision stress the momentum budget is closed only when the entire particle is considered.

\subsubsection{Stress balance of the fluid phase in the $y$-direction}
\label{sec:single_momy_fluid}

\begin{figure}
\centering
\placeThreeSubfiguresDown{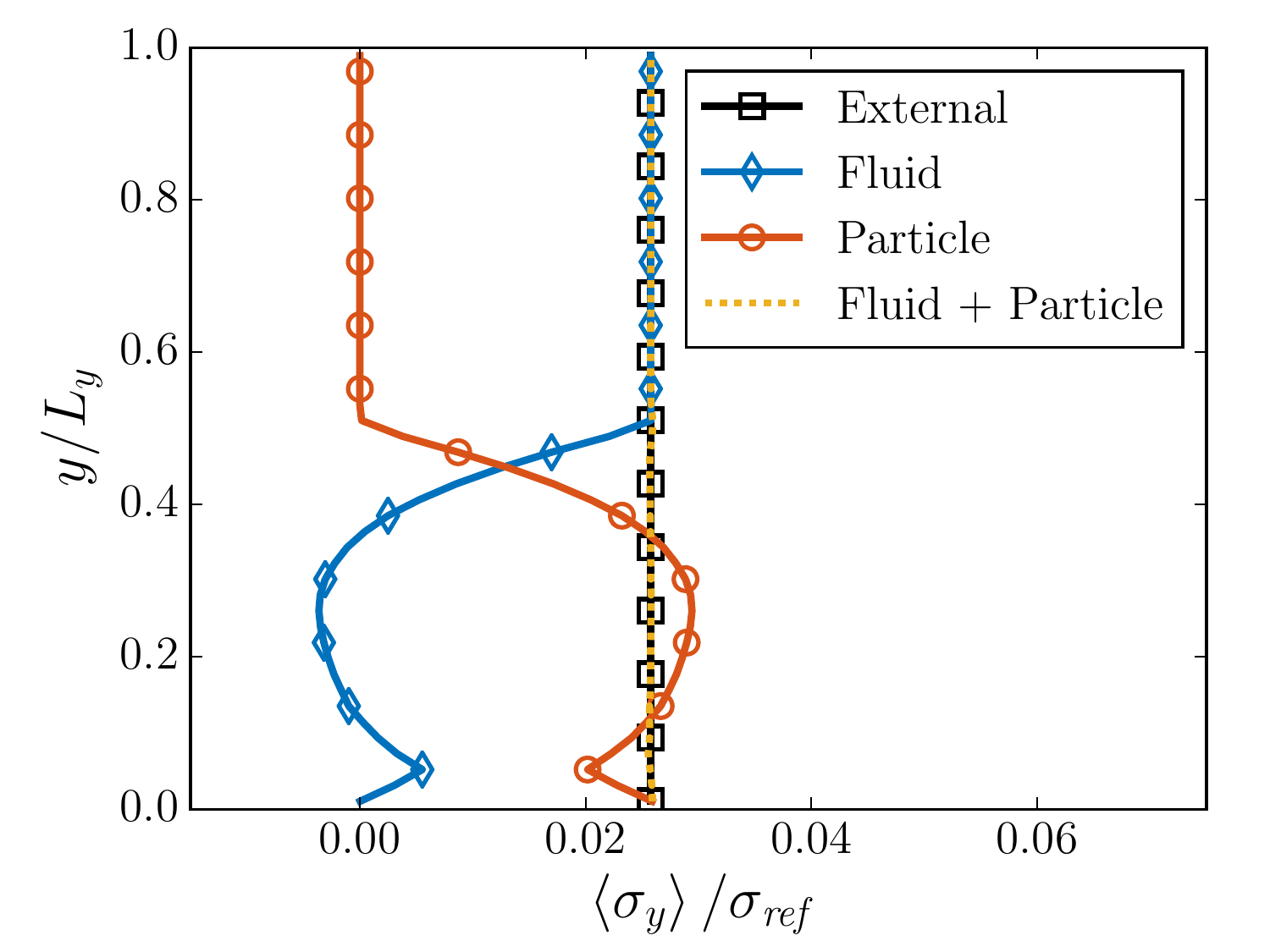}{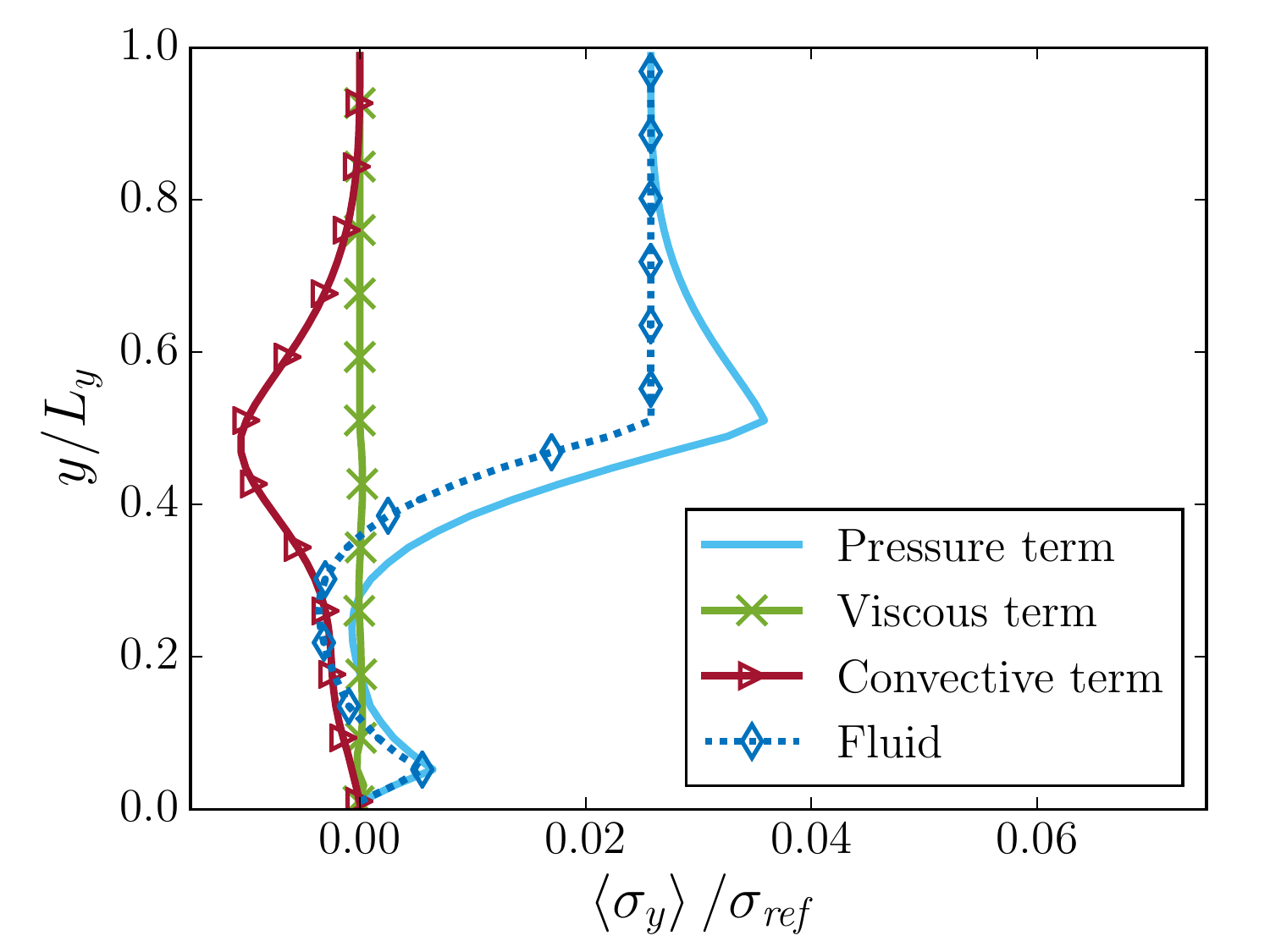}{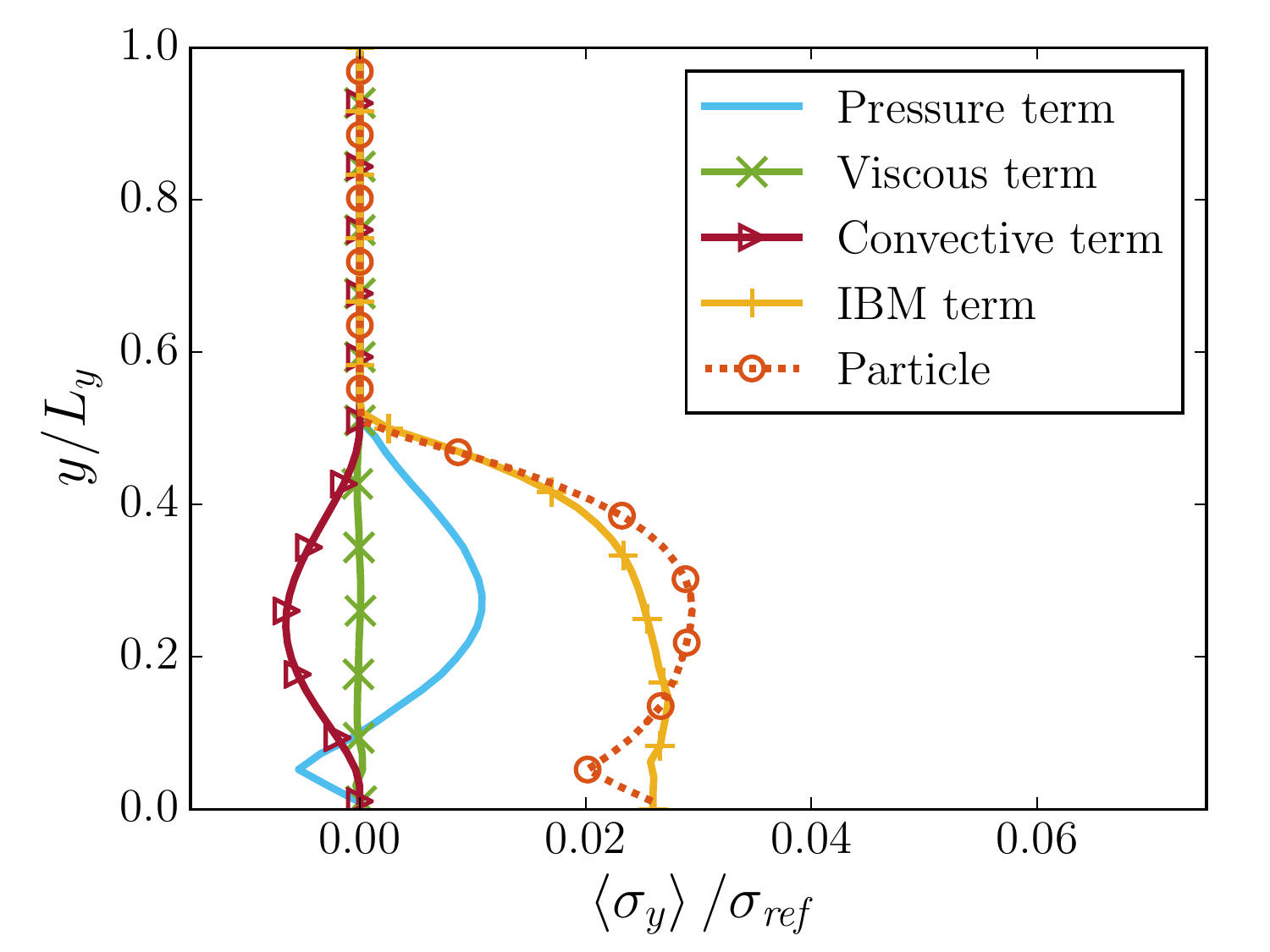}{0.75}{0pt}{0pt}{0pt}
\caption{Single rolling particle: stress balance of the fluid phase in the $y$-direction according to \eqref{eq:fy_stress}.  The components of (a) are further broken down for (b) the fluid stress and (c) the particle stress.  As shown in (a), the sum of the fluid and particle stresses is in equilibrium with the external stress, the fluid stress accounts for the stress in the upper half of the domain, and the particle stress accounts for the stress in the lower half of the domain.  As shown in (b), the pressure term ($\sigma_\mathit{Fpres,y}$) and convective term ($\sigma_\mathit{Fconv,y}$) account for the fluid stress.  (c) indicates that the pressure term ($\sigma_\mathit{Ppres,y}$), convective term ($\sigma_\mathit{Pconv,y}$), and IBM term ($\sigma_\mathit{PIBM,y}$) account for the particle stress.}
\label{fig:single_momy_fluid}
\end{figure}

We present the results for the momentum balance of the fluid phase in the $y$-direction, given by \eqref{eq:fy_stress}, in figure~\ref{fig:single_momy_fluid}.  Figure~\ref{fig:single_momy_fluid}a shows the balance between the external stress, comprised of the pressure and viscous terms at the top wall, and the sum of fluid and particle stresses. Different from the $x$-momentum balance, the external stress for the $y$-momentum fluid phase does not depend on the $y$-coordinate, but instead maintains a constant value.  This stress is carried exclusively by the fluid in the upper half of the domain and is then almost completely transferred to the particle in the lower half of the domain. The particle stress represents the lift force acting on the particle phase, and its value at the lower wall represents the total lift acting on the particle. The majority of the lift stress occurs along the top of the particle ($0.4 < y/L_y < 0.5$). The particle stress at the lower wall is equivalent to the external stress or the fluid stress at the upper wall.  The increasing fluid stress towards the upper wall indicates a decreasing fluid pressure according to the definition of the fluid stress in \eqref{eq:fy_stress}.  As a reminder, the hydrostatic pressure has been subtracted out from the fluid pressure. Thus, the lift force on the particle is supported by the lower pressure in the fluid at the upper wall. However, the maximum fluid stresses for the $y$-momentum balance in figure~\ref{fig:single_momy_fluid}a are up to two orders of magnitude smaller than those for the $x$-momentum balance in figure~\ref{fig:single_momx_fluid}a.

Figure~\ref{fig:single_momy_fluid}b decomposes the fluid stress in figure~\ref{fig:single_momy_fluid}a into the pressure, viscous, and convective terms, given by $\sigma_\mathit{Fpres,y}$, $\sigma_\mathit{Fvisc,y}$, and $\sigma_\mathit{Fconv,y}$ in \eqref{eq:fy_stress}, respectively.  Similarly, figure~\ref{fig:single_momy_fluid}c decomposes the particle stress in figure~\ref{fig:single_momy_fluid}a into the IBM, pressure, viscous, and convective terms, given by $\sigma_\mathit{PIBM,y}$, $\sigma_\mathit{Ppres,y}$, $\sigma_\mathit{Pvisc,y}$, and $\sigma_\mathit{Pconv,y}$ in \eqref{eq:fy_stress}, respectively.  Contrary to the $x$-momentum balance, we find the pressure and convective terms to be significant and the viscous term to be negligible.  We can thus simplify \eqref{eq:fy_stress} to obtain the approximate balance
\begin{IEEEeqnarray}{rCl} \label{eq:fy_stress_simplified}
\underbrace{
-\left<\left.p\right|_{L_y}\right>
}_\text{External stress}
&=& \underbrace{
-\left<\left.\gamma p\right|_y \right>
- \rho_f \left<\left.\gamma vv\right|_y \right>
}_\text{Fluid stress} \nonumber\\
&&\underbrace{
- \int_y^{L_y} \left<f_{\mathit{IBM,y}}\right> {\mrm{d}y}
- \left<\left.\phi p\right|_y \right>
- \rho_f \left<\left.\phi vv\right|_y \right>
}_\text{Particle stress} .
\end{IEEEeqnarray}
Note that the IBM term matches the particle stress only at the lower wall; accounting for the pressure and convective terms inside the particle is important for resolving the particle stress throughout the domain.

\subsubsection{Stress balance of the particle phase in the $y$-direction}
\label{sec:single_momy_particle}

\begin{figure}
\centering
\includegraphics[width=0.5\textwidth]{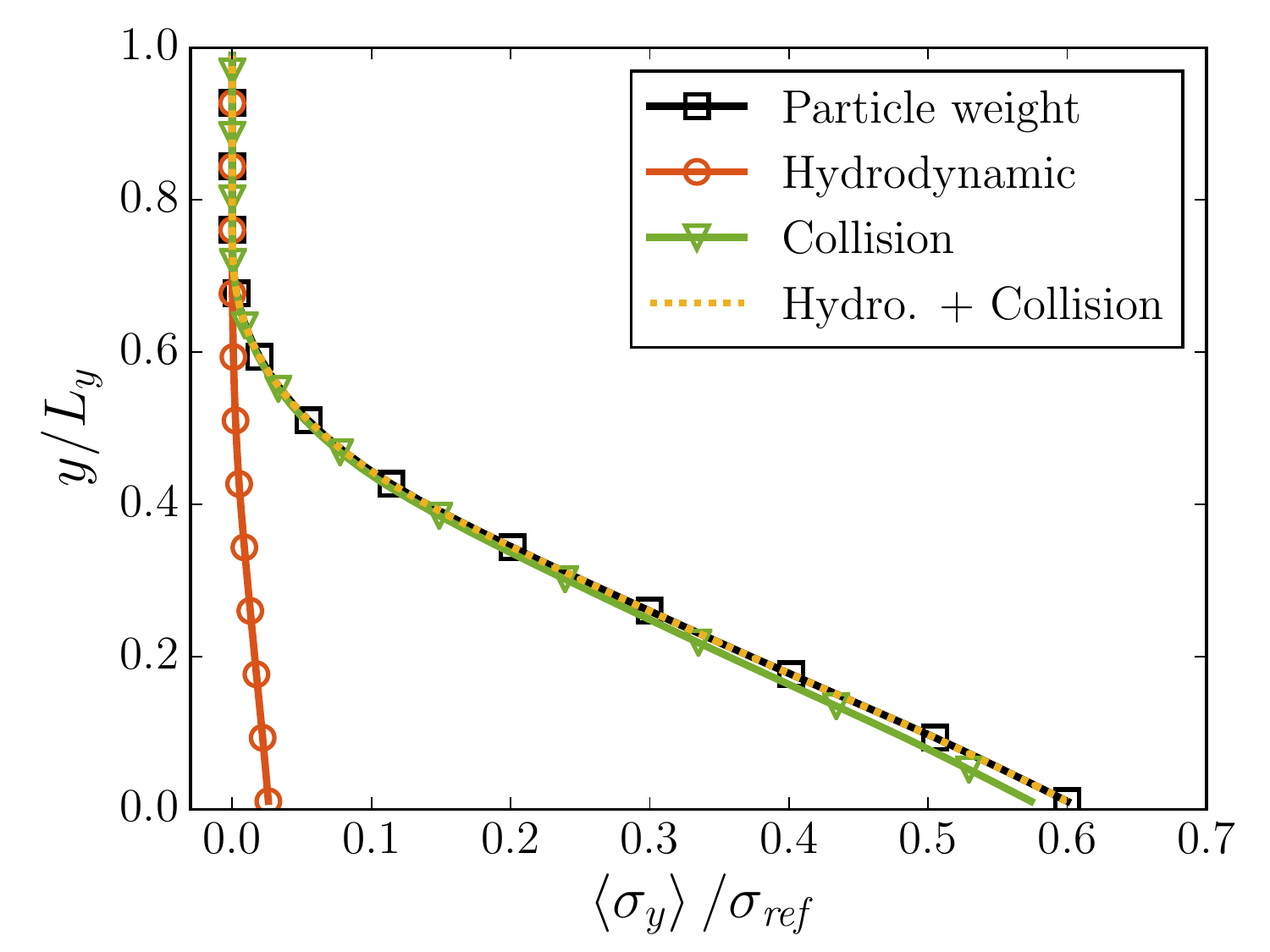}
\caption{Single rolling particle: stress balance of the particle phase in the $y$-direction according to \eqref{eq:py_stress}. The sum of the hydrodynamic and collision stresses is in equilibrium with the particle weight. The collision force with the lower wall largely supports the particle's weight, but a small hydrodynamic lift force also contributes.}  \label{fig:single_momy_particle}
\end{figure}

Figure~\ref{fig:single_momy_particle} shows the stress balance for the particle phase given by \eqref{eq:py_stress}, in which the particle weight is in equilibrium with the sum of the hydrodynamic stress and the collision stress.  In this case, the particle weight represents the gravitational force, $\mbs{F}_{g,p}$ acting on the single particle, smeared by the coarse-graining method.  The fluid exerts a positive lift force on the particle, but the vast majority of the particle's weight is supported by the collision force with the lower wall. Comparing figure~\ref{fig:single_momx_particle} to figure~\ref{fig:single_momy_particle}, we can see that the lift force (related to the $y$-momentum hydrodynamic stress) is a fraction of the drag force (related to the $x$-momentum hydrodynamic stress), but the $y$-momentum collision force is six times larger than the drag force. In fact, the particle weight and collision stress are the only terms in the $y$-momentum balance that are comparable in magnitude to the stresses in the $x$-momentum balance. Having analyzed the momentum balances for the simple test case of a rolling particle in a pressure-driven channel flow, we now proceed to the more complex case of a sediment bed involving many particles.
\subsection{Stress balance of a sheared particle bed} \label{sec:bed_mom}

\subsubsection{Time evolution of the particle bed} \label{sec:bed_time}

\begin{figure}
\placeThreeSubfiguresLine{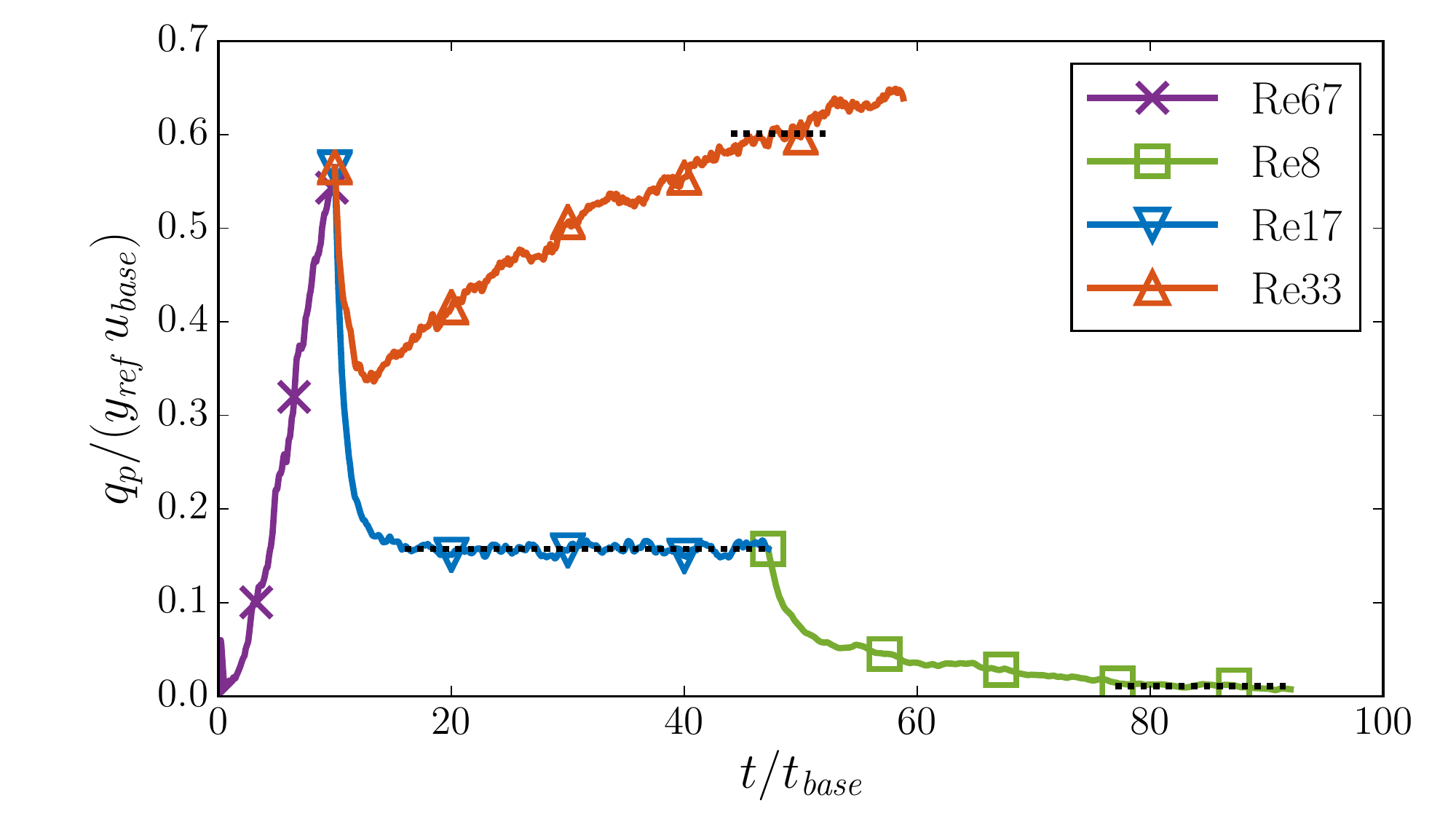}{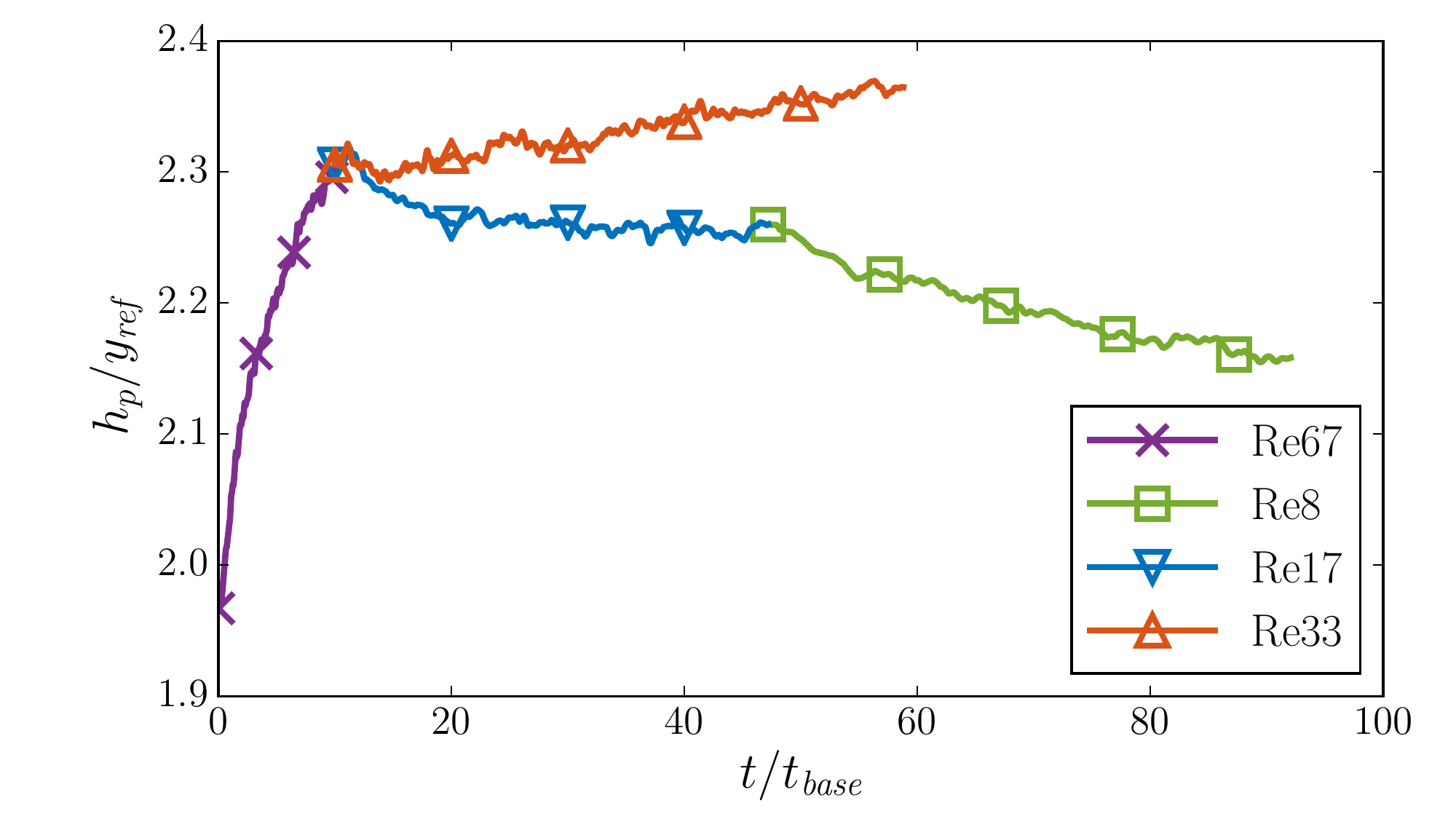}{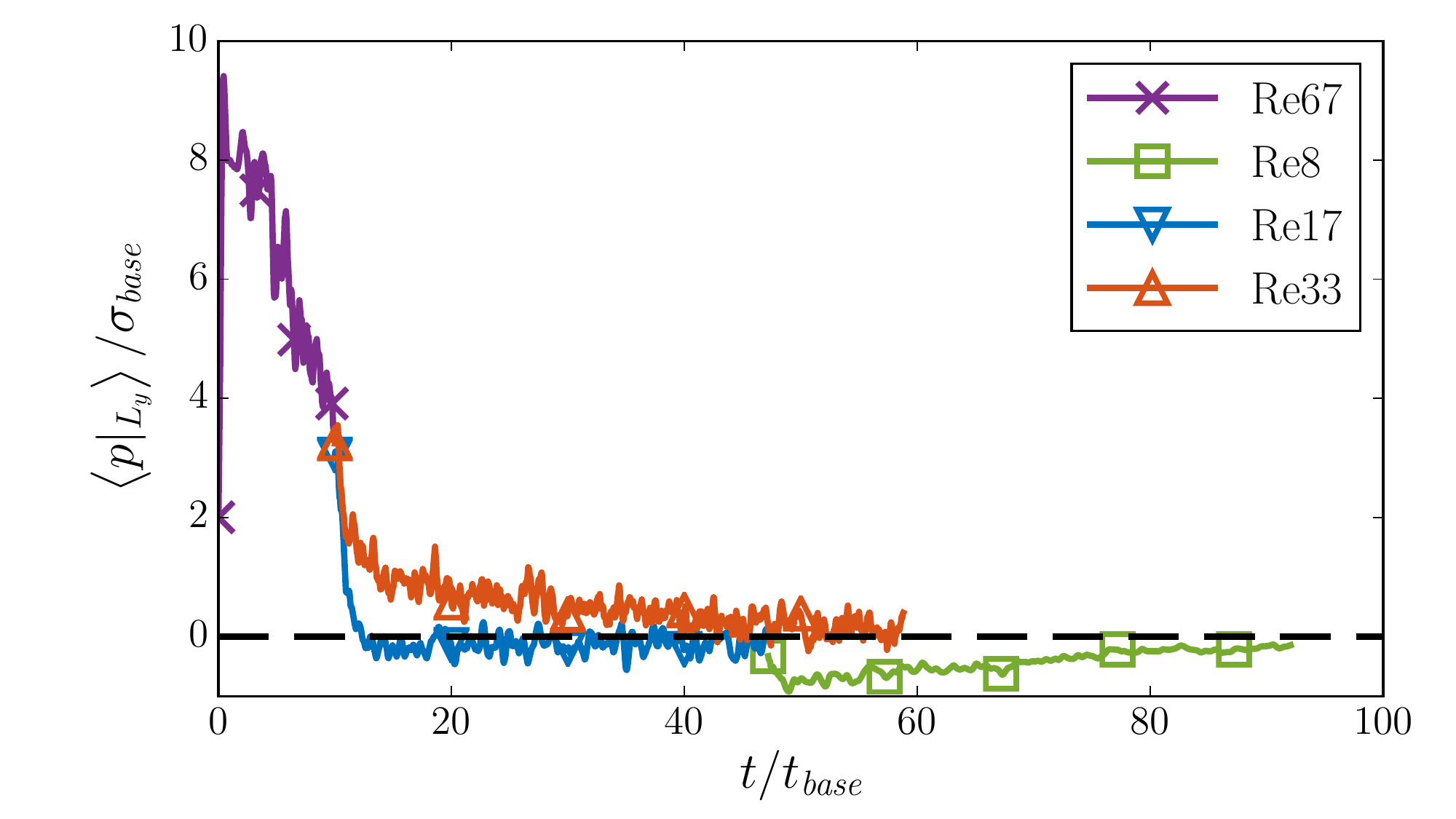}{0.65\textwidth}{0.5625}{0pt}{0pt}{0pt}
\caption{Sheared particle bed: (a) Particle volumetric flux, given by \eqref{eq:p_flux}, for the different simulation runs.  Dotted lines indicate the average particle flux over the averaging time interval for each simulation. (b) Bed height, defined at $\left<\phi\right> = 0.05$. (c) Spatially-averaged fluid pressure at the top wall, relative to the lower wall and with the hydrostatic pressure subtracted out. The particle beds for runs Re67 and Re33 exhibit dilation while the bed for run Re8 contracts. Run Re17 has a constant particle flux and bed height during the time-averaging window.}
\label{fig:bed_p_flux}
\end{figure}

\begin{figure}
\centering
\placeTwoSubfigures{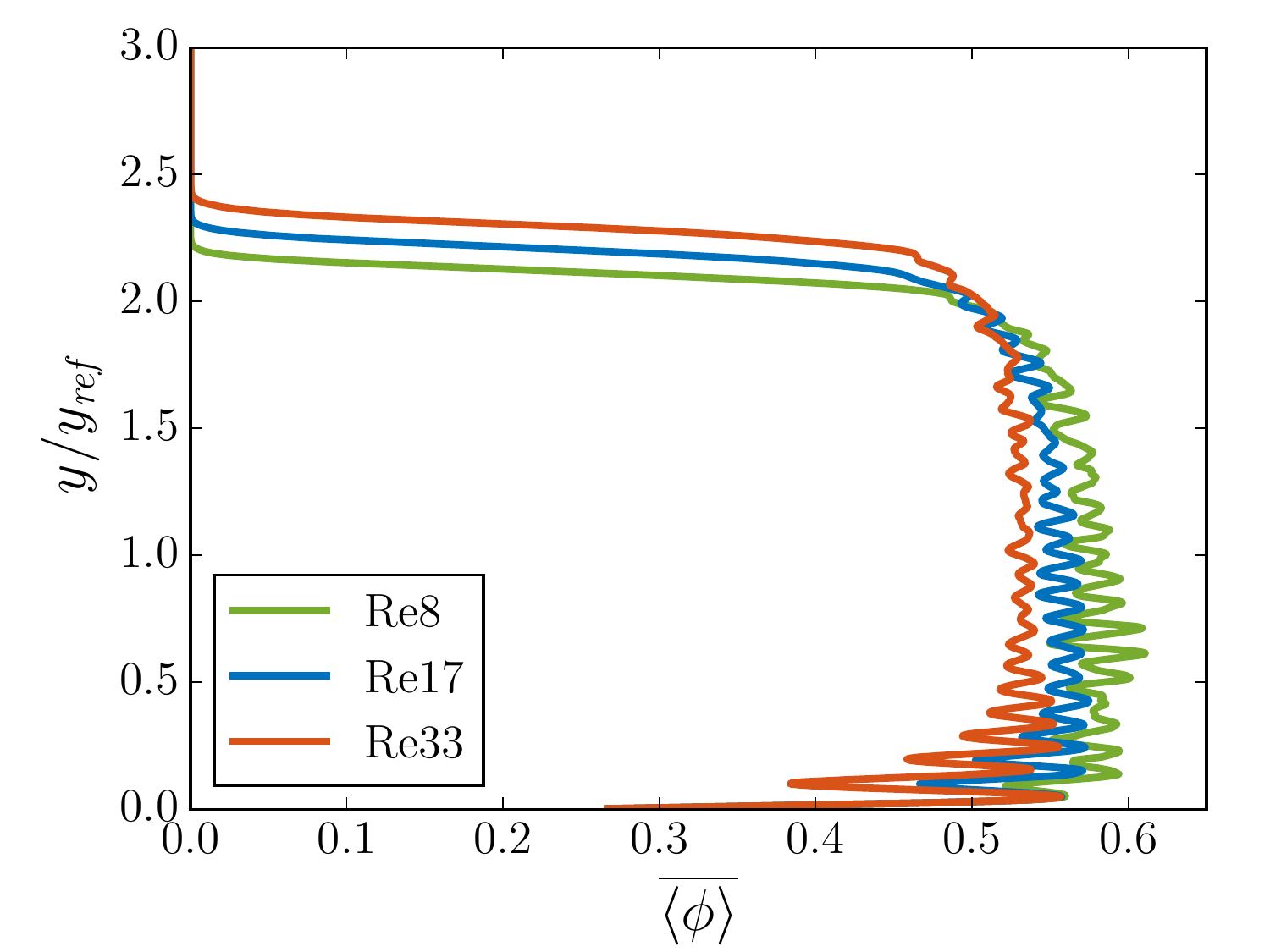}{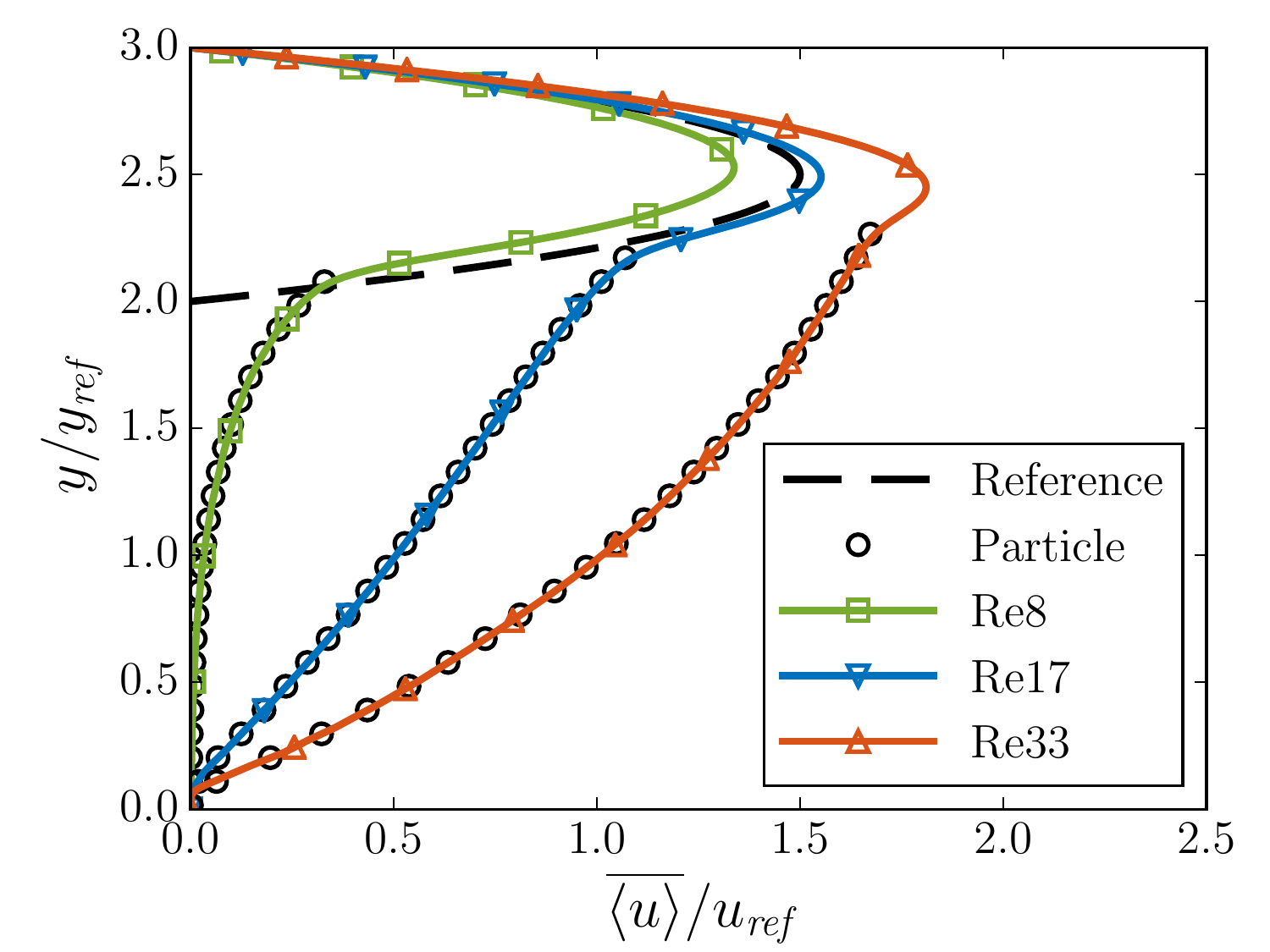}{0.75}{0pt}{0pt}{0pt}
\caption{Sheared particle bed: profiles for the different simulation runs averaged horizontally and in time for (a) the particle volume fraction and (b) the streamwise velocity.  The average fluid velocity is given by \eqref{eq:u_fluid}, while the average coarse-grained particle velocity is given by \eqref{eq:cg_u}.}  \label{fig:bed_flow}
\end{figure}

We conducted simulations of a Poiseuille flow over a particle bed at four different flow rates: one to initialize the bed (Re67) and three to study the bed under different flow conditions (Re8, Re17, Re33).  The time evolution of various bulk quantities for these simulations is shown in figure~\ref{fig:bed_p_flux}. Figure~\ref{fig:bed_p_flux}a shows the particle flux, $q_p$, over time for the different simulation runs, where we use the volumetric particle flux per unit width
\begin{equation} \label{eq:p_flux}
q_p = \frac{1}{L_x \, L_z} \sum_{p=1}^{N_p} V_p u_{p,x} .
\end{equation}
The particle flux rapidly increases during run Re67, accompanied by an increase in the bed height, $h_p$, or dilation of the particle bed (figure~\ref{fig:bed_p_flux}b).  Here we define the bed height to be the location at which the local particle volume fraction becomes $\left<\phi\right>=0.05$ to be consistent with our definition in \citet{Biegert2017a}. Upon resuming run Re67 at a reduced pressure gradient, we find that run Re17 quickly reaches a steady-state configuration, characterized by a constant particle flux and bed height. Run Re33, on the other hand, does not reach a steady-state during the simulation time, and the particle flux and bed height continue to increase with time. Run Re8, which was resumed from run Re17, experiences a continuing decline in the particle flux and bed height. Figure~\ref{fig:bed_flow}a shows the volume fractions of the beds for the three simulations averaged in space and time, where the bed height increases for the higher flow rates.  Accompanying this dilation is a decrease in the volume fraction of particles within the bed.
Consider the pressure at the top wall as a function of time, shown in figure~\ref{fig:bed_p_flux}c. Recall from section~\ref{sec:single_momy_fluid} that a negative pressure at the top wall corresponds to a lift force acting on the particle phase. Comparing figures~\ref{fig:bed_p_flux}b and~\ref{fig:bed_p_flux}c, we can see a clear correlation between a positive pressure during bed dilation (increasing $h_p$), a negative pressure during bed contraction (decreasing $h_p$), and a slightly-negative pressure during steady-state. Thus, when the particle bed tries to dilate, the fluid immediately responds with a negative lift force, and in turn responds to bed contraction with a positive lift force. Alternatively, we can imagine that, when the bed dilates, a positive pressure forms above the bed as fluid flows into the bed to fill the void space, while, when the bed contracts, a negative pressure forms above the bed as the particles squeeze fluid out of the bed.
These three simulations provide an opportunity to study the forces governing the particle bed evolution and to explore the imbalances that cause the bed to dilate or contract in order to reach a steady state. Furthermore, we note that measuring the porosity of a sediment bed is straightforward within our numerical framework, but it is by no means trivial in an experimental setup as noted by \cite{Aussillous2013}. Nevertheless, the porosity remains a crucial parameter for continuum modeling, which has so far been a serious impediment for these type of models.

We will investigate the momentum balances of these simulations in part to understand these bed transitions. Although neither one of these two simulations is at a steady-state, we focus on runs Re8 and Re33 to explore the mechanisms behind bed contraction and dilation, as well as the origins of the different fluid and particle velocity profile shapes, shown in figure~\ref{fig:bed_flow}b.
As described in section~\ref{sec:bed_setup}, we use time-averaging to smooth out the fluctuations due to particle-particle interactions. These time averages, which were used to generate figure~\ref{fig:bed_flow} as well as the stress balance results, are given in table~\ref{tab:bed_runs} and shown graphically by the dotted lines in figure~\ref{fig:bed_p_flux}a. According to \cite{Jenkins2017}, the three cases represent three distinctively different regimes. The sediment bed of Re8 approaches a ``glassy" regime, whereas the sediment motion in Re17 and Re33 can be considered ``layered" and ``collisional," respectively.
The simulation data show that there is very little slip between the fluid and particle phases and that the reference velocity, $u_\mathit{ref}$, provides a reasonable estimate for the fluid velocity in the clear fluid layer above the particles for Re17 ($y/y_\mathit{ref} > 2.3$), even when the entire particle bed is in motion. However, increasing the flow rate does increase the velocity profile relative to the reference case. There is a clear qualitative difference between run Re8, whose velocity profile is concave and goes to zero within the bed at $y/y_\mathit{ref} \approx 0.5$, and run Re33, whose velocity profile is convex and goes to zero only at the fixed particles at the lower wall. For brevity, we will neglect the momentum balance for Re17, which is similar in bed morphology to Re33. The fact that the fluid velocity is equal to the particle velocity is consistent with the observation of \cite{Aussillous2013}. This has important implications for our perspective on continuum modeling such as $\mu(I)$-rheology, as the data shown in figure \ref{fig:bed_flow}b can be used to compute the shear rate $\partial u / \partial y$.

\subsubsection{Stress balance of the fluid phase in the $x$-direction}
\label{sec:bed_momx_fluid}

We now investigate the momentum balance for the simulations involving a bed of mobile particles, focussing on runs Re8 and Re33 to get a sense of the results for different flow conditions. In order to obtain steady-state results, we apply the time-averaging operator \eqref{eq:time_average} to the $x$-momentum balances \eqref{eq:fx_stress} and \eqref{eq:px_stress}, resulting in double-averaged equations akin to \citet{Nikora2013} and \citet{Vowinckel2017b}.

\begin{figure}
\placeSixSubfigures{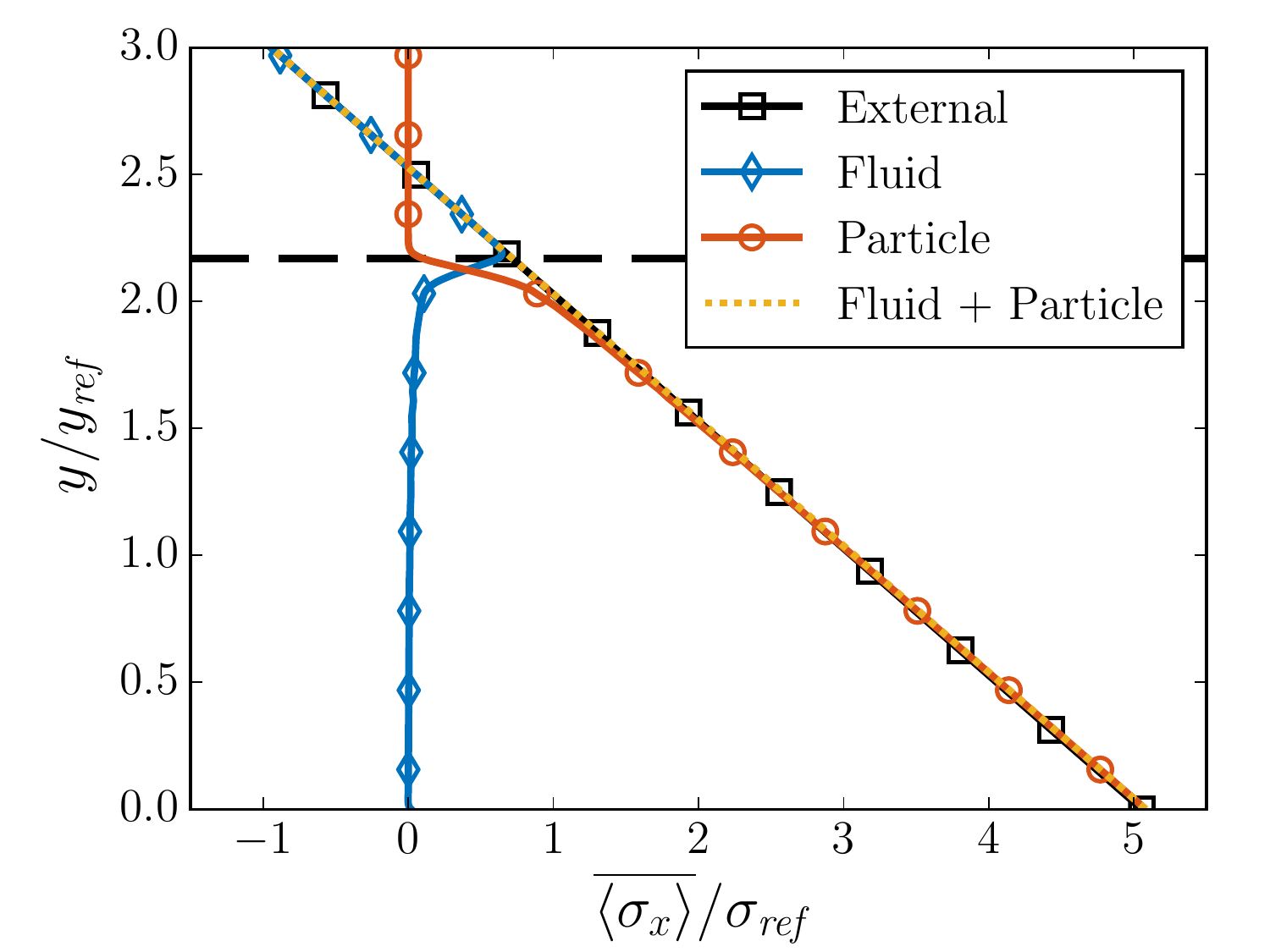}{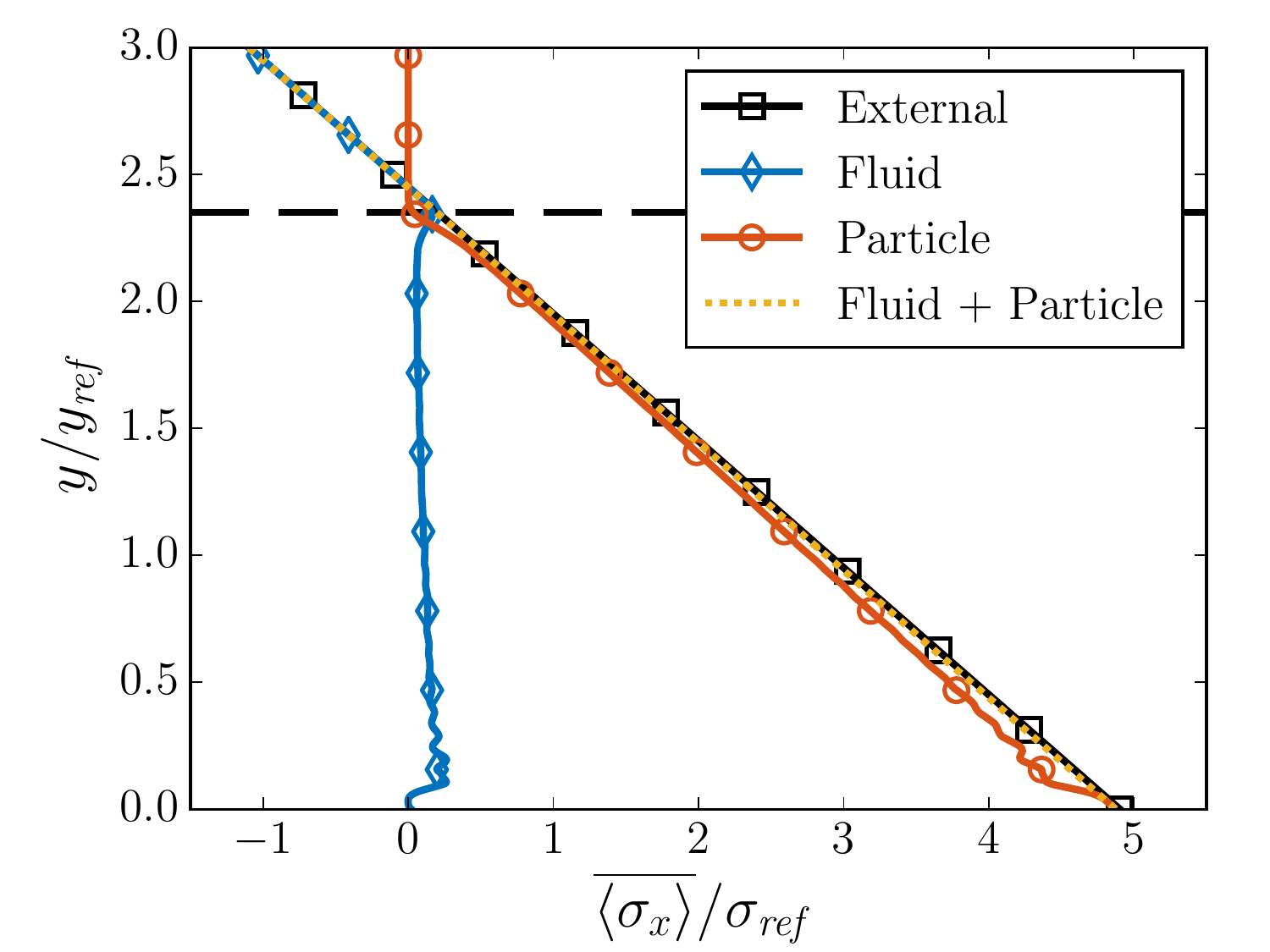}{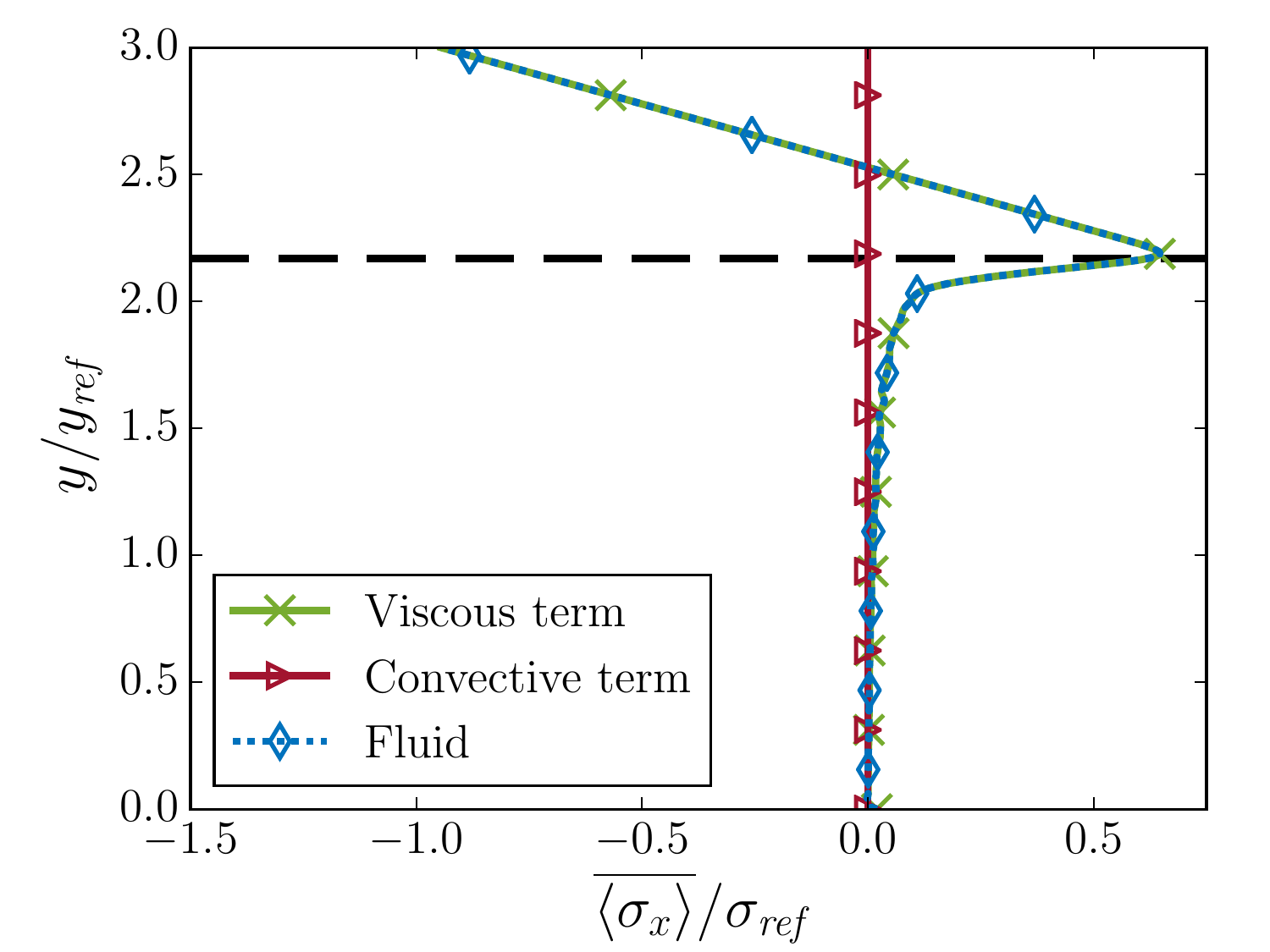}{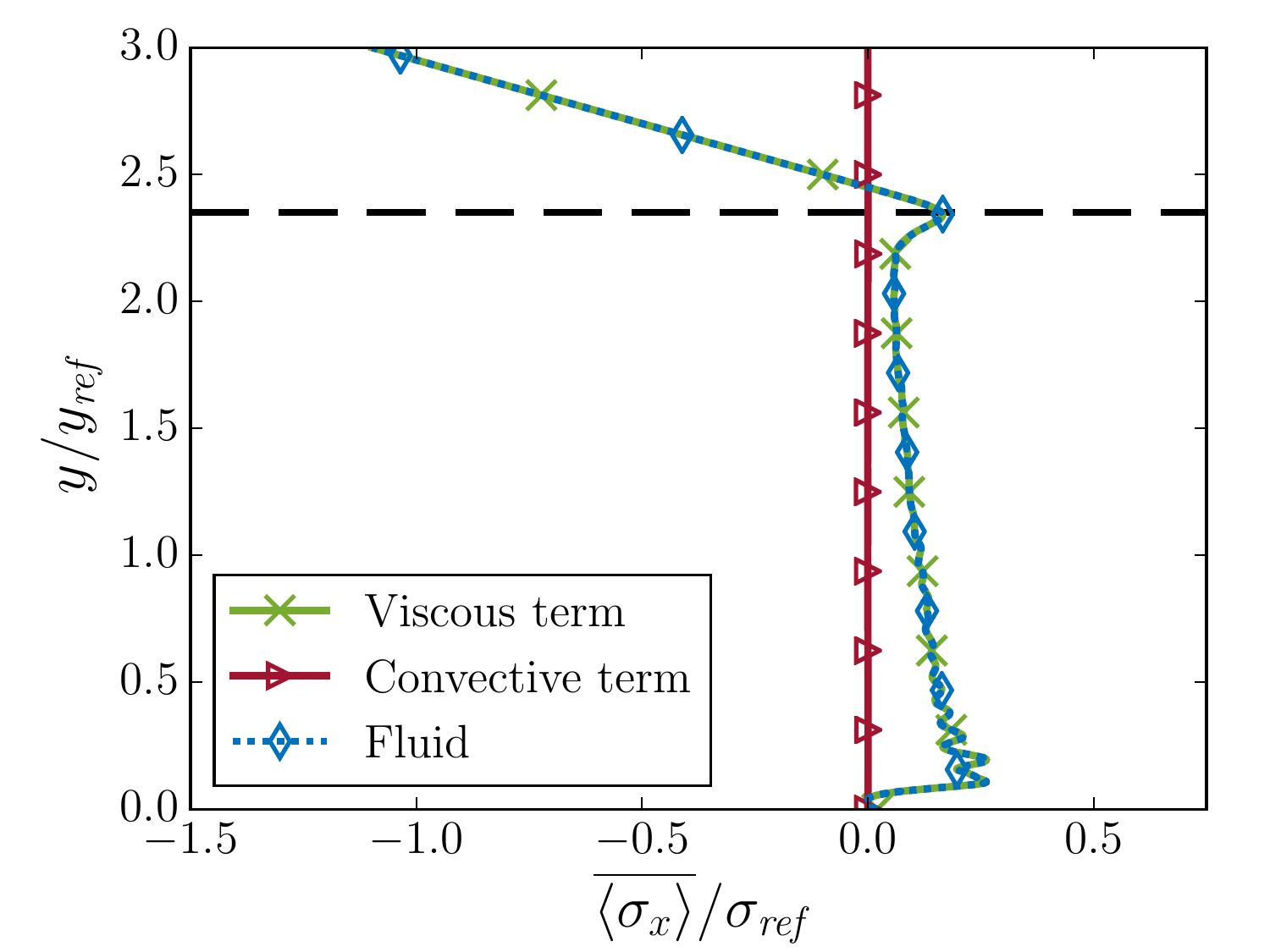}{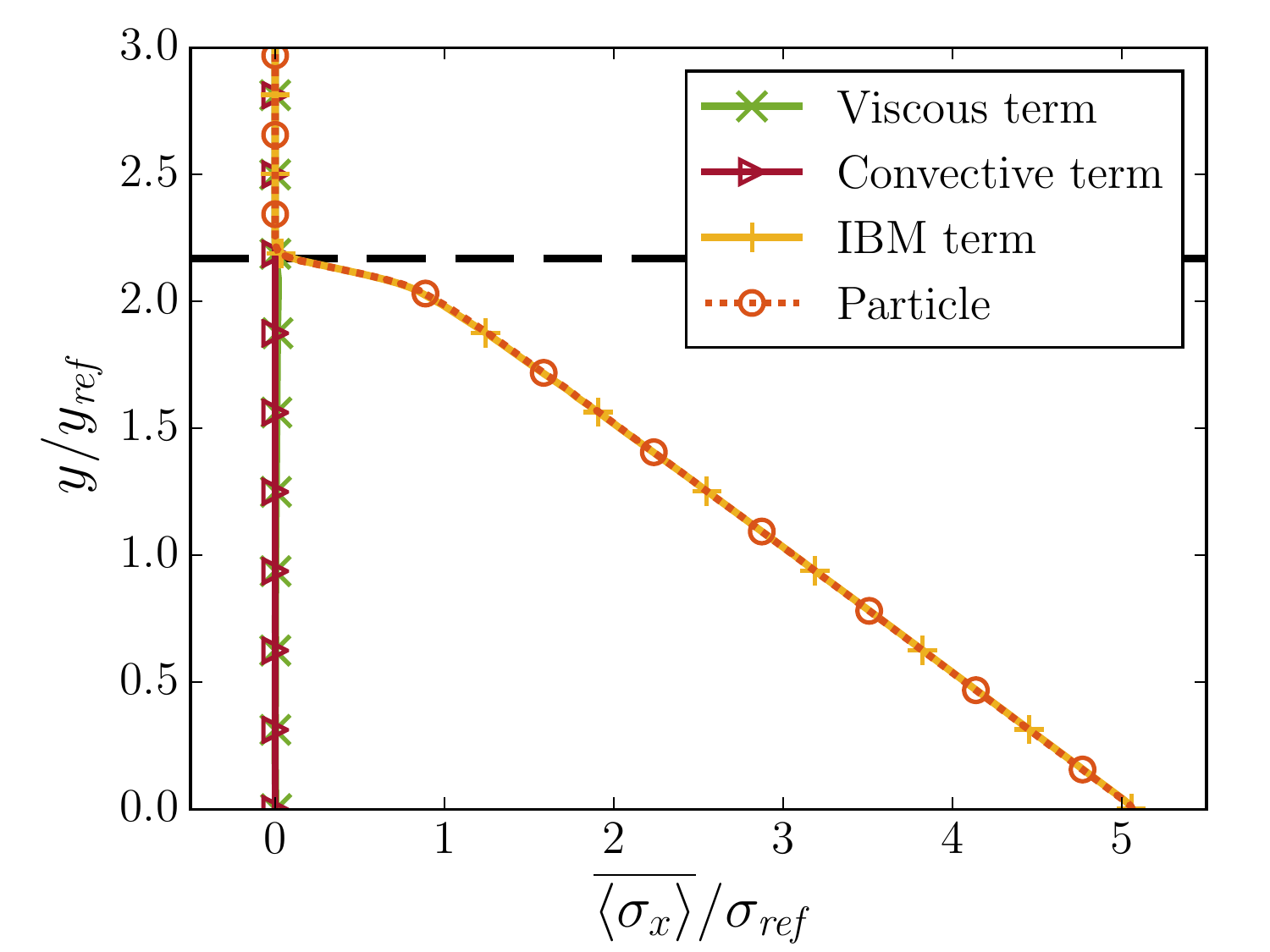}{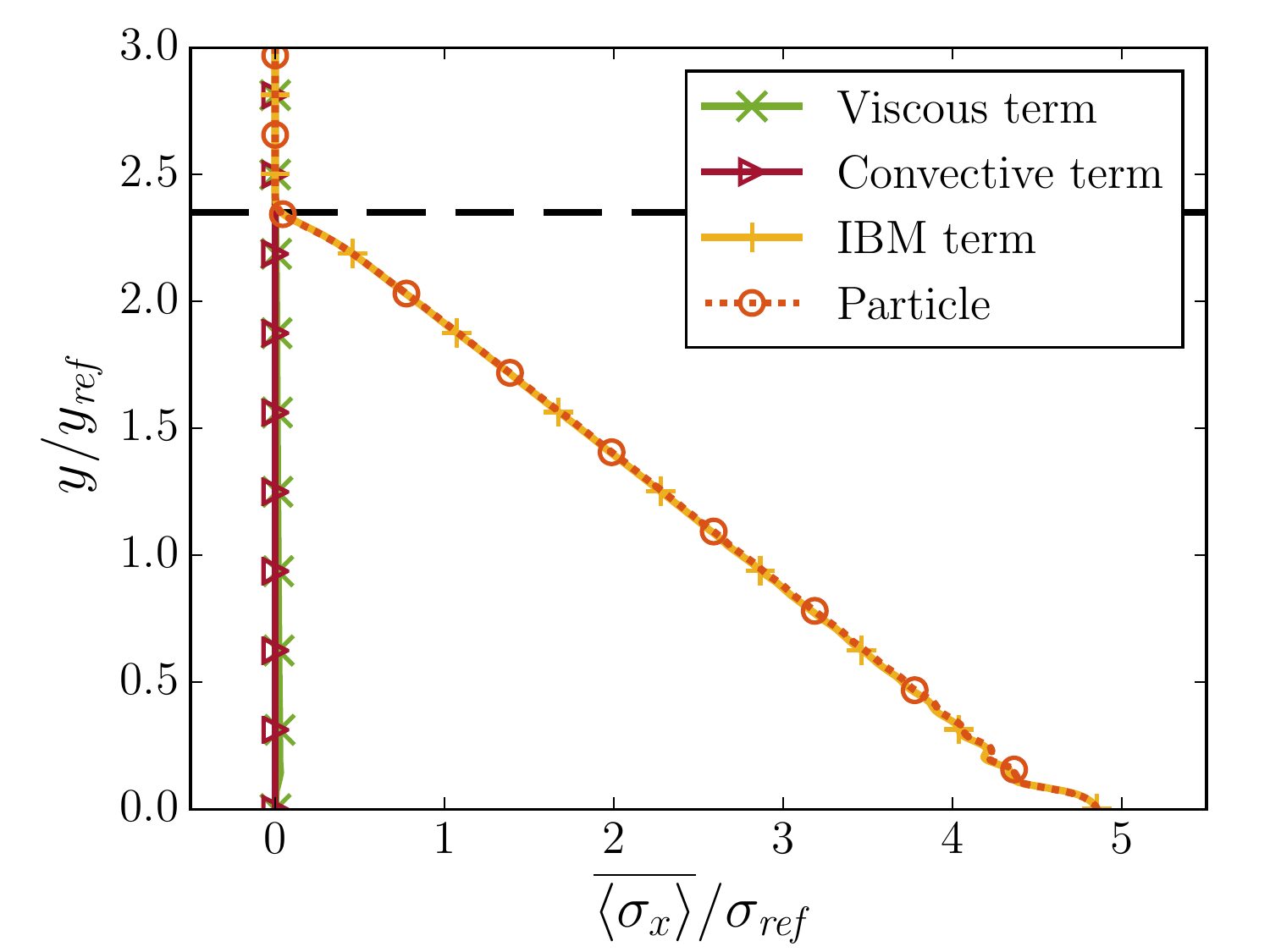}{0.75}{0pt}{0pt}{0pt}
\caption{Sheared particle bed: stress balance of the fluid phase in the $x$-direction according to \eqref{eq:fx_stress}. Frames (a), (c), and (e) correspond to run Re8 while (b), (d), and (f) correspond to run Re33. The components of (a) and (b) are further broken down in (c) and (d) for the fluid stress, and in (e) and (f) for the particle stress. The horizontal dashed line marks the height of the particle bed, $h_p$. As shown in (a) and (b), the sum of the fluid and particle stresses is in equilibrium with the external stress and consists of mostly the particle stress within the bed and the fluid stress above the bed. As shown in (c) and (d), the viscous term ($\sigma_\mathit{Fvisc,x}$) accounts for almost all of the fluid stress. Frames (e) and (f) demonstrate that the IBM term ($\sigma_\mathit{PIBM,x}$) makes up nearly the entire particle stress.}
\label{fig:bed_momx_fluid}
\end{figure}

Figure~\ref{fig:bed_momx_fluid} shows the momentum balance of the fluid phase, given by \eqref{eq:fx_stress}, for runs Re8 (left column) and Re33 (right column), in which we expect the external stress to match the sum of the fluid and particle stresses.  In figures~\ref{fig:bed_momx_fluid}a and~\ref{fig:bed_momx_fluid}b, the external stress at the top wall is close to $\overline{\left<\sigma_x\right>}/\sigma_\mathit{ref}=-1$, which is the stress at the top wall we would expect from the reference case.  This result is consistent with the observation that the velocity profiles in figure~\ref{fig:bed_flow}b are similar to that of the reference case, so that the chosen scaling seems appropriate.  In the upper part of the flow ($y/y_\mathit{ref} > 2.3$), there are no particles, and the fluid stress matches all of the external stress.  Within the particle bed ($y/y_\mathit{ref} < 2.3$), however, the majority of the external stress is taken up by the particles. As expected, the total stress comes out to be a linear profile, which would make it conceptually easy from the perspective of continuum modeling.

Figures~\ref{fig:bed_momx_fluid}c and~\ref{fig:bed_momx_fluid}d show the terms in \eqref{eq:fx_stress} that contribute to the fluid stress.  The viscous term, $\sigma_\mathit{Fvisc,x}$, alone contributes to the fluid stress, which is consistent with the observations for the single rolling sphere. Run Re8 differs from Re33 in that the fluid stress reaches a higher positive value above the particle bed and quickly drops to zero within the particle bed.  The fluid stress for Re33, on the other hand, reaches a somewhat constant value within the particle bed, increasing towards the lower wall. These results are consistent with the velocity profiles in figure~\ref{fig:bed_flow}b, where the concavity of the profile for Re8 results in a high shear stress at the fluid/particle bed interface and low stresses within the bed, while the convexity of the profile for Re33 leads to a large shear stress at the lower wall.

Figures~\ref{fig:bed_momx_fluid}e and~\ref{fig:bed_momx_fluid}f show the terms in \eqref{eq:fx_stress} that contribute to the particle stress. Similar to the single sphere simulation, the IBM term $\sigma_\mathit{PIBM,x}$ accounts for practically the entire particle stress. The major differences in these curves between the two simulations is that the stress for Re8 increases rapidly at the fluid/particle bed interface, then more gradually within the bed, whereas the stress for Re33 increases gradually at the fluid/particle interface and within the bed, and then more rapidly at the lower wall. This result is consistent with the locations of the sharp gradients in the fluid stress balance, so that the fluid and particle stresses together close the $x$-momentum balance.

\subsubsection{Stress balance of the particle phase in the $x$-direction}
\label{sec:bed_momx_particle}

\begin{figure}
\placeFourSubfigures{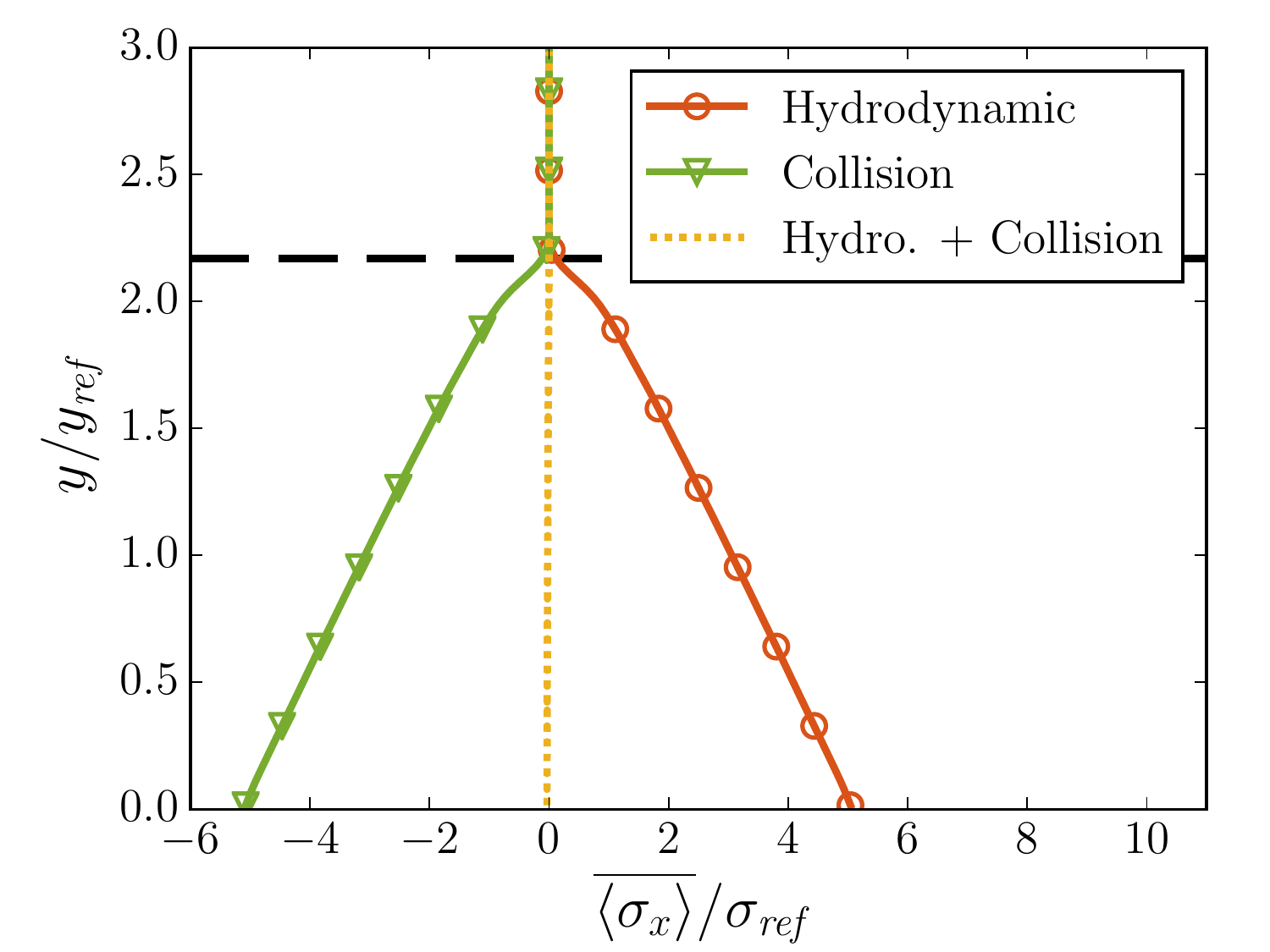}{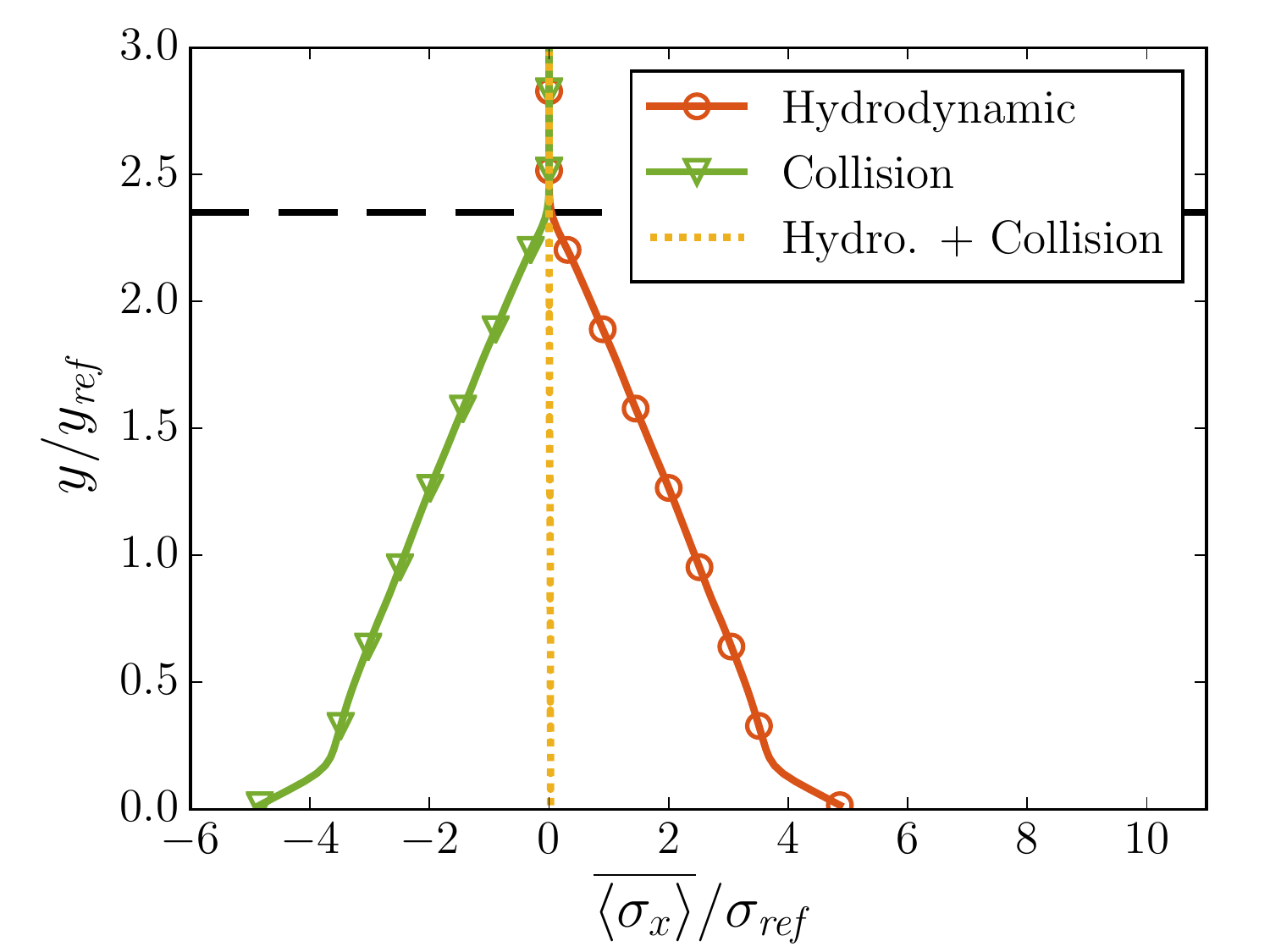}{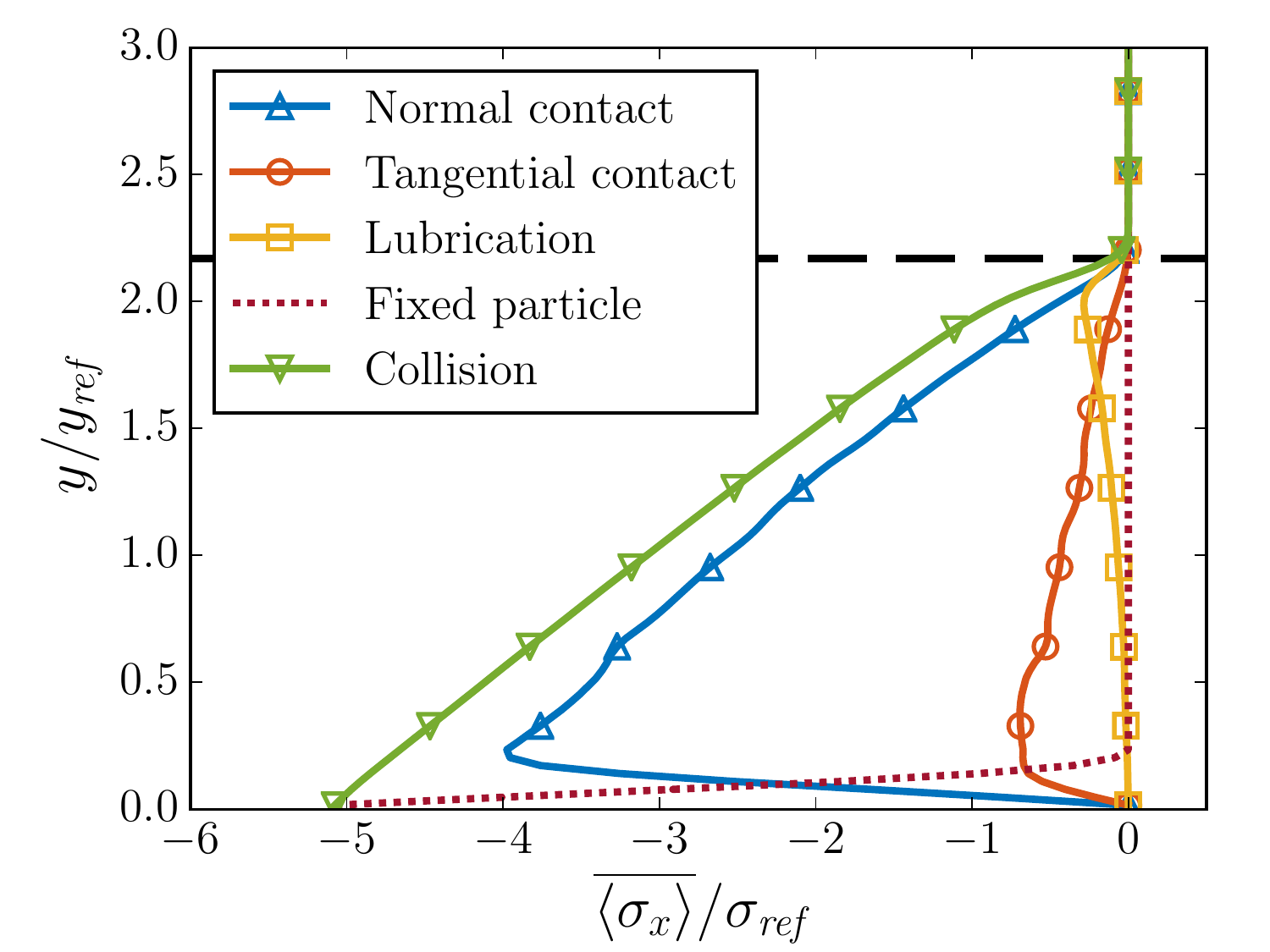}{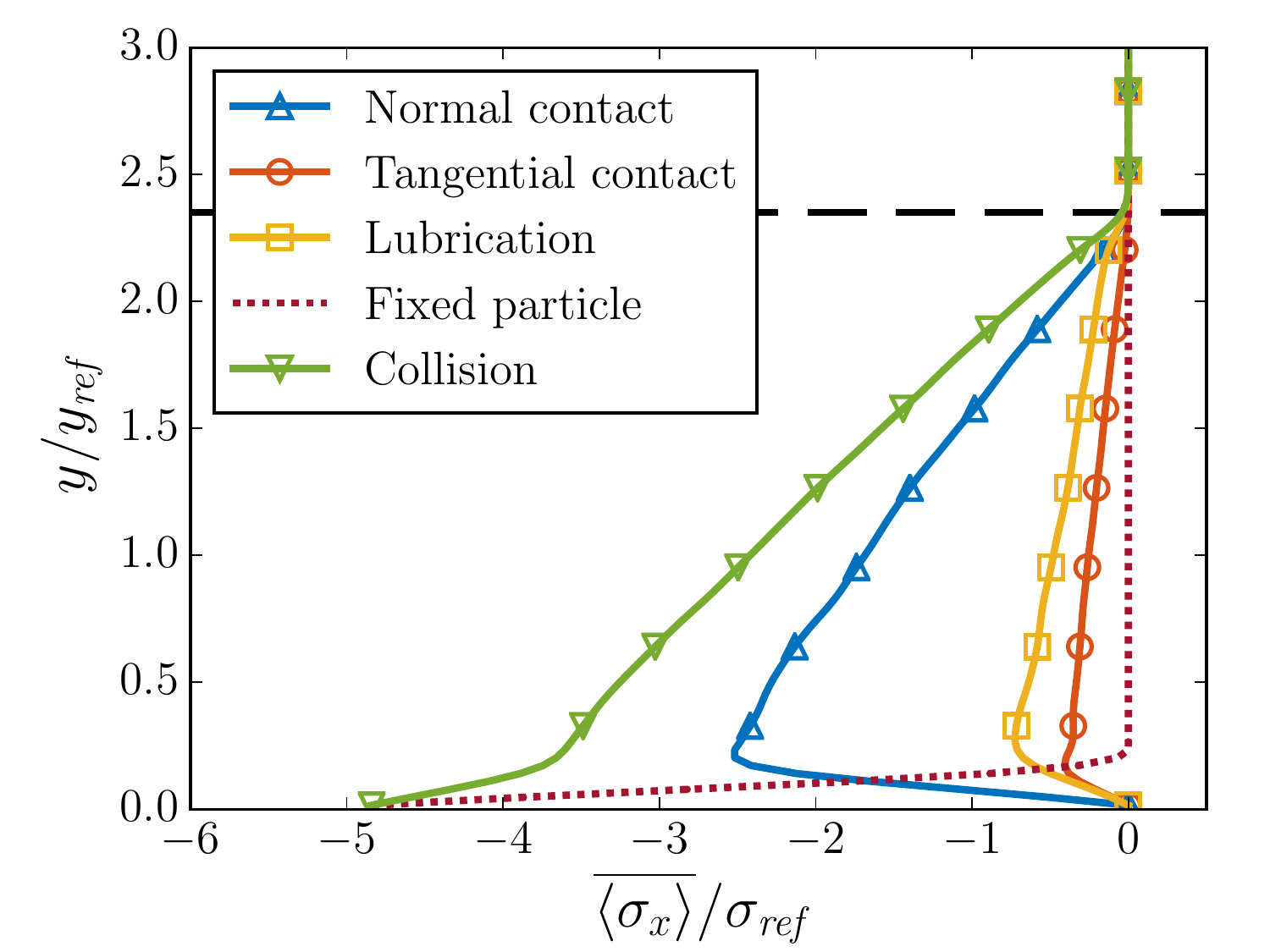}{0.75}{0pt}{0pt}{0pt}
\caption{Sheared particle bed: stress balance of the particle phase in the $x$-direction according to \eqref{eq:px_stress}. Frames (a) and (c) correspond to run Re8, while (b) and (d) correspond to run Re33. The components of the collision stresses in (a) and (b) are further broken down in (c) and (d), according to \eqref{eq:p_collision}. The horizontal dashed line marks the height of the particle bed, $h_p$. As shown in (a) and (b), the hydrodynamic stress, which propels the particles in the positive $x$-direction, is in equilibrium with the collision stress, which hinders the particles. (c) and (d) indicate that all different types of collisions contribute to the collision stress, although normal contact stresses dominate. Lubrication stresses correlate with the local shear rate.}
\label{fig:bed_momx_particle}
\end{figure}

Figure~\ref{fig:bed_momx_particle} shows the coarse-grained particle phase stresses, given by the time average of \eqref{eq:px_stress} for runs Re8 (left side) and Re33 (right side). In figures~\ref{fig:bed_momx_particle}a and~\ref{fig:bed_momx_particle}b, the hydrodynamic stress propelling the particles in the positive $x$-direction and the collision stress slowing the particles in the negative $x$-direction are both zero above the particle bed ($y/y_\mathit{ref} > 2.3$). They increase in magnitude deeper within the particle bed. The sum of the hydrodynamic and collision stresses is zero, indicating that the particle phase stress balance is also in equilibrium, even for the ``unsteady" case, Re33. The hydrodynamic and collision stresses have larger gradients at the fluid/particle bed interface for Re8, and at the lower wall for Re33, which is consistent with the locations of the larger particle stress gradients in figures~\ref{fig:bed_momx_fluid}e and~\ref{fig:bed_momx_fluid}f.

Figures~\ref{fig:bed_momx_particle}c and~\ref{fig:bed_momx_particle}d show the terms in \eqref{eq:p_collision} that contribute to the collision stresses for runs Re8 and Re33, respectively. In both simulations, normal contacts dominate the collision stress, but tangential contacts (friction) and lubrication do play important roles as well. Comparing these figures to the velocity profiles in figure~\ref{fig:bed_flow}b, we can see that the lubrication stress correlates with the shear rate, which is largest at the fluid/bed interface for Re8 and at the lower wall for Re33. This result is consistent with the fact that the lubrication force is dissipative and scales with the relative velocity between particles, similar to a viscous stress. In fact, for this reason the fluid stresses within the particle beds in figures~\ref{fig:bed_momx_fluid}c and~\ref{fig:bed_momx_fluid}d compare remarkably well qualitatively to the lubrication stresses. Tangential contacts play a larger role in the more static bed of run Re8, where the lubrication stress approaches zero at some intermediate depth, than it does in run Re33, where the lubrication stress exists throughout the bed. Finally, the fixed particle stress, representing the force required to hold the fixed particles in place, is similar for both simulations, but there is a steep drop in the total collision stress just above the lower wall for run Re33.  As shown through the fluid phase balance, this result is due to the large shear rate causing a large fluid stress at the lower wall.

\subsubsection{Stress balance of the fluid/particle mixture in the $x$-direction}
\label{sec:bed_momx_both}

\begin{figure}
\centering
\placeTwoSubfigures{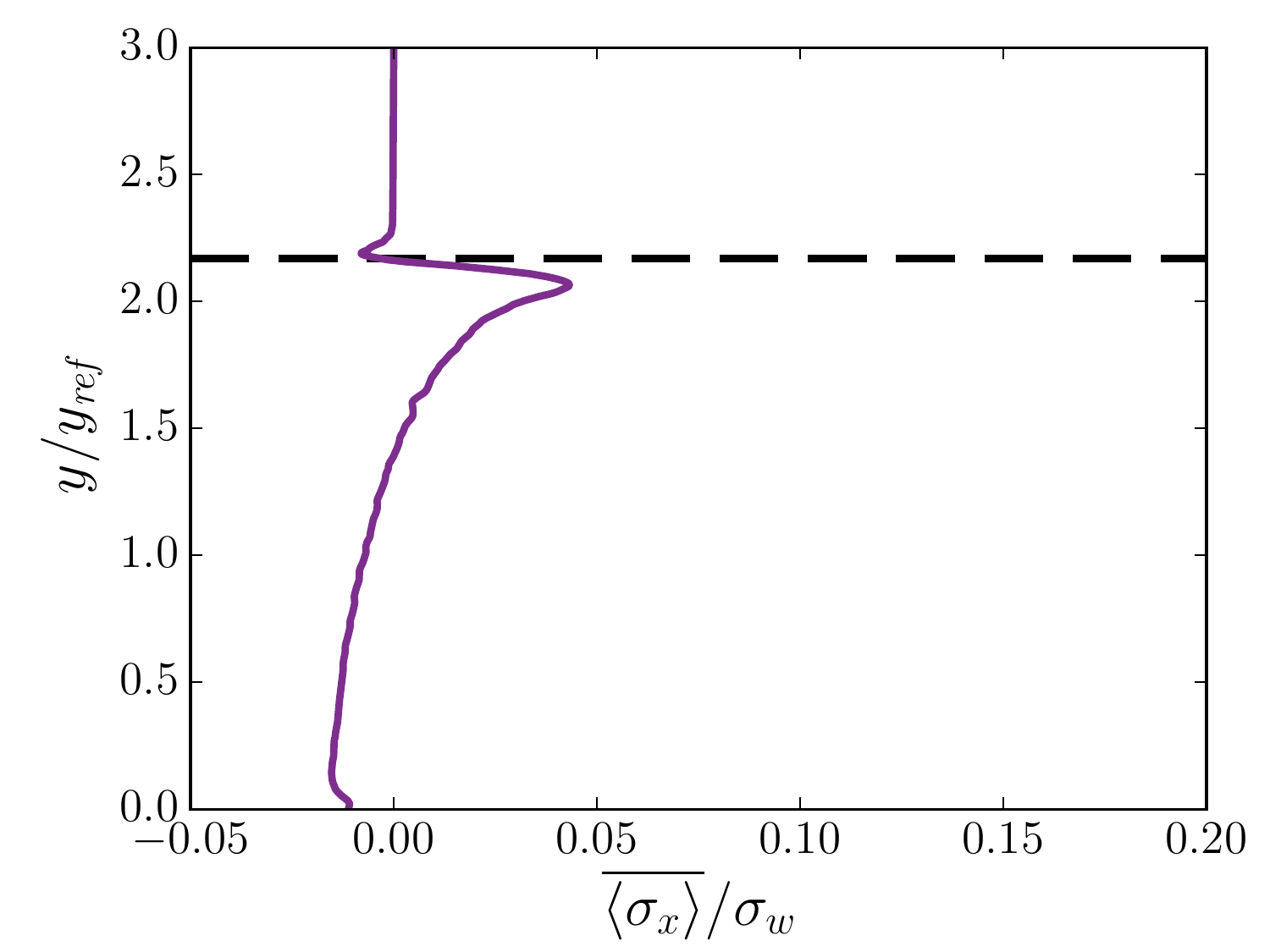}{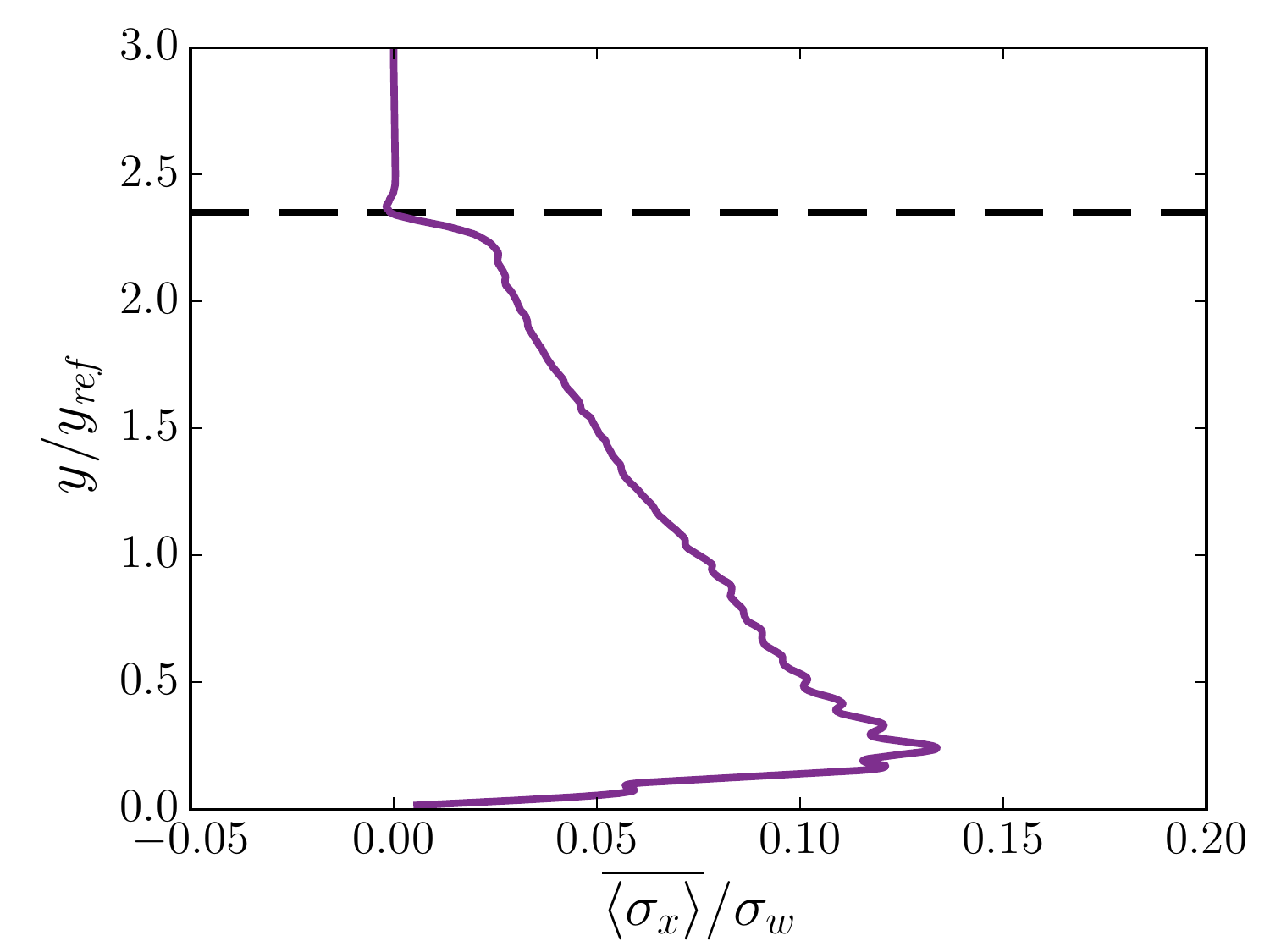}{0.75}{0pt}{0pt}{0pt}
\caption{Sheared particle bed: stress balance in the $x$-direction for the fluid/particle mixture, given by \eqref{eq:bx_stress}, for (a) Run Re8 and (b) Run Re33. Shown is the difference between the external stress and the sum of the fluid and collision stresses normalized by $\sigma_w$, the stress at the lower wall. The horizontal dashed line marks the height of the particle bed, $h_p$. The stress imbalance correlates with the local shear rate and does not exceed 14\% of the lower wall shear stress.}  \label{fig:bed_momx_both}
\end{figure}

As for the single rolling sphere case, we now analyze the stress balance for the fluid/particle mixture, given by \eqref{eq:bx_stress}. Again the mixture balance does not close. We present the imbalances (external minus fluid and collision stresses) for runs Re8 and Re33 in figure~\ref{fig:bed_momx_both}, where we use the normalization $\sigma_w$, which is the external stress at the lower wall, to illustrate the discrepancy. Considering the magnitude of the stresses in the $x$-direction (figure~\ref{fig:bed_momx_fluid}), these imbalances result in errors on the order of 10\%. Figure~\ref{fig:bed_momx_both}a shows that the imbalance for Re8 is greatest in the upper portion of the particle bed. Figure~\ref{fig:bed_momx_both}b, on the other hand, demonstrates a significant imbalance between the sum of the fluid and collision stresses and the external stress throughout the particle bed for Re33. Larger imbalances appear to correlate with larger shear rates in the fluid/particle velocity profiles (figure~\ref{fig:bed_flow}b). Therefore, one possible explanation for the imbalance is the same one we found for the single rolling sphere: the collision stress balances the net fluid force acting on the center of mass of the particles, and it is not resolved along the particle surface. We expect this effect to be more pronounced in regions with higher shear rates, where the upper and lower portions of the particles can experience stronger stress differences. This is consistent with the observation in figure~\ref{fig:bed_momx_both}. Finally, just as in the case of the single rolling sphere, the stress balance does close for both simulations when the entire domain is included within the control volume.

The $x$-momentum balance results for run Re17 (not shown for brevity) are both qualitatively and quantitatively very similar to those for run Re33. While we might perhaps have expected this result, given the similarities in their velocity profiles (figure~\ref{fig:bed_flow}b), it is nevertheless interesting, given the unsteadiness in the Re33 simulation. As we will see in the following, however, the major differences between steady and unsteady beds lie in the $y$-momentum balance, rather than in the $x$-balance.

\subsubsection{Stress balance of the fluid phase in the $y$-direction}
\label{sec:bed_momy_fluid}

\begin{figure}
\placeSixSubfigures{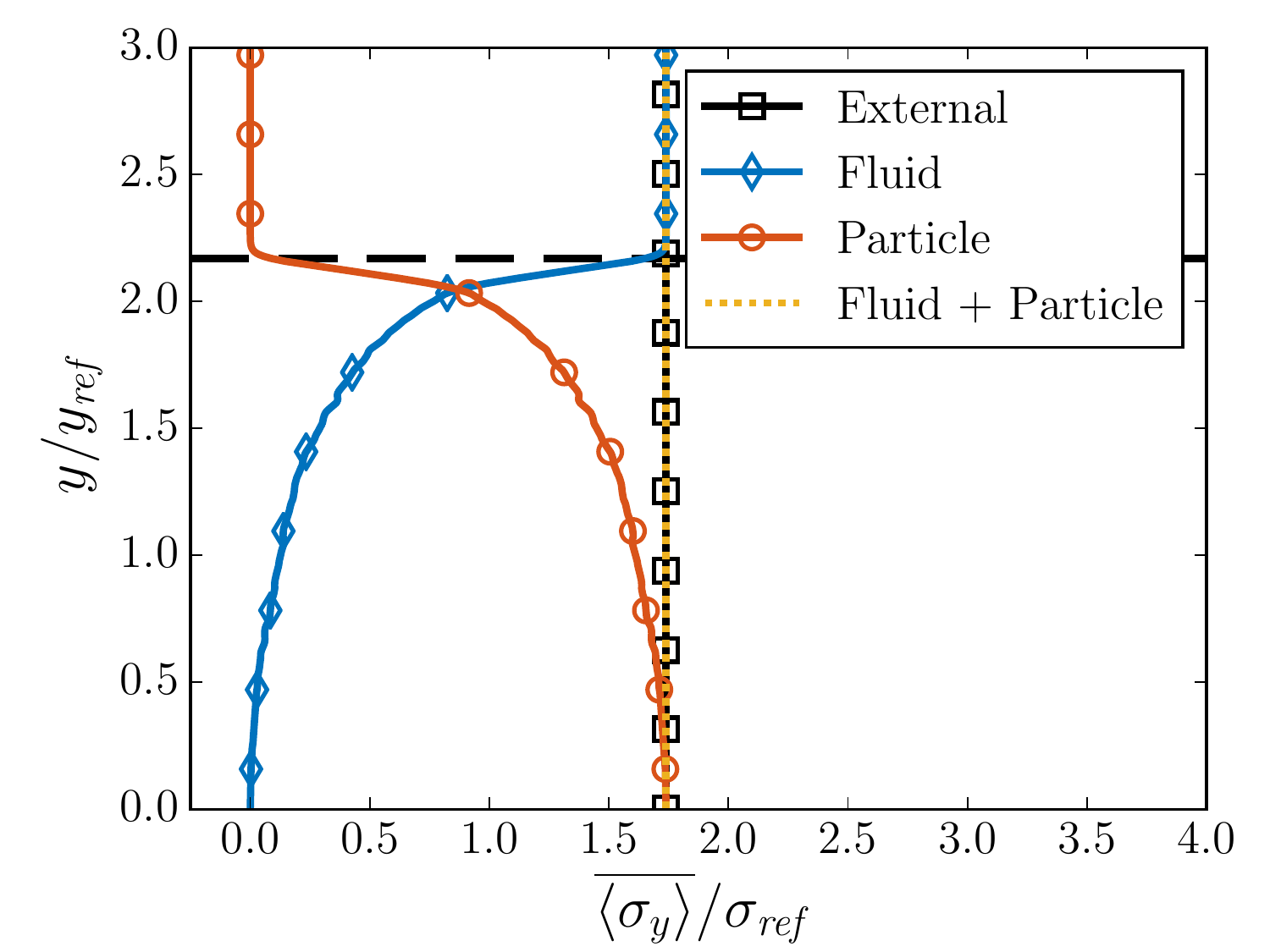}{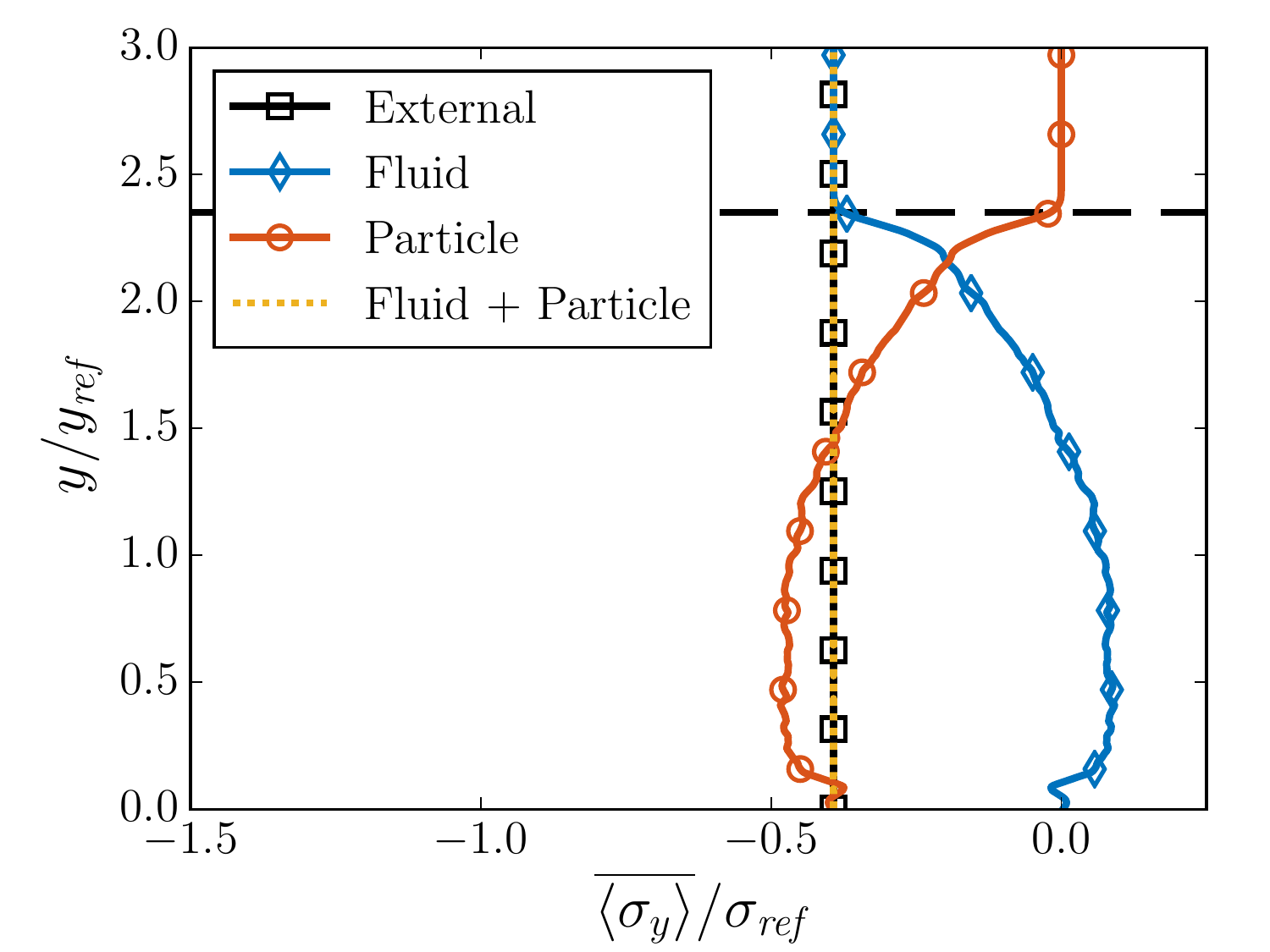}{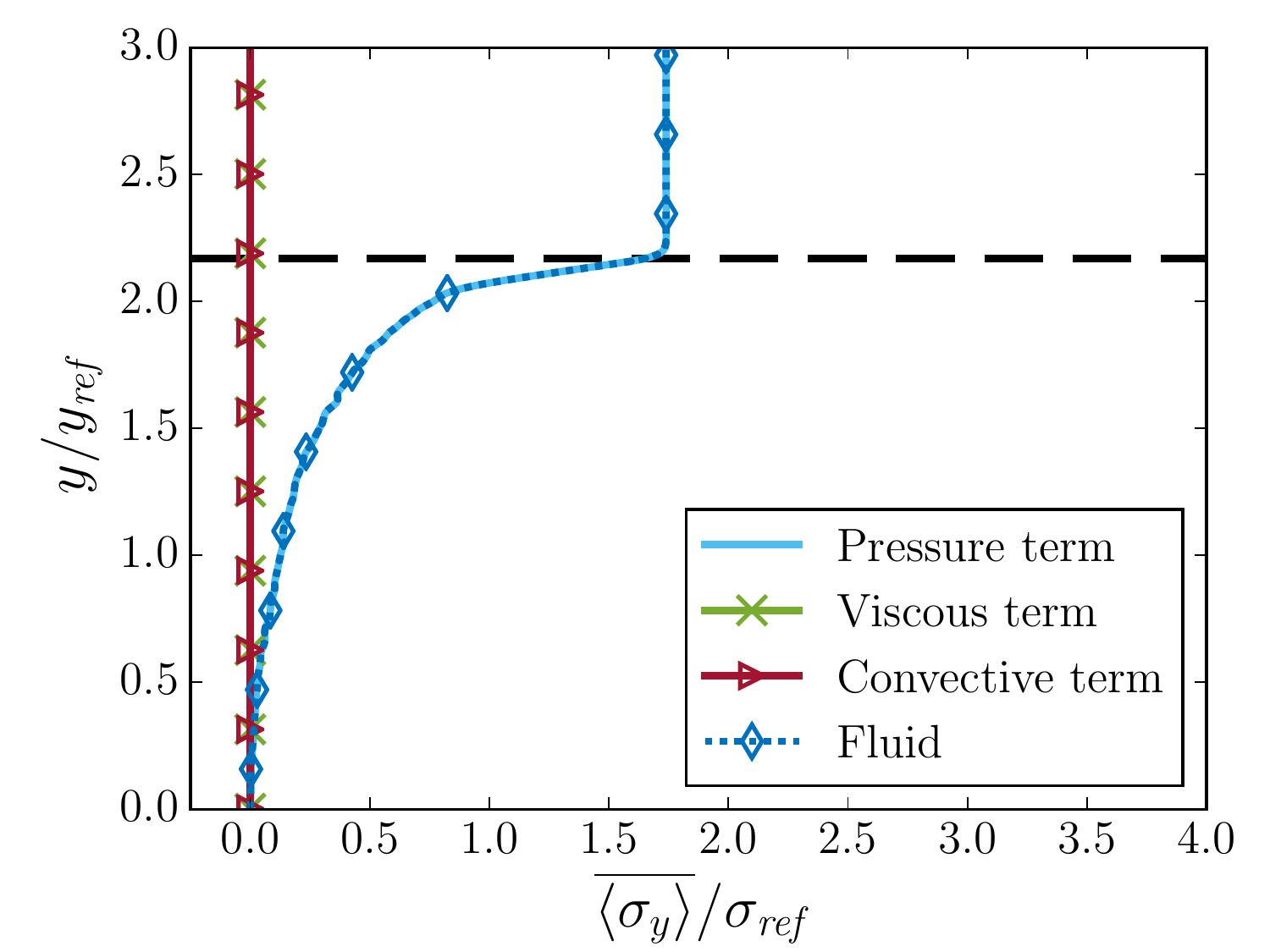}{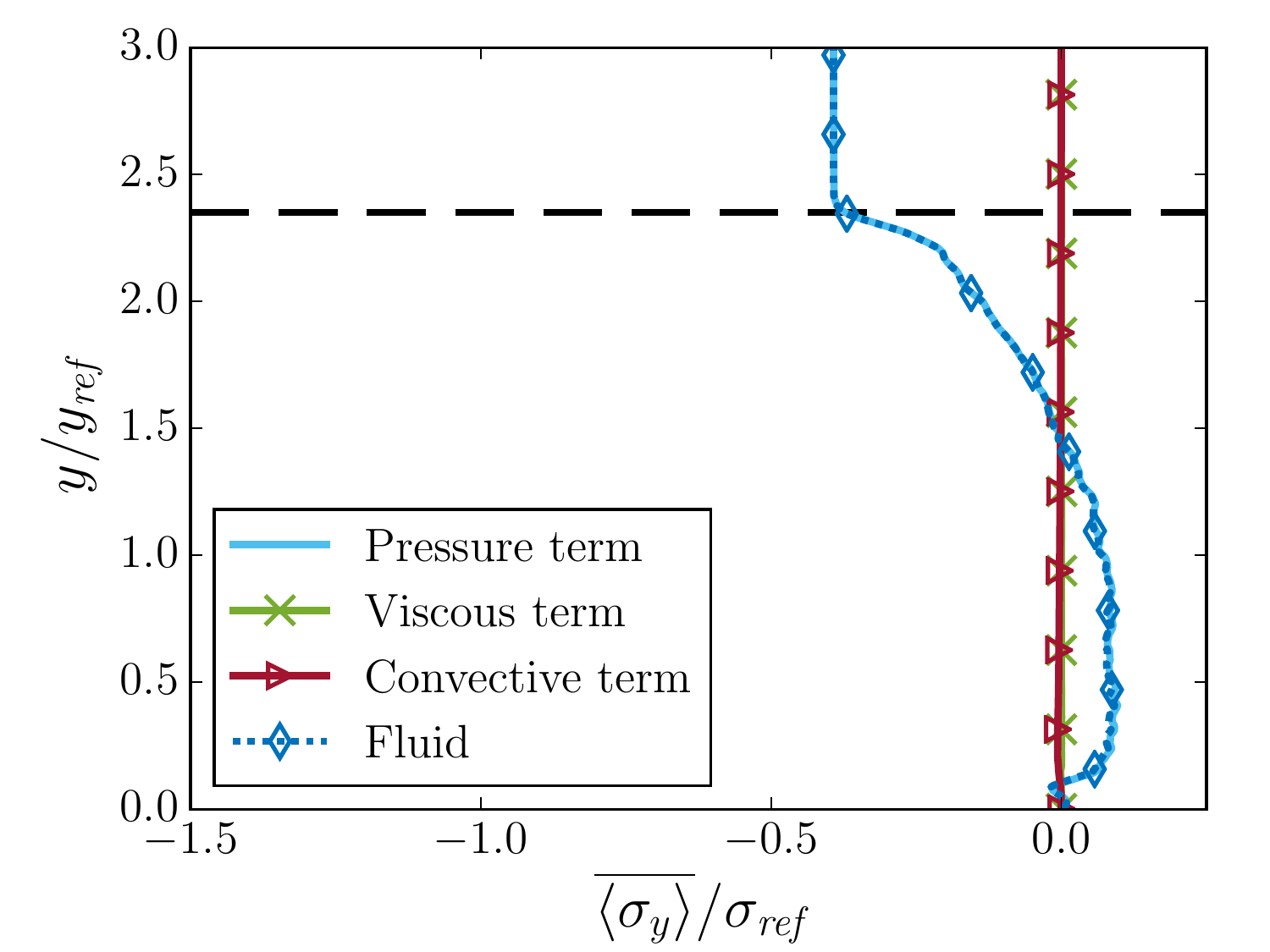}{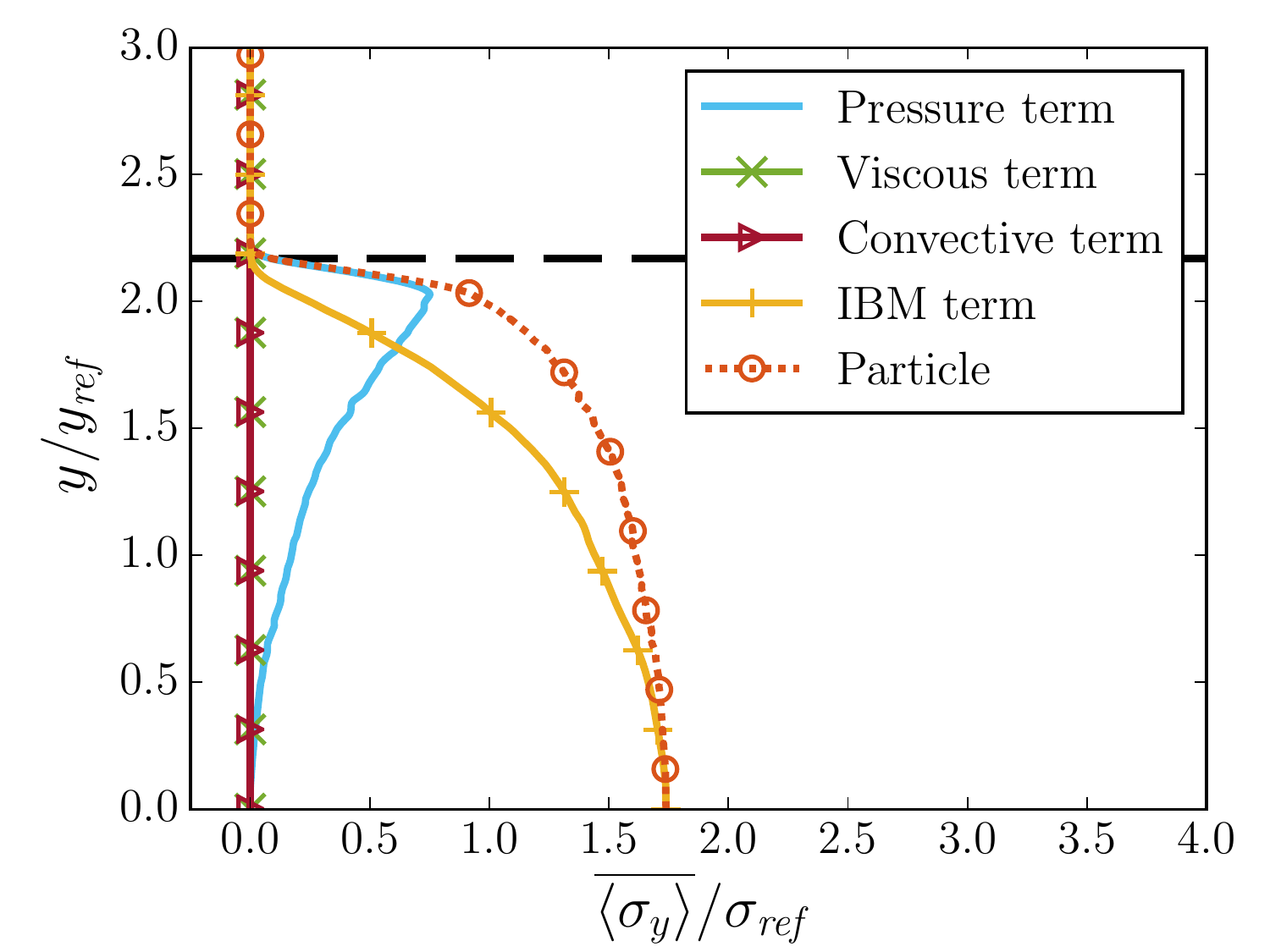}{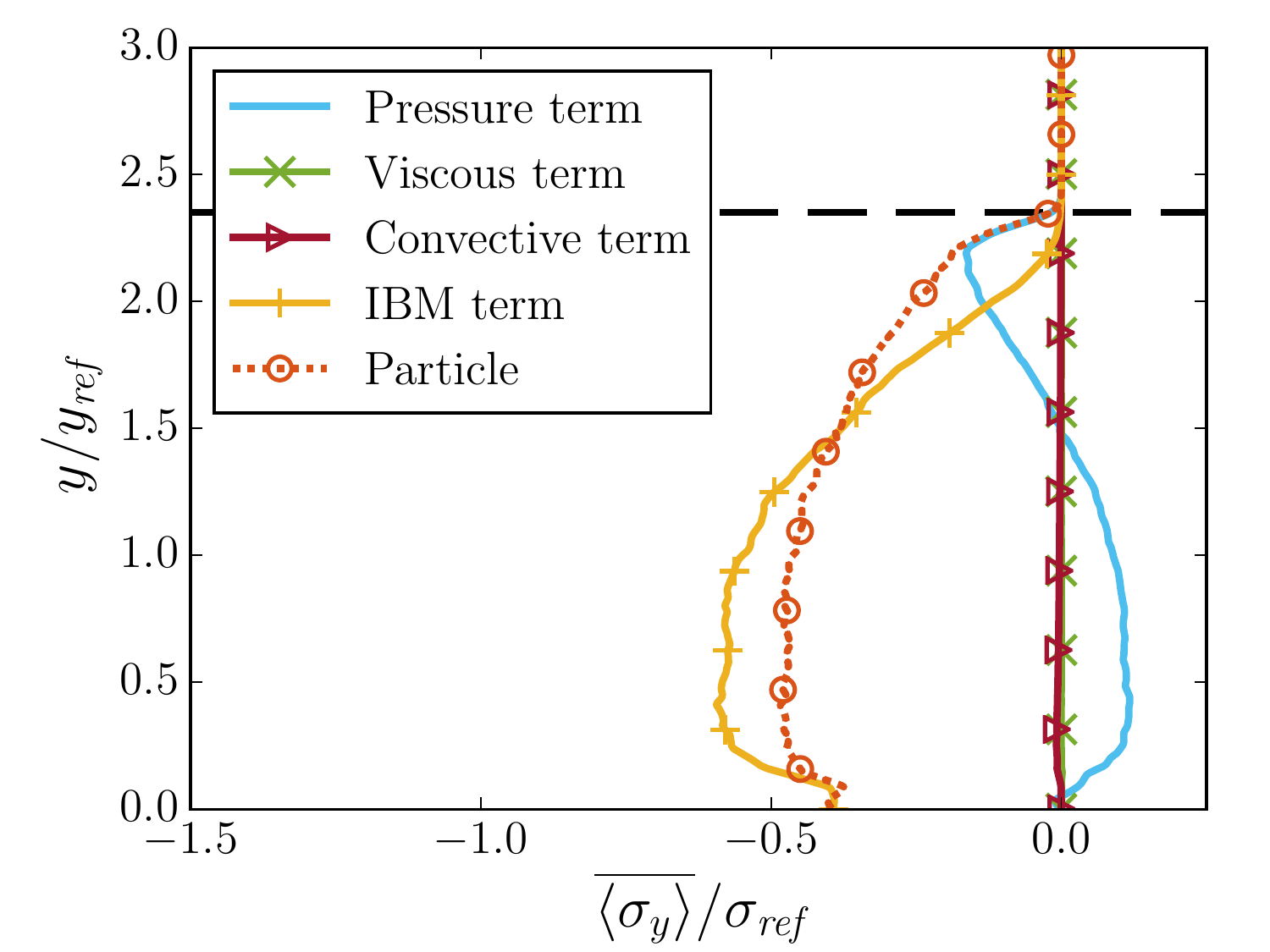}{0.75}{0pt}{0pt}{0pt}
\caption{Sheared particle bed: stress balance of the fluid phase in the $y$-direction according to \eqref{eq:fy_stress}. Frames (a), (c), and (e) correspond to run Re8, while (b), (d), and (f) show run Re33.  The components of (a) and (b) are further broken down in (c) and (d) for the fluid stress, and in (e) and (f) for the particle stress.  The horizontal dashed line marks the height of the particle bed, $h_p$. As shown in (a) and (b), the sum of the fluid and particle stresses is in equilibrium with the external stress, the fluid stress accounts for the stress above the particle bed, and the particle stress accounts for the stress within the particle bed. The positive external stress in (a) corresponds to a lower pressure at the top wall relative to the bottom wall, while the negative external stress in (b) corresponds to a higher top wall pressure. Frames (c) and (d) show that the fluid stress is nearly identical to the pressure term ($\sigma_\mathit{Fpres,y}$) while viscous and convective terms are negligible. Similarly, frames (e) and (f) demonstrate that the particle stress is given by the sum of the pressure term ($\sigma_\mathit{Ppres,y}$) and the IBM term ($\sigma_\mathit{PIBM,y}$).}
\label{fig:bed_momy_fluid}
\end{figure}

We now apply the time-averaging operator \eqref{eq:time_average} to the $y$-momentum balances of the fluid phase, \eqref{eq:fy_stress}, and the particle phase, \eqref{eq:py_stress}. Figure~\ref{fig:bed_momy_fluid} shows the stress balance of the fluid phase for runs Re8 (left column) and Re33 (right column). Figures~\ref{fig:bed_momy_fluid}a and~\ref{fig:bed_momy_fluid}b show the balance between the sum of the fluid and particle stresses and the external stress, which represents the stress at the upper wall since we do not impose a body force on the fluid in the $y$-direction. Once again, the sum of the fluid and particle stresses is in balance with the external stress, even for the unsteady simulation, Re33. Similar to the results for the single sphere, the particle stress quickly takes up the stress from the fluid within the particle bed ($y/y_\mathit{ref} < 2.3$). The stresses for run Re8 (figure~\ref{fig:bed_momy_fluid}a) are positive, while those for run Re33 (figure~\ref{fig:bed_momy_fluid}b) are negative, as a result of the transient dilation and contraction of the respective beds, as explained with the next subfigures.

Figures~\ref{fig:bed_momy_fluid}c and~\ref{fig:bed_momy_fluid}d show the terms in \eqref{eq:fy_stress} that contribute to the fluid stress. The viscous term is seen to be near zero, and in contrast to the single sphere simulation, the convective term is negligible for these simulations as well, so that only the pressure term plays a role. Recall that, according to \eqref{eq:fy_stress}, a positive fluid stress corresponds to a lower fluid pressure relative to the lower wall while a negative fluid stress corresponds to a higher fluid pressure, where the hydrostatic pressure has been subtracted out.
Thus, simulation Re8 has a lower fluid pressure above the particle bed, while run Re33 has a higher fluid pressure above the bed. This observation is consistent with figure~\ref{fig:bed_p_flux}, which revealed that the fluid pressure at the top wall (relative to the lower wall) is negative above contracting beds (e.g. Re8) and positive above dilating beds (e.g. Re33). However, another interesting feature in figure~\ref{fig:bed_momy_fluid}d is that, while the fluid pressure is higher above the bed, it is lower within the lower portion of the bed ($y/y_\mathit{ref} < 1.5$) than at the lower wall. This may be due to the unsteady nature of the flow.

Figures~\ref{fig:bed_momy_fluid}e and~\ref{fig:bed_momy_fluid}f show the terms in \eqref{eq:fy_stress} that contribute to the particle stress. Again, the convective terms do not contribute to the particle stress like they did for the single rolling sphere, meaning only the pressure and IBM terms play a significant role. The pressure terms for the particle and fluid stresses behave similarly, and are important at the fluid/bed interface. Thus, only accounting for the IBM term can lead to an incorrect evaluation of the particle stress.

\subsubsection{Stress balance of the particle phase in the $y$-direction}
\label{sec:bed_momy_particle}

\begin{figure}
\placeFourSubfigures{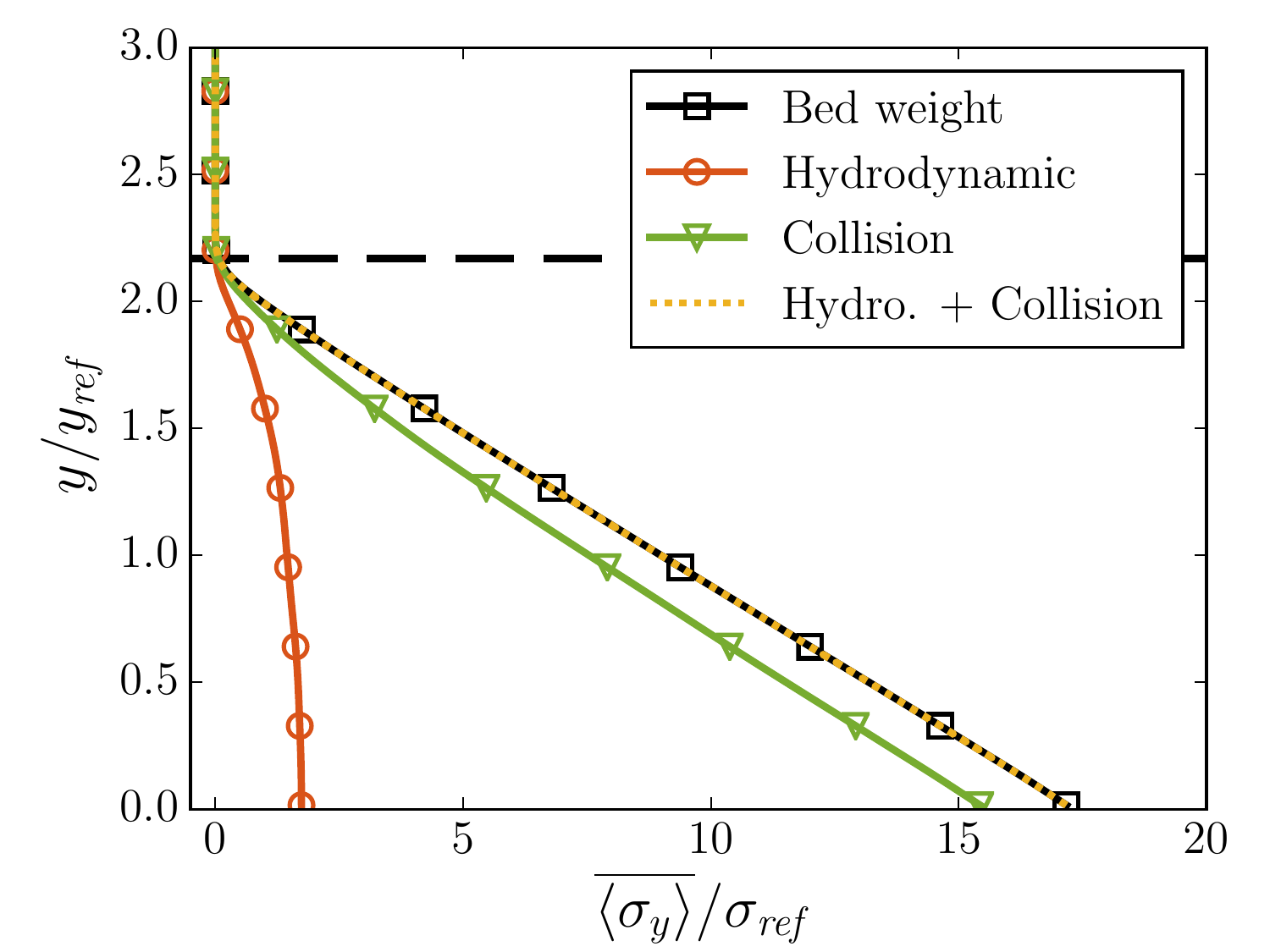}{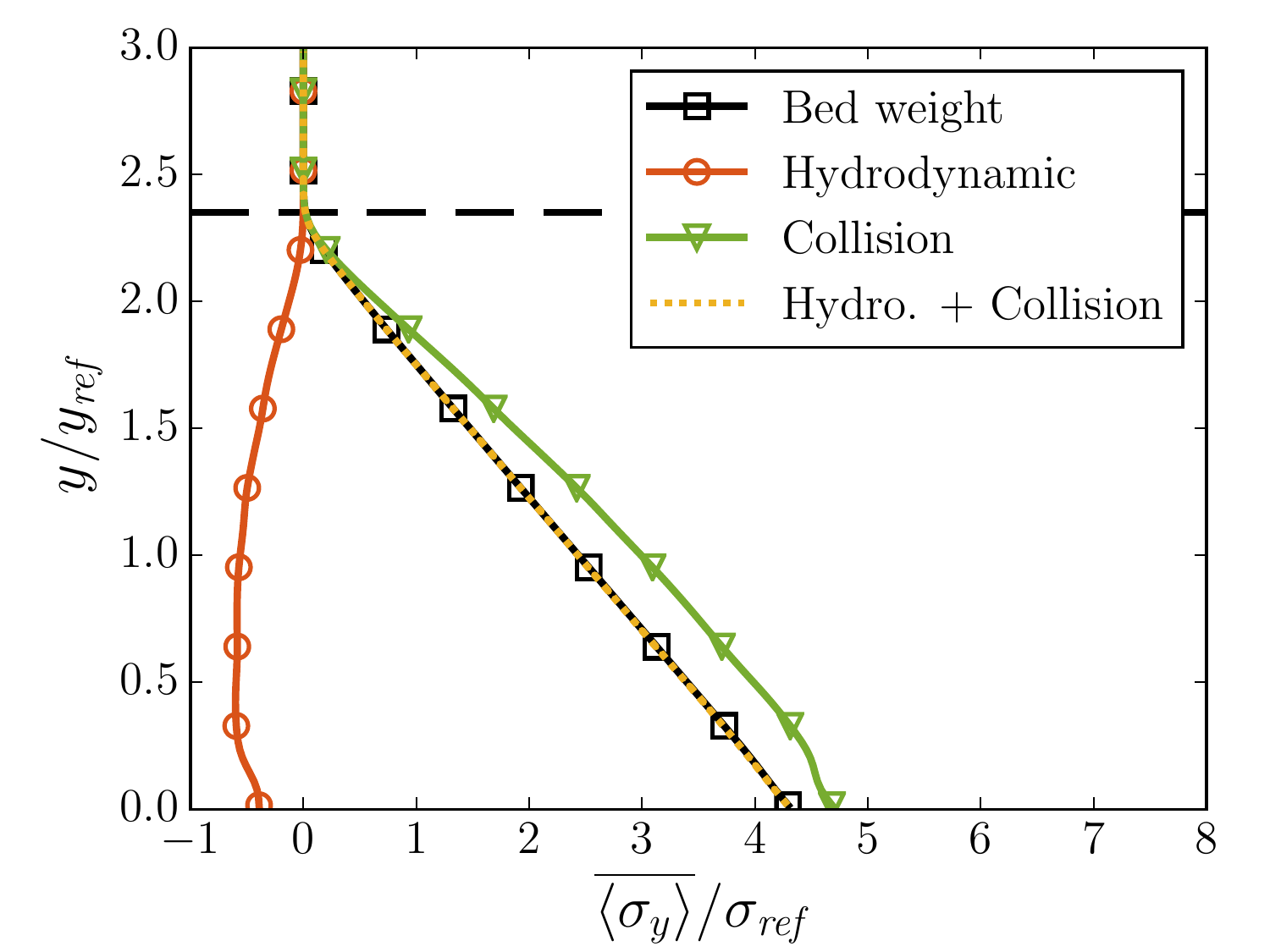}{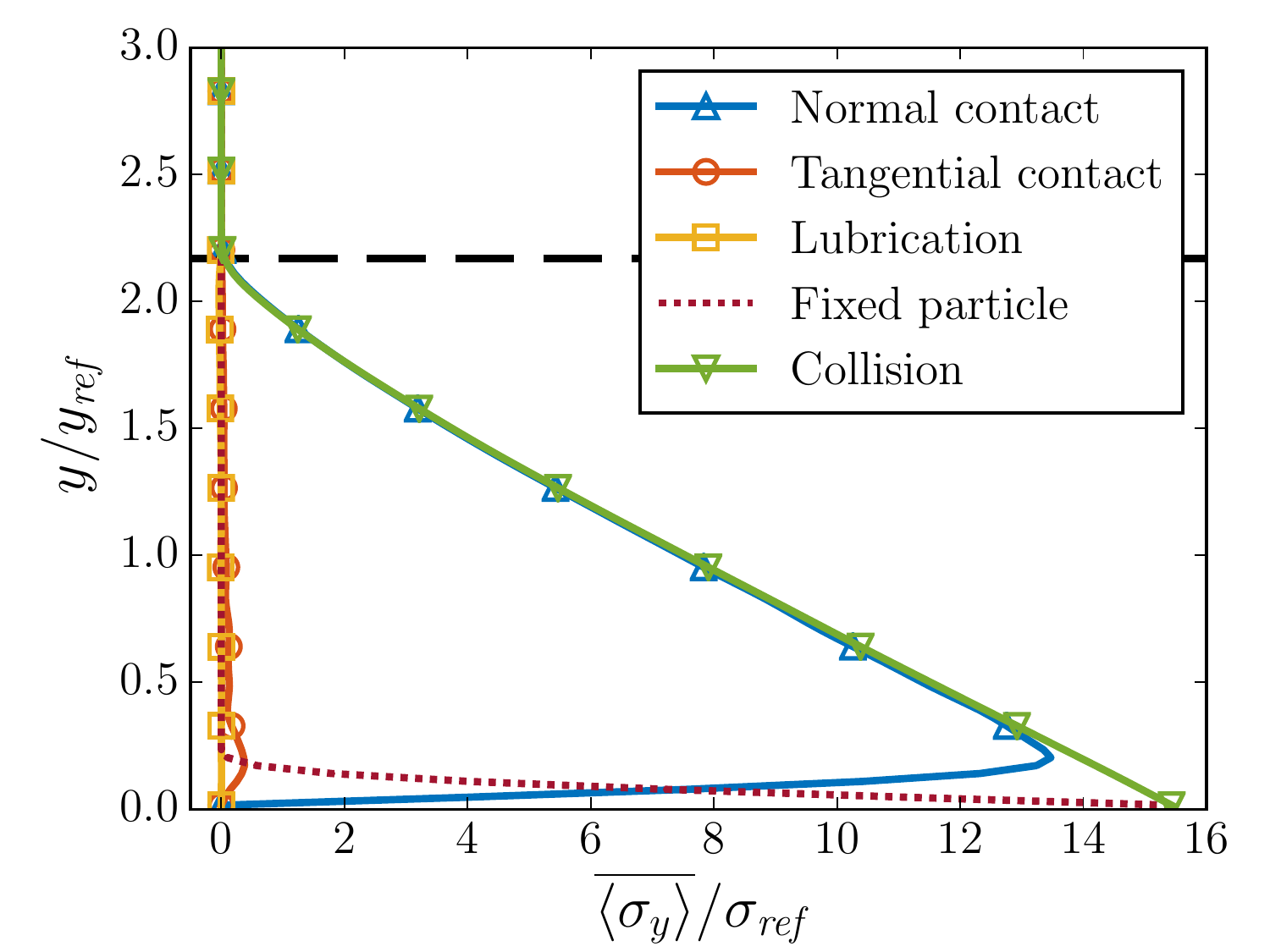}{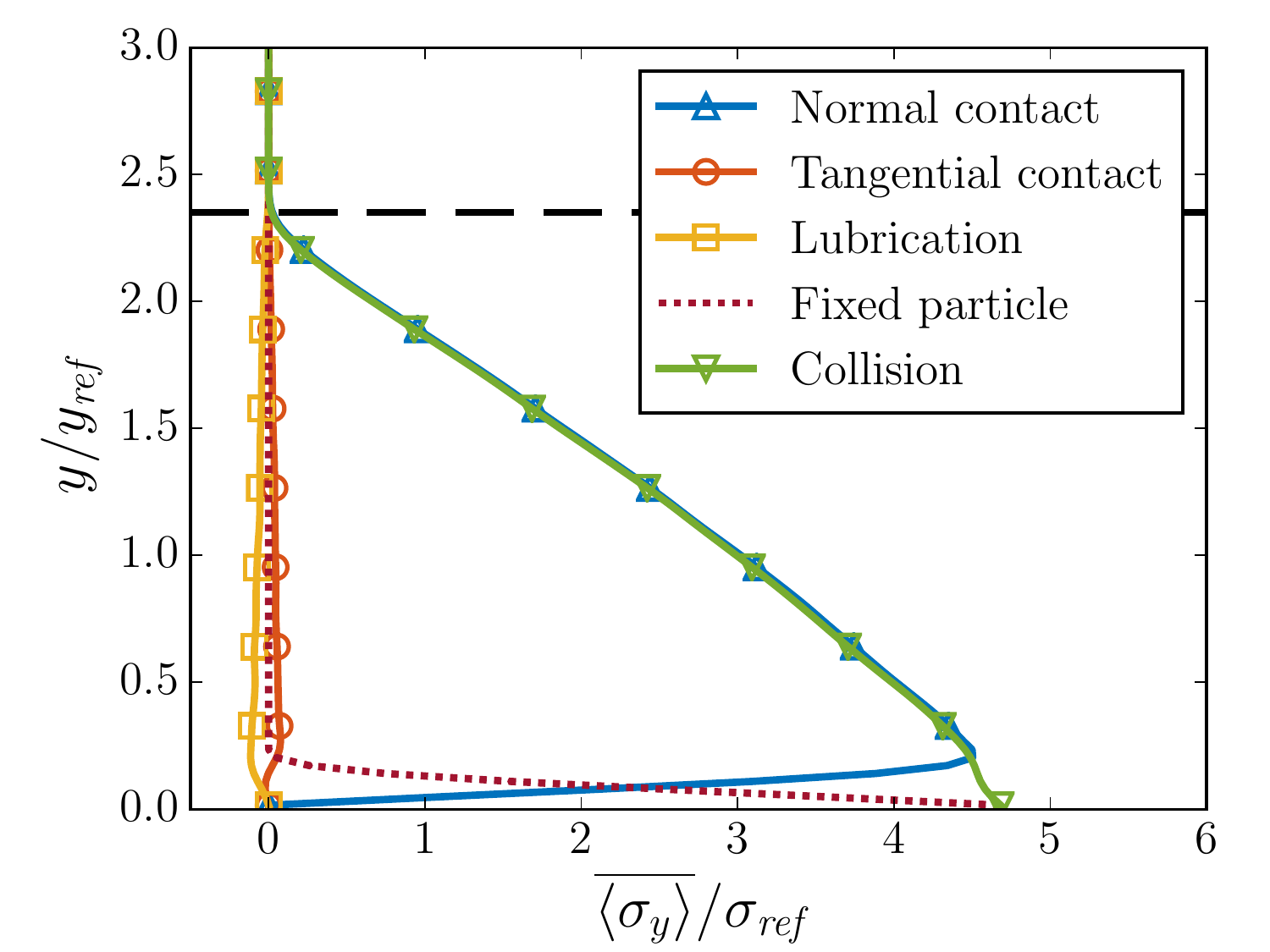}{0.75}{0pt}{0pt}{0pt}
\caption{Sheared particle bed: stress balance of the particle phase in the $y$-direction according to \eqref{eq:py_stress}. Frames (a) and (c) correspond to run Re8, while (b) and (d) correspond to run Re33.  The components of the collision stresses in (a) and (b) are further broken down in (c) and (d) according to \eqref{eq:p_collision}. The horizontal dashed line marks the height of the particle bed, $h_p$. Frames (a) and (b) show that the sum of the hydrodynamic and collision stresses is in equilibrium with the bed weight, with most of the weight supported by the collision stress. The hydrodynamic lift force acting on the particles can be (a) positive or (b) negative, and the hydrodynamic stress at the lower wall matches the external stress in figure~\ref{fig:bed_momy_fluid}. As shown in (c) and (d), the normal contact stress alone accounts for the collision stress.}
\label{fig:bed_momy_particle}
\end{figure}

Figure~\ref{fig:bed_momy_particle} shows the coarse-grained particle phase stresses, given by the time average of \eqref{eq:py_stress} for runs Re8 (left column) and Re33 (right column). In figures~\ref{fig:bed_momy_particle}a and~\ref{fig:bed_momy_particle}b, the bed weight increases almost linearly from the top of the particle bed down to the lower wall, balanced by the sum of the hydrodynamic and collision stresses. Again, this observation is consistent with the results of \cite{Aussillous2013}. Another way to interpret the bed weight is to think of it as the granular pressure $P^p$. Indeed, this has been done by \cite{Boyer2011} and \cite{Stickel2005} for continuum modeling. The fact that we have found a linear profile for this physical quantity again simplifies the situation from a modeling perspective. In contrast to the $x$-momentum particle phase results, the $y$-momentum results show clear differences between runs Re8 and Re33.
First, the hydrodynamic stress is positive for Re8 and negative for Re33, so that the collision stresses are less than and greater than the bed weight, respectively, for the stress balance to be in equilibrium. This difference is directly related to the pressure and steady-state differences discussed previously.
Second, while the collision stresses for Re8 and Re33 are similar in magnitude in the $x$-direction, the $y$-momentum collision stress is three times larger for Re8 than for Re33.  In other words, the collision stress for Re8 is three times larger in the $y$-direction than in the $x$-direction, but the collision stress for Re33 is nearly equal in the two directions.  This observation implies differences in collisional geometries; particles in Re8 collide such that they direct most of the collision force in the $y$-direction while particles in Re33 collide at a lower angle such that they evenly split the collision force between the $x$- and $y$-directions.

Figures~\ref{fig:bed_momy_particle}c and~\ref{fig:bed_momy_particle}d show the terms in \eqref{eq:p_collision} that contribute to the collision stresses for runs Re8 and Re33, respectively.  In both simulations, the collision stress is almost completely a result of normal contacts, which contrasts with the collision stresses in the $x$-direction, which also had significant contributions from tangential contacts and lubrication.  Tangential contacts contribute slightly to support the bed weight near the lower wall for run Re8 (figure~\ref{fig:bed_momy_particle}c) and throughout the bed for run Re33 (figure~\ref{fig:bed_momy_particle}d).  A slight negative lubrication force is present throughout the Re33 bed, indicating a net motion of particles away from each other in the $y$-direction due to the fact that the lubrication force is dissipative and proportional to the relative velocity between particles.  This observation is consistent with a dilating particle bed, where the space between particles increases.

From figures~\ref{fig:bed_momx_particle} and~\ref{fig:bed_momy_particle}, we have seen that the normal contact forces play a dominant role in both the $x$- and $y$-momentum balances for the particles, but these forces are coupled by the geometry of the particle bed. For instance, two particles colliding have a single normal contact force between them, but the relative force directed in the $x$-direction or $y$-direction depends on where the point of contact occurs in the coordinate system. In these simulations, the particle phase is driven in the $x$-direction by the pressure gradient and hydrodynamic forces. At equilibrium, collisions balance the driving force, and by geometry also provide a particle pressure in the $y$-direction opposing the weight of the bed. A collision stress larger than the bed weight, as seen for Re33 in figure~\ref{fig:bed_momy_particle}b, then causes the particle bed to dilate upwards. However, as seen in the same figure, the hydrodynamic stress balances the excess collision stress, keeping the system in equilibrium and slowing the rate of dilation. Thus, the negative hydrodynamic stress for the particle phase, or negative pressure measured at the top wall for the fluid phase, indicates that a particle bed is still dilating in order to attain a collision geometry that allows collision stresses to balance in both $x$- and $y$-directions.

\section{Conclusions}

We have derived a momentum balance for particle-resolved IBM simulations in order to understand the stresses governing the motion of sheared particle beds.  This balance differs from previous efforts in that it fully accounts for the particle stress using the fluid contained within the particles.  We then validated the method against a simulation of a single rolling sphere, showing that it works for situations that do not have a statistically-significant assemblage of particles.
From this simulation, we found that the momentum balances for the fluid phase, \eqref{eq:fx_stress} and \eqref{eq:fy_stress}, and the particle phase, \eqref{eq:px_stress} and \eqref{eq:py_stress}, are valid for instantaneous flow fields at a steady-state relative to the particle.  We have also shown that these momentum balances are valid on a particle-resolved scale in which the control volumes cut through a particle that is large relative to the total domain size.  In contrast, the method of \citet{Zhang2010} requires control volumes that enclose many particles, functioning under a statistical-averaging framework.
From these momentum balances, we determined that the collision between the particle and the wall played a large role in the flow, with friction slowing the flow from the reference Poiseuille case and with normal contact supporting the particle's weight.
We also explored the significance of the various terms comprising the fluid phase balance, \eqref{eq:fx_stress} and \eqref{eq:fy_stress}, which allowed us to simplify them into \eqref{eq:fx_stress_simplified} and \eqref{eq:fy_stress_simplified}.  The balances are thus roughly given by $\sigma_\mathit{Evisc,x}+\sigma_\mathit{Ebody,x} = \sigma_\mathit{Fvisc,x} + \sigma_\mathit{PIBM,x}+\sigma_\mathit{Pvisc,x}$ for the $x$-direction and $\sigma_\mathit{Epres,y} = \sigma_\mathit{Fpres,y}+\sigma_\mathit{Fconv,y} + \sigma_\mathit{PIBM,y}+\sigma_\mathit{Ppres,y}+\sigma_\mathit{Pconv,y}$ for the $y$-direction.
Finally, we investigated the momentum balance for a fluid-particle mixture, given by \eqref{eq:bx_stress}, and found that it did not close because coarse-grained particle quantities cannot resolve changes in stress along the particle surface.

We applied time-averaging to the momentum balances for the fluid phase, \eqref{eq:fx_stress} and \eqref{eq:fy_stress}, and the particle phase, \eqref{eq:px_stress} and \eqref{eq:py_stress}, finding that they close for simulations involving flows with many particles, even those that did not attain a statistical steady state.  A reference case, a Poiseuille flow in the upper third of the domain where no particle are present initially, provided a reasonable scaling of the velocities and stresses in the $x$-direction, even when the entire particle bed was in motion.

We also investigated the terms comprising \eqref{eq:fx_stress} and \eqref{eq:fy_stress} for the simulations of the particle beds, finding that the simplified expressions from the single particle balance, \eqref{eq:fx_stress_simplified} and \eqref{eq:fy_stress_simplified}, would be equally valid for these simulations, and could even be further simplified to
$\sigma_\mathit{Evisc,x}+\sigma_\mathit{Ebody,x} = \sigma_\mathit{Fvisc,x} + \sigma_\mathit{PIBM,x}$ for the $x$-direction and $\sigma_\mathit{Epres,y} = \sigma_\mathit{Fpres,y} + \sigma_\mathit{PIBM,y}+\sigma_\mathit{Ppres,y}$ for the $y$-direction.
Therefore, simulations involving similar flow conditions would be justified in using only $\sigma_\mathit{PIBM,x}$ to calculate the $x$-momentum stress, as was done in \citet{Kidanemariam2017}, \citet{Vowinckel2014}, and \citet{Vowinckel2017b}.  In our experience, the viscous term contributes to the $x$-direction particle stress only for much more viscous flows.  However, the pressure term must be included in the particle stress in the $y$-direction.

We also investigated the terms comprising the collision force, given by \eqref{eq:p_collision}.  For the flow conditions for the present simulations, the normal contact force dominates the $x$-momentum collision force and is nearly solely responsible for the $y$-momentum collision force.  Lubrication and tangential contact forces contribute similarly small amounts to the collision forces in the $x$-direction.

With our scaling based on the reference Poiseuille flow, the $x$-momentum balances were very similar qualitatively and quantitatively for the various flows over a particle bed, even though the simulations were in a transient state.  We found that the $y$-momentum balances were crucial in revealing transient behavior of the particle beds.  The fluid pressure at the top wall relative to the bottom wall (neglecting hydrostatics) indicated whether the bed was dilating (positive pressure) or contracting (negative pressure).  Analyzing the forces on the particles within the bed also revealed fluid forces acting to oppose the upward motion of the particles in a dilating bed and the downward motion of the particles in the contracting bed.

We also applied the $x$-momentum balance of the fluid/particle mixture, \eqref{eq:bx_stress}, to the sheared bed of particles.  As with the case for the single sphere, we found that this balance does not close unless the entire domain is considered, but we did find that the gap in the closure is related to the local shear rate: higher shear rates led to larger gaps.  This gap may exist for any flow that has significant shear acting across a particle diameter, but further studies should be conducted to understand this dependence, which may allow for a closure of the mixture stress balance.  It would then prove a powerful tool for measuring the stresses in dense particle-laden flows.

Another extension of this work would be to include the time-dependent terms.  The real power in the accurate measurement of particle-fluid stresses would be to analyze unsteady flows on short time scales.  This could, for instance, allow us to study the rheology of transient particle-laden flows, such as the onset or cessation of erosion.  Finally, extending the time-averaged equations for use with turbulent flows would permit its use for a broad range of important sediment transport problems.

\section{Acknowledgements}

This research is supported in part by the Department of Energy Office of Science Graduate Fellowship Program (DOE SCGF), made possible in part by  the American Recovery and Reinvestment Act of 2009, administered by ORISE-ORAU under contract no. DE-AC05-06OR23100.  It is also supported by the Petroleum Research Fund, administered by the American Chemical
Society, grant number 54948-ND9.  BV gratefully acknowledges the Feodor-Lynen scholarship provided by the Alexander von Humboldt foundation, Germany, and EM thanks Petrobras for partial support.  Computational resources for this work used the Extreme Science and Engineering Discovery Environment (XSEDE), which was supported by the National Science Foundation, USA, Grant No. TG-CTS150053.

\appendix

\section{Time derivative for a rolling sphere} \label{sec:time_derivative}

In section~\ref{sec:balance}, we discussed the momentum balance for a rolling sphere and mentioned that, at steady-state,
\begin{equation} \label{eq:time_derivative}
\int\limits_{\Omega_\mathit{CV}^+} \rho_f \frac{\partial{\mbs{u}}}{\partial{t}} \,\mrm{d}V +
\int\limits_{\Gamma_\mathit{CV}^p} \rho_f (\mbs{u}\mbs{u}) \cdot \mbs{n}^+ \,\mrm{d}A = 0 .
\end{equation}
To show this equivalence, we utilize a reference frame that moves with the sphere.  The fluid velocity field for this reference frame is given by $\widetilde{\mbs{u}}\left(\widetilde{x},\widetilde{y},\widetilde{z},\widetilde{t}\,\right)$.  At steady-state, the particle translates to the right with velocity $\mbs{u}_p = a \hat{\mbs{\imath}}$, where $\hat{\mbs{\imath}}$ is the unit vector in the $x$-direction.  Also at steady-state, the fluid velocity field in the moving reference frame does not vary in time, so that $\partial \widetilde{\mbs{u}}/\partial \widetilde{t} = 0$.  The velocity field in the laboratory reference frame, $\mbs{u}$, is related to the velocity field in the moving reference frame, $\widetilde{\mbs{u}}$, by
\begin{equation} \label{eq:u_tilde}
\mbs{u}(x,y,z,t) = \widetilde{\mbs{u}}(x-at,y,z,t) + a\hat{\mbs{\imath}} .
\end{equation}
At the particle surface ($\Gamma^p$ in the laboratory reference frame, $\widetilde{\Gamma}^p$ in the moving reference frame), the fluid velocity matches the rigid body velocity of the particle
\begin{IEEEeqnarray}{rCl's}
\mbs{u} &=& \mbs{u}_p + \mbs{\omega}_p \times (R_p \mbs{n}^-) & at $\Gamma^p$ \label{eq:u_rigid}\\
\widetilde{\mbs{u}} &=& \mbs{\omega}_p \times (R_p \widetilde{\mbs{n}}^-) & at $\widetilde{\Gamma}^p$ .
\end{IEEEeqnarray}

\begin{figure}
\centering
\includegraphics[width=0.5\textwidth]{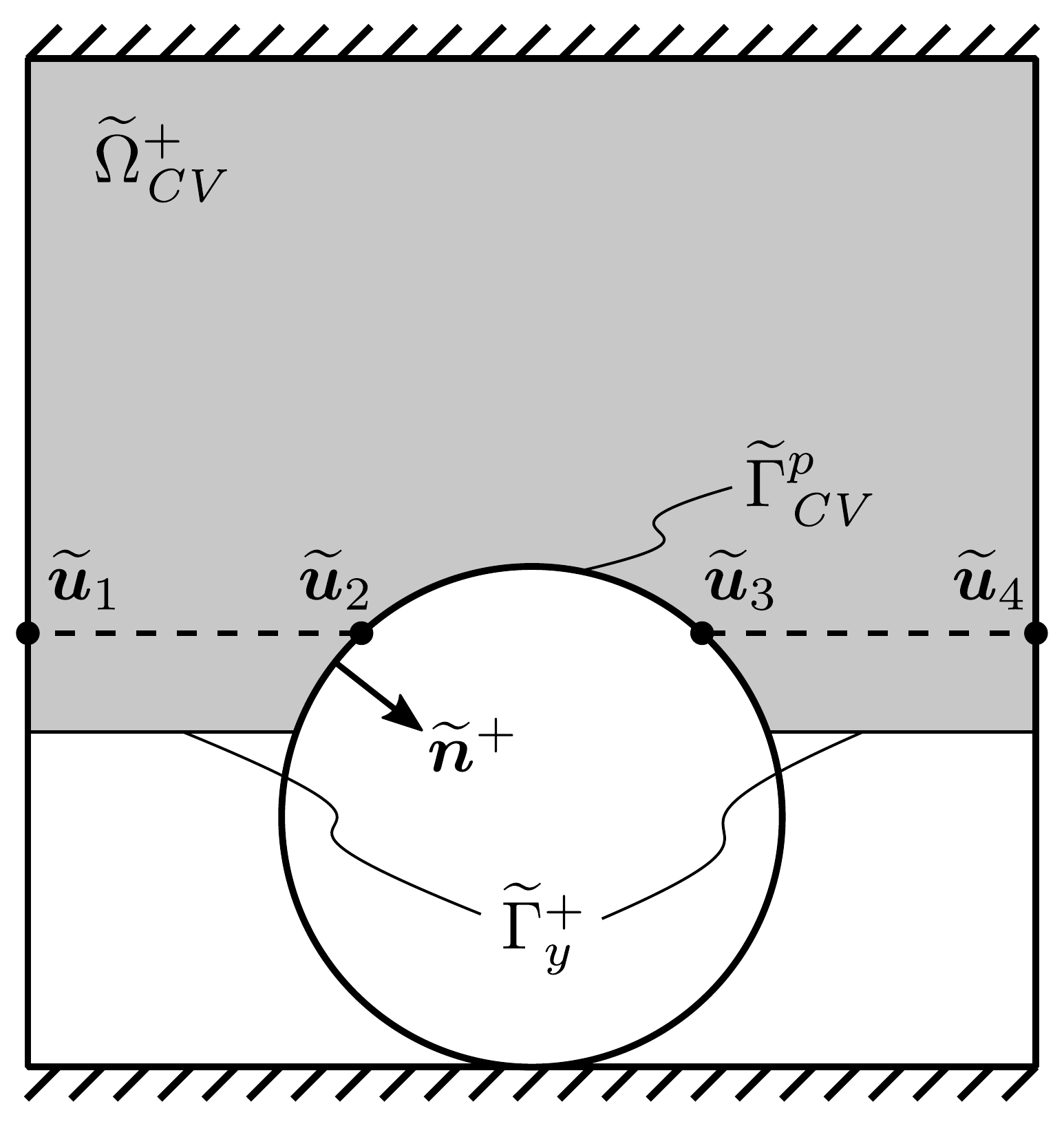}
\caption{Line integral of fluid velocity field in a reference frame moving with the particle.}  \label{fig:cv_integrate}
\end{figure}

Consider now the time derivative of the fluid within the control volume $\Omega_\mathit{CV}^+$, which is present in \eqref{eq:NS_int2}.  We can transform this quantity into the moving reference frame though the following steps.  First, we use the fact that the Jacobian determinant $\left|\partial\widetilde{\mbs{x}}/\partial\mbs{x}\right| = 1$ together with \eqref{eq:u_tilde} to obtain
\begin{equation}
\int\limits_{\Omega_\mathit{CV}^+} \frac{\partial\mbs{u}}{\partial t} \, \mrm{d}V =
\int\limits_{\widetilde{\Omega}_\mathit{CV}^+} \frac{\partial}{\partial t} \left[
\widetilde{\mbs{u}}(x-at,y,z,t) + a\hat{\mbs{\imath}} \right] \, \mrm{d}\widetilde{V} .
\end{equation}
Next, we evaluate the time derivative, using the fact that $\partial (a\hat{\mbs{\imath}})/\partial t = \mbs{0}$ to obtain
\begin{equation}
\int\limits_{\Omega_\mathit{CV}^+} \frac{\partial\mbs{u}}{\partial t} \, \mrm{d}V =
\int\limits_{\widetilde{\Omega}_\mathit{CV}^+} \left(
-a \frac{\partial \widetilde{\mbs{u}}}{\partial \widetilde{x}} +
\frac{\partial \widetilde{\mbs{u}}}{\partial \widetilde{t}} \right) \, \mrm{d}\widetilde{V} .
\end{equation}
Due to the steady-state conditions, this expression simplifies to
\begin{equation}
\int\limits_{\Omega_\mathit{CV}^+} \frac{\partial\mbs{u}}{\partial t} \, \mrm{d}V =
-a \int\limits_{\widetilde{\Omega}_\mathit{CV}^+}
\frac{\partial \widetilde{\mbs{u}}}{\partial \widetilde{x}} \, \mrm{d}\widetilde{V} .
\end{equation}
We can split this integral into line integrals along the $x$-direction for a given $y$ and $z$ coordinate, illustrated by figure~\ref{fig:cv_integrate}, which evaluate to
\begin{IEEEeqnarray}{rCl}
\int\limits_{\Omega_\mathit{CV}^+} \frac{\partial\mbs{u}}{\partial t} \, \mrm{d}V
&=& -a \int_0^{L_z}\!\!\!\int_y^{L_y} \left(\widetilde{\mbs{u}}_2 - \widetilde{\mbs{u}}_1 \right)
\, \mrm{d}\widetilde{y}\,\mrm{d}\widetilde{z}
-a \int_0^{L_z}\!\!\!\int_y^{L_y} \left(\widetilde{\mbs{u}}_4 - \widetilde{\mbs{u}}_3 \right)
\, \mrm{d}\widetilde{y}\,\mrm{d}\widetilde{z} \\
&=& a \int_0^{L_z}\!\!\!\int_y^{L_y} \left(\widetilde{\mbs{u}}_3 - \widetilde{\mbs{u}}_2 \right)
\, \mrm{d}\widetilde{y}\,\mrm{d}\widetilde{z} ,
\end{IEEEeqnarray}
where we used the property $\widetilde{\mbs{u}}_1 = \widetilde{\mbs{u}}_4$ resulting from the periodic boundaries.  Finally, we rewrite this integral in terms of one over the surface of the sphere within the control volume
\begin{equation} \label{eq:dudt_final}
\int\limits_{\Omega_\mathit{CV}^+} \frac{\partial\mbs{u}}{\partial t} \, \mrm{d}V
= a \! \int\limits_{\widetilde{\Gamma}_\mathit{CV}^p} \widetilde{\mbs{u}} \widetilde{n}_x^-
\, \mrm{d}\widetilde{A} ,
\end{equation}
where $n_x^-$ is the $x$-component of $\mbs{n}^-$, the unit vector pointing outward from the particle.  This component of the normal vector accounts for the change of variables from $\mrm{d}\widetilde{y}\,\mrm{d}\widetilde{z}$ to $\mrm{d}\widetilde{A}$.

Now consider the convective term, where we can replace the fluid velocity with the rigid body velocity \eqref{eq:u_rigid} because we are evaluating the integral over the particle surface
\begin{IEEEeqnarray}{rCl}
\int\limits_{\Gamma_\mathit{CV}^p} (\mbs{u}\mbs{u}) \cdot \mbs{n}^+ \,\mrm{d}A
&=& \int\limits_{\Gamma_\mathit{CV}^p} \mbs{u}\left(\mbs{u}_p \cdot \mbs{n}^+\right) \,\mrm{d}A \label{eq:uun_eq1}\\
&=& \int\limits_{\Gamma_\mathit{CV}^p} \mbs{u}\left(a n_x^+\right) \,\mrm{d}A .
\end{IEEEeqnarray}
We used the orthogonality of $\mbs{\omega}_p \times (R_p \mbs{n}^-)$ to $\mbs{n}^+$ in \eqref{eq:uun_eq1}.  We then apply a change of variables into the moving reference frame to get
\begin{equation}
\int\limits_{\Gamma_\mathit{CV}^p} (\mbs{u}\mbs{u}) \cdot \mbs{n}^+ \,\mrm{d}A
= a \int\limits_{\widetilde{\Gamma}_\mathit{CV}^p} (\widetilde{\mbs{u}} + a\hat{\mbs{\imath}}) \widetilde{n}_x^+ \,\mrm{d}\widetilde{A} .
\end{equation}
Due to the symmetry of $\widetilde{\Gamma}_\mathit{CV}^p$, integrating $\widetilde{n}_x^+$ over this surface evaluates to zero.  This property together with $\widetilde{\mbs{n}}^+ = -\widetilde{\mbs{n}}^-$ finally gives us
\begin{equation} \label{eq:uun_final}
\int\limits_{\Gamma_\mathit{CV}^p} (\mbs{u}\mbs{u}) \cdot \mbs{n}^+ \,\mrm{d}A
= -a \int\limits_{\widetilde{\Gamma}_\mathit{CV}^p} \widetilde{\mbs{u}} \widetilde{n}_x^- \,\mrm{d}\widetilde{A} .
\end{equation}

Together with a constant fluid density, \eqref{eq:dudt_final} and \eqref{eq:uun_final} show that \eqref{eq:time_derivative} is valid.

\section{Coarse-graining} \label{sec:cg}

\subsection{Coarse-graining method} \label{sec:cg_method}

Our desire to compare our results to continuum models for the particle phase requires us to analyze our simulation results, such as particle velocities and forces, from a continuum viewpoint.  Binning, or averaging these values within control volumes based on the location of the particle center, is a simple method that conserves the measured quantities, but it requires a large sample size of particles, either using large bins, which reduces the spatial resolution, or large time averages, which only works well for steady-state configurations and can be computationally-expensive to obtain.

Instead, we employ the coarse-graining method based on the works of \citet{Goldhirsch2010} and \citet{Weinhart2012}.  This coarse-graining method also conserves quantities of interest, but additionally smooths out the resulting continuum field.  Thus, while the binning method might be sensitive to particles jumping from one bin to the next, there is no such sensitivity in the coarse-graining framework.  In fact, this method can represent smooth fields even for instantaneous data.  While the coarse-graining method may smear information at the fluid/particle interface, we have chosen to use it for analyzing information within the particle bed.

We will first define a few coarse-grained quantities.  For instance, we can obtain a continuum density field $\rho^\mathit{cg}$, which is defined at every point in space $\mbs{x}$ and time $t$:
\begin{equation} \label{eq:cg_rho}
\rho^\mathit{cg}(\mbs{x},t) = \sum_{p=1}^{N_p} m_p \mathcal{W}(\mbs{x} - \mbs{x}_p(t)) ,
\end{equation}
where $N_p$ is the number of particles, $m_p$ and $\mbs{x}_p(t)$ are the mass and position of the center of particle $p$, and $\mathcal{W}$ is the conservative coarse-graining function, described further in section~\ref{sec:cg_function}.
We can similarly define a coarse-grained volume fraction,
\begin{equation} \label{eq:cg_phi}
\phi^\mathit{cg}(\mbs{x},t) = \sum_{p=1}^{N_p} V_p \mathcal{W}(\mbs{x} - \mbs{x}_p(t)) ,
\end{equation}
where $V_p$ is the volume of particle $p$, and momentum density,
\begin{equation} \label{eq:cg_p}
\mbs{p}^\mathit{cg}(\mbs{x},t) = \sum_{p=1}^{N_p} m_p \mbs{u}_p(t) \mathcal{W}(\mbs{x} - \mbs{x}_p(t)) ,
\end{equation}
where $\mbs{u}_p(t)$ is the translational velocity of particle $p$.  From this momentum density we can define a macroscopic velocity field,
\begin{equation} \label{eq:cg_u}
\mbs{u}^\mathit{cg}(\mbs{x},t) = \frac{\mbs{p}^\mathit{cg}(\mbs{x},t)}{\rho^\mathit{cg}(\mbs{x},t)} .
\end{equation}

For other quantities acting at the particle center, such as forces, we define the coarse-grained quantity to be
\begin{equation} \label{eq:cg_F}
\mbs{F}^\mathit{cg}(\mbs{x},t) = \sum_{p=1}^{N_p} \mbs{F}_p(t) \mathcal{W}(\mbs{x} - \mbs{x}_p(t)) .
\end{equation}

For this analysis, we have coarse-grained the forces acting on the particle centers.  We could alternatively coarse-grain the collision forces in a manner similar to that of \citet{Weinhart2012}, which allows us to evaluate the entire stress tensor for collisions.  We are limited in our analysis of simulation results, however, to using the information on hydrodynamic forces at the particle centers.  Hence, for consistency we have to limit our analysis to all quantities acting at the particle centers.

\subsection{Coarse-graining function} \label{sec:cg_function}

The coarse-graining function $\mathcal{W}$ plays a very similar role to that of the delta functions used in the immersed boundary method (IBM): smoothly spreading a quantity from one mesh to another.  The main properties identified by \citet{Weinhart2013} are that $\int_{\mathbb{R}^3} \mathcal{W}(\mbs{r}) \, \mrm{d}\mbs{r} = 1$, which conserves the spread quantity, and that $\mathcal{W}(\mbs{r})$ has two continuous derivatives, which allows one to evaluate gradients of the resulting coarse-grained fields analytically.  While \citet{Weinhart2012} used a Gaussian coarse-graining function and \citet{Weinhart2013} used a polynomial coarse-graining function, we instead implement one based on the delta function of \citet{Roma1999}:
\begin{equation} \label{eq:cg_function}
\mathcal{W}(\mbs{r}) = \frac{1}{w^3} \delta(r_x / w) \, \delta(r_y / w) \, \delta(r_z / w) ,
\end{equation}
where $w$ sets the coarse-graining width and
\begin{equation}
\delta(r) = \begin{cases}
	\frac{1}{3} \left(1 + \sqrt{-3 r^2 + 1} \right) & |r| \leq 0.5 \\
	\frac{1}{6} \left[5 - 3|r| - \sqrt{-3 (1-|r|)^2 + 1} \right] & 0.5 < |r| \leq 1.5 \\
	0 & |r| > 1.5 .
\end{cases}
\end{equation}
Thus, $\mathcal{W}(\mbs{r})$ has a radius of influence of $1.5w$ and one continuous derivative.  We chose this function because it exhibits good conservation properties and because we do not evaluate the coarse-graining expressions analytically and hence do not need multiple continuous derivatives.  In order to implement the coarse-graining method, we must create an Eulerian mesh on which to spread the Lagrangian (particle-centered) quantities.  We could set the coarse-grained mesh width, $h^{cg}$, to match that of the fluid grid, i.e. $h^{cg} = h$, or we could set it to a coarser value, i.e. $h^{cg} > h$, to reduce the computational cost of the coarse-graining evaluation.  The function we selected allows us to perfectly conserve quantities when using coarser values of $h^{cg}$.  More precisely, this coarse-graining function conserves quantities as long as $w$ is an integer multiple of the coarse-graining mesh size, $h^{cg}$, i.e. $w = n h^{cg}$ for $n \in \mathbb{Z}$.  We have used $w = 3 h^{cg}$ in our analysis, whereas the Gaussian or polynomial functions, on the other hand, would require smaller values for $h^{cg}$, such as $w = 10 h^{cg}$, in order to get closer to conserving the spread quantities.

The coarse-graining width, $w$, determines the distance over which the particle-centered quantities are spread.  \citet{Weinhart2013} studied the sensitivity of results to $w$, finding that they did not change appreciably under two regimes: the sub-particle scale $w \approx 0.05 D_p$ and the particle scale $w \approx D_p$.  We used the latter scale in order to generate smooth continuum fields from the particle quantities.

\subsection{Handling boundaries} \label{sec:cg_boundaries}

\begin{figure}
\centering
\includegraphics[width=0.5\textwidth]{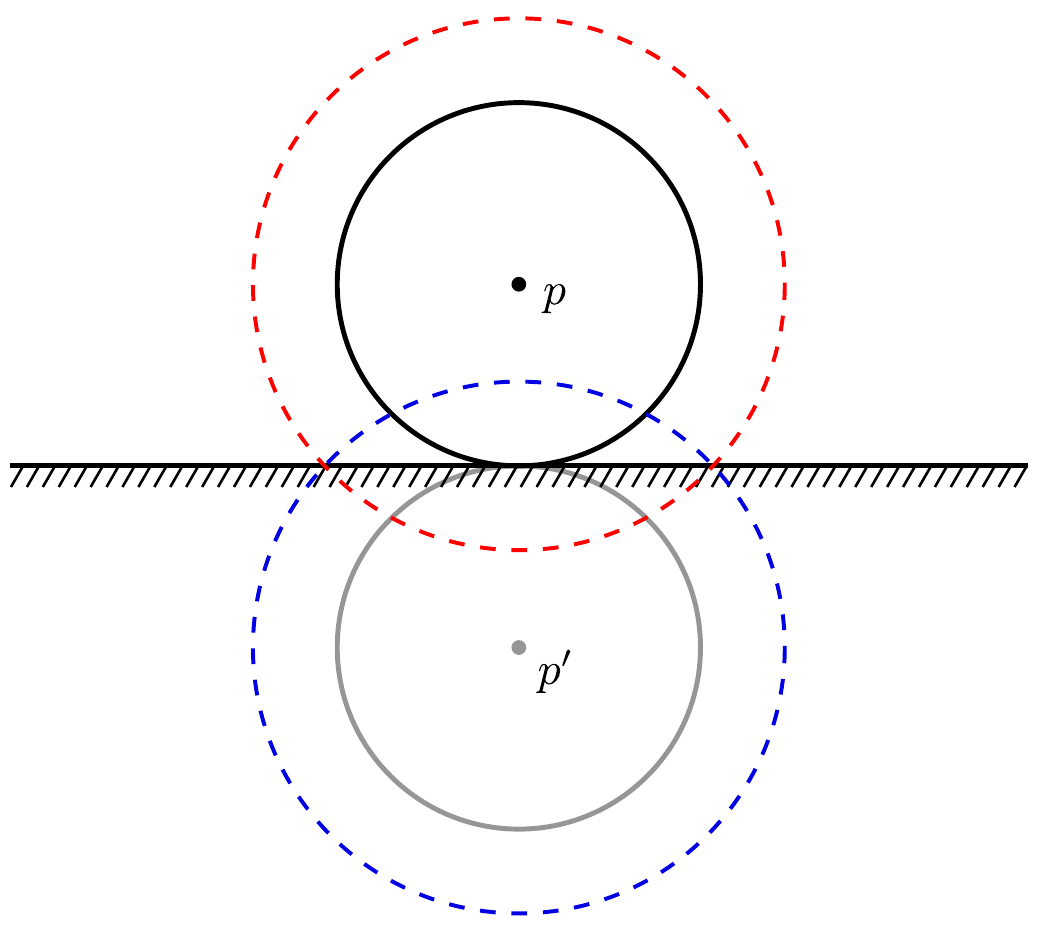}
\caption{Reflection of coarse-grained quantities near a wall, whose radii of influence are represented by dashed lines.}  \label{fig:cg_wall}
\end{figure}

When particles approach boundaries, some of their coarse-grained data can be lost due to the coarse-graining function \eqref{eq:cg_function} spreading information beyond the wall.  For example, consider the coarse-grained representation of the volume fraction for particle $p$, which is sitting on the wall and has a coarse-grained width $1.5w > R_p$, as shown in figure~\ref{fig:cg_wall}.  The red dashed line shows the area over which the mass, and hence volume fraction, will be spread.  Because a portion of the mass is spread below the wall, it will not be accounted for when taking spatial averages within the domain, and the overall volume fraction will be underrepresented near the wall.
We can account for this lost mass using the method of \citet{Zhu2002} and \citet{Sun2015}, who extended it to consider corners of boundaries, by reflecting this particle across the wall (represented by the gray particle) and including the coarse-grained values from this reflected particle (represented by the blue dashed circle).  We employ this method at the particle/fluid interface as well, creating an artificial wall at $y=y_p$, where $y_p$ is the height of the particle bed, only when calculating the coarse-grained particle velocity field $\mbs{u}^\mathit{cg}$.

\bibliographystyle{jfm}
\bibliography{library}

\end{document}